\documentclass[12pt,aps,tightenlines]{revtex4}
\pdfoutput=1

\usepackage{amsmath}
\usepackage{amsfonts}
\usepackage{bm}
\usepackage{euscript}
\usepackage{wasysym}
\usepackage{graphicx}
\usepackage{subfigure}
\usepackage{color}
\usepackage{lscape}
\usepackage{rotating}
\usepackage{framed}
\usepackage[english]{babel}

\graphicspath{{figs/}}

\newcommand{\Ttensor}{\mathsf{T}}

\newcommand{\Ltensor}{\mathsf{L}}
\newcommand{\Mtensor}{\mathsf{M}}
\newcommand{\rtensor}{\mathsf{r}}
\newcommand{\ntensor}{\mathsf{n}}

\newcommand{\mathd}{\mathrm{d}}

\newcommand{\Diff}{\mathrm{D}}

\begin{document}
%+Title
\title{Interfacial instability of turbulent two-phase stratified flow:
 Multi-equation turbulent modelling with rapid distortion}
\author{Lennon \'O N\'araigh$\,^1$\footnote{Corresponding author. Email:lennon.o-naraigh05@imperial.ac.uk},
P. D. M. Spelt$\,^1$, O. K. Matar$\,^1$, and T. A. Zaki$\,^2$}
\affiliation{Departments of Chemical$\,^1$ and Mechanical$\,^2$ Engineering,
South Kensington Campus, Imperial College London, SW7 2AZ}
%
% \author{Lennon \'O N\'araigh$\,$\footnote{Corresponding author. Email:lennon.o-naraigh05@imperial.ac.uk},\\
% Department of Chemical Engineering,\\South Kensington Campus,\\
% Imperial College London, SW7 2AZ}
%
%
\date{\today}
%-Title
%
%

\begin{abstract}
We investigate the linear stability of a flat interface that separates a
liquid layer from a fully-developed turbulent gas flow.  In this context,
linear-stability analysis involves the study of the dynamics of a small-amplitude
wave on
the interface, and we develop a model that describes wave-induced perturbation
turbulent stresses (PTS).  We demonstrate the effect of the PTS on the stability
properties of the system in two cases: for a laminar thin film, and for deep-water
waves.  In the first case, we find that the PTS have little effect on the
growth rate of the waves, although they do affect the structure of the perturbation
velocities.  In the second case, the PTS enhance the maximum growth rate,
although the overall shape of the dispersion curve is unchanged.  Again,
the PTS modify the structure of the velocity field, especially at longer
wavelengths.  Finally, we demonstrate a kind of parameter tuning that enables
the production of the thin-film (slow) waves in a deep-water setting.
\end{abstract}

\maketitle
\section{Introduction}
\label{sec:intro}

Understanding the stability of a flat interface separating the liquid
and gas phases in a turbulent boundary layer is a long-standing problem~\cite{OceanwavesBook}.
  In this paper, we focus in detail on one aspect of this problem, namely
  modelling the perturbation
 turbulent stresses that arise when a wave on the interface interacts with
 the turbulence and generates Reynolds stresses.  The approach we take has
 multiple facets: we derive a model velocity field describing the base turbulent
 flow in the system, and we then perform a linear-stability analysis around
 this base state using the Orr--Sommerfeld formalism.  We construct a detailed
 model to describe the turbulent stresses that appear in these linear-stability
 equations.  Motivated by the paper of Ierley and Miles~\cite{Miles2001},
 we use a model that linearly couples the streamfunction to the turbulent
 kinetic energy and Reynolds stresses, and divides the gas domain into near-equilibrium
 and rapid-distortion regions.

The purpose of our investigation is twofold.  In the present paper, we develop
a model that describes in detail the averaged effects of turbulence on the
linear stability of the gas-liquid interface.  We carry out some preliminary
calculations for two archetypal cases: the thin liquid film sheared by a
turbulent gas flow, and the generation of interfacial waves over deep water.  This model forms the basis for a more detailed study of the thin-film case (in a pressure-driven scenario), which is reported elsewhere~\cite{Onaraigh2_2009}.
%
%
%  In Part II, we use the turbulence model to carry out detailed calculations
%  on the stability properties of the thin-film case (in a pressure-driven
%  scenario), and compare the results
%  with experimental data.  There, we also develop a classification for `slow'
%  and `fast' waves.
%
%
Since the present work focusses on turbulent gas-liquid instability in general,
we place our work in context by describing the
studies that have been performed on similar gas-liquid systems in the past.
\paragraph*{The generation of waves by wind:} The problem of computing the growth rate of an interfacial instability for
a turbulent gas-liquid system was first considered by Jeffreys~\cite{Jeffreys1925},
Phillips~\cite{Phillips1957}, and Miles~\cite{Miles1957}.
 Phillips studied a resonant interaction between turbulent pressure fluctuations
 and capillary-gravity waves on an interface, while Miles focussed on a particular
 kind
 of shear inviscid flow that produces linear instability.
In the paper of Miles, in later works by the same author~\cite{Miles1959,Miles1962b},
and in the work of Benjamin~\cite{Benjamin1958},
the liquid layer is neglected, and is replaced by a wavy wall, and the
problem is reduced to computing the stress exerted on the interface by the
turbulent gas flow. Energy transfer from the gas to the interface is governed
by the second derivative of the mean flow. Indeed, the growth rate of the
instability is determined by the sign of the second derivative of the mean
flow at the critical height -- the height at which the mean flow and wave
speed are equal.
Recent work of Lin \textit{et al.}~\cite{lin2008} indicates that the Phillips mechanism may be important in an initial regime, leading to wave growth that is approximately linear in time. Their results show that this is followed by an exponential growth regime, which is primarily governed by the disturbances in the flow induced by the waves themselves. In this paper, we focus on this latter stage; we shall investigate the effects of the presence of short waves (possibly induced by a Phillips-type mechanism) on this exponential wave growth regime in a later, more detailed, study.

\paragraph*{Modelling Reynolds stresses:}  In these early works~\cite{Miles1957,Miles1959,Miles1962b,Benjamin1958}, the turbulent nature of the gas flow is taken into account
through the prescription of a logarithmic mean profile in the gas. Nevertheless,
the Reynolds stress terms that enter into the stability equations are ignored.
This problem is rectified by Van Duin and Janssen~\cite{Janssen1992}, and by Belcher and co-workers
in a series of papers~\cite{Belcher1993, Belcher1994,Belcher1998}.
The latter group studies the interfacial stability of a sheared
air-water interface, and specialize to an air-water system for oceanographical
applications.  With the exception of~\cite{Belcher1994}, they treat the interface in a manner
similar to Miles.  Particular care is taken in
developing an understanding of the structure of the turbulent
shear stresses inherent in the problem through the use of scaling arguments
and a
truncated mixing-length model.  This is representative of an approximation
of a Reynolds-averaged ensemble of realizations of the turbulent flow for
a given phase of a small-amplitude interfacial wave.  In this approach, an
eddy viscosity is formulated in terms of the typical scale of a turbulent
eddy, which depends
on the distance between the eddy itself and the air-water interface. Far
from the interface, the turbulent eddies are advected quickly over an interfacial
undulation, and have insufficient time to equilibrate, and so-called rapid-distortion
theory is needed.  Such an approach was developed by Townsend~\cite{Townsend1979}.
In the papers of Belcher and co-workers, however, the far-field region is simply modelled by the Rayleigh
equation.  Thus, the mixing-length is truncated: it is a simple function
of the vertical co-ordinate close to the interface, and is set to zero far
from the interface.  We propose instead to follow the approach of Townsend~\cite{Townsend1979}
and Ierley and Miles~\cite{Miles2001}: not only do we interpolate between
the turbulent domains (a common factor between all these papers), but we
also explicitly model the rapid-distortion region.  Our model is based on
the linearized closure model of the Reynolds-averaged equations pioneered
by Launder \textit{et al.}~\cite{LRR1975} and discussed in the book by Pope~\cite{TurbulencePope}.

\paragraph*{Numerical approaches:} The work discussed so far uses asymptotic
techniques for the determination of stability.  Numerical methods provide
more information on the stability, since they are valid over all parameter
ranges, and since they readily enable the re-construction of the velocity and pressure
fields from the solution of an eigenvalue problem.
The paper of Boomkamp and Miesen~\cite{Boomkamp1996} provides yet more information, since they classify interfacial instabilities
according to an energy budget: the contributions to the time-change
in kinetic energy are derived from the linearized Navier--Stokes equations,
and several destabilizing factors are identified. Of interest in the present
application are the critical-layer mechanism and the viscosity-contrast mechanism.
In this approach~\cite{Miesen1995,Boomkamp1996}, and in similar work by
\"Ozgen \cite{Ozgen1998,Ozgen2008},
the viscous liquid layer and gas layers are modelled co-equally; these added
complications
necessitate a numerical solution of the equation. The interfacial
wave speed is then determined, along with the growth rate, as the solution
of
an eigenvalue problem. Turbulence enters the problem only though the logarithmic
mean-flow profile chosen, and the Reynolds stress terms are ignored.

\paragraph*{Modelling of the base flow:}  The papers mentioned so far
use a model of the base flow that captures its logarithmic shape.  These
base models contain a free parameter, however, since the interfacial friction velocity
$U_{*\mathrm{i}}$ is undetermined.  In the approach we take, we use a closed model that
fixes this parameter as a function of the Reynolds number.  Our formalism
is based on that introduced by Biberg~\cite{Biberg2007}, although it differs
in one key respect: we pay close attention to the near-interface
modelling (and also to the modelling near the upper wall, if present), and remove the singularities that arise when the logarithm in
the velocity is evaluated an upper wall or an interface.  This is accomplished using a Van Driest
interpolation~\cite{TurbulencePope, demekhin2006}, in which the velocity field smoothly transitions between the
log layer and the viscous near-wall (interfacial) region.

These are the ingredients of our model, which we develop in the following
way.  In Sec.~\ref{sec:model} we provide a detailed description of the rapid-distortion
and near-equilibrium turbulence which we constitute in terms of the streamfunction.
 We introduce the Orr--Sommerfeld equation for the streamfunction, which
 couples to the turbulent stresses.
  In Sec.~\ref{sec:numerics} we describe the numerical method used to solve the model
  equations.  In Sec.~\ref{sec:thif} we describe the effects of the turbulent
  modelling on a thin-film flow (this problem was studied in the absence
  of perturbation turbulent stresses by Miesen and Boersma~\cite{Miesen1995}).  We find that the growth rate is
  not significantly affected by the turbulence modelling, although the structure
  of the flow field is.  Nevertheless, for deep-water waves (Sec.~\ref{sec:deep_water}),
  both the growth rate and the flow structure are modified by the turbulence
  (the growth rate is enhanced).  In this deep-water study,
  we compare our results with direct numerical simulation, and investigate
  a mechanism
  for generating thin-film type waves in a deep water setting.  Finally,
  in Sec.~\ref{sec:conc} we present our conclusions.

\section{Theoretical formulation}
\label{sec:model}

\begin{figure}[htb]
\centering\noindent
\includegraphics[width=0.7\textwidth]{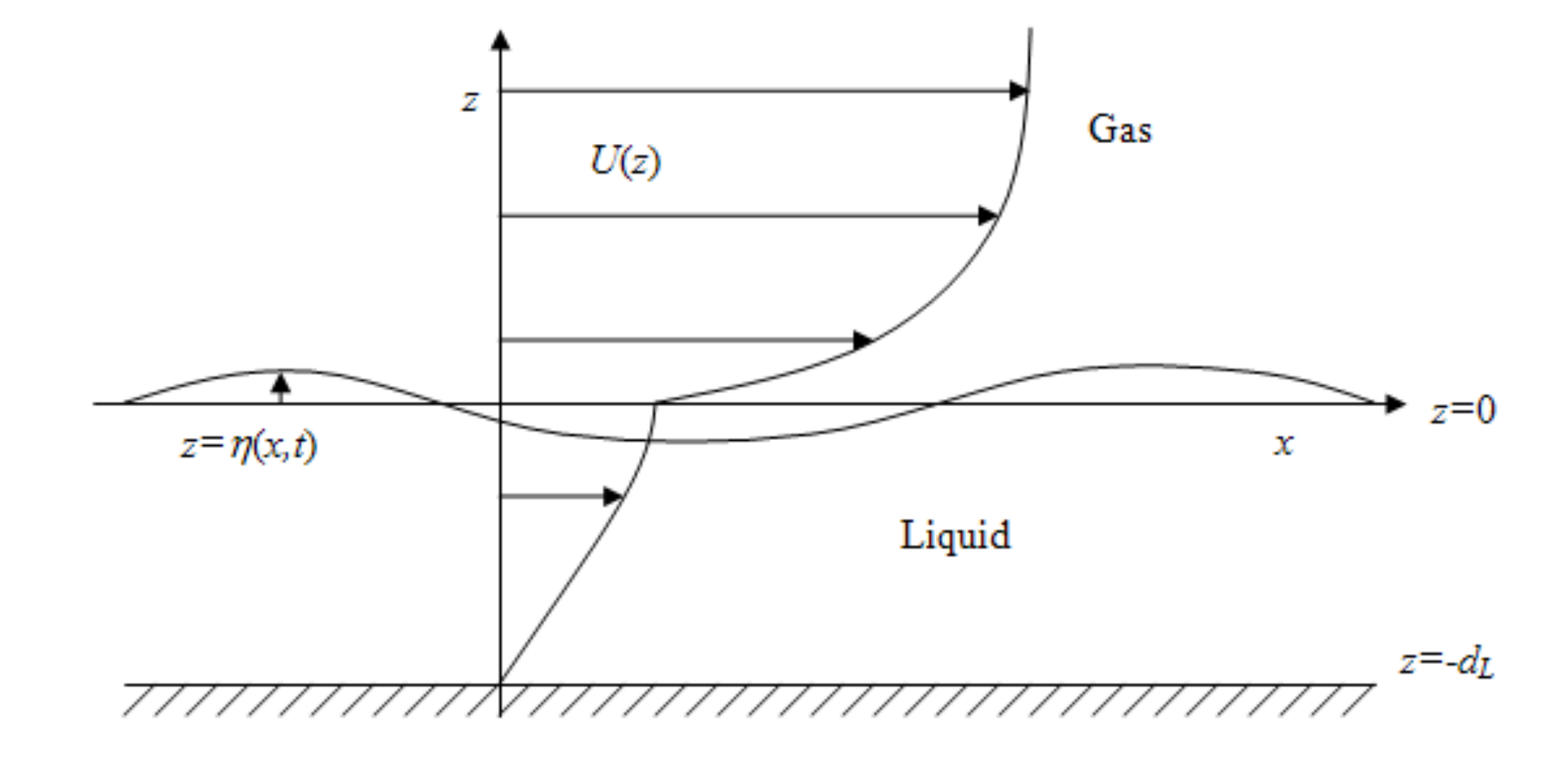}
\caption{A schematic diagram showing the system of interest.  There is turbulent
shear flow in the gas layer, while the liquid layer either represents a thin
laminar film, or deep-water flow, as in oceans.}
\label{fig:schematic}
\end{figure}

In this section we provide a problem description and develop a turbulence
model
that takes account
of rapid distortion and near-equilibrium regions in the turbulent stratified
two-phase flow.  Our approach is inspired by the paper of Miles~\cite{Miles2001}.
 It takes into account the interactions between interfacial waves and the
 turbulence.  The work here is, however, multi-faceted: there is a model
 of the flat-interface turbulent state, together with a linear-stability
 analysis around this state, in which the effects of the turbulence are modelled in
 detail.
The two-layer system we consider has the same structure as boundary-layer
flow, and is
described schematically in Fig.~\ref{fig:schematic}.  A rectangular co-ordinate
system is used to model the flow, with the plane $z=0$ coinciding with the
undisturbed location of the interface. Since the interfacial
waves that develop are two-dimensional, we restrict to a two-dimensional
system, in which the bottom layer is a liquid of depth $d_L$, which in this paper will either be
a thin laminar layer, or a deep-water flow ($d_L=\infty$).  The top
layer is gaseous, turbulent and fully-developed. A shear stress
is applied along the channel. The mean profile is thus a
uni-directional flow in the horizontal, $x$-direction.
In the gas layer, near the gas-liquid interface and the gas-wall
boundary, the flow is dominated by molecular viscosity, since here the viscous scale exceeds
the characteristic length scale of the turbulence~\cite{TurbulenceMonin,TurbulencePope}.
In the bulk of the gas region, the flow possesses a logarithmic
profile~\cite{TurbulenceMonin,TurbulencePope}.
Finally, we assume that the gas-liquid interface is smooth.
\subsection{The perturbation equations}

Following standard practice dating back to Miles~\cite{Miles1957}, we base
the dynamical equations for the interfacial motion on the Reynolds-averaged
Navier--Stokes (RANS) equations.  The turbulent velocity is decomposed into averaged
and fluctuating parts $\bm{U}=\left(U,W\right)$ and $\left(u',w'\right)$ respectively.
 The averaged velocity depends on space and time through the RANS equations:
\begin{subequations}
\begin{equation}
\rho\left(\frac{\partial U_i}{\partial t}+\bm{U}\cdot\nabla U_i\right)=-\frac{\partial
P}{\partial x_i}+\mu\Delta
U_i-\rho\left(\frac{\partial}{\partial x}\langle u' u_i'\rangle+\frac{\partial}{\partial
z}\langle w' u_i'\rangle\right),\qquad i=x,z,
\label{eq:rans}
\end{equation}
\begin{equation}
\nabla\cdot\bm{U}=0,
\end{equation}%
\end{subequations}%
where $\langle\cdot\rangle$ denotes the averaging process.  The constants
$\rho$ and $\mu$ denote the density and viscosity respectively.  In addition
to viscous stress, Eq.~\eqref{eq:rans} contains a stress term that arises
due to the interaction between the turbulence and the mean flow.  We use
these equations to model a flat-interface or base state of the two-phase
system shown in Fig.~\ref{fig:schematic}.  Next, we introduce a small
disturbance that shifts
%
%
%
% The first step in our analysis is to define the flat-interface or base
% state and obtain  dynamical equations that describe the perturbations around
% this state.  A small disturbance shifts
%
%
the flat interface
at $z=0$ to $z=\eta$, where $|\eta|\ll1$, and this induces a change in the
average velocity and pressure fields, denoted as follows:
\[
\left(U,W,P\right)=\left(U_0\left(z\right)+\delta u\left(x,z,t\right),\delta
w\left(x,z,t\right),P_0\left(x,z\right)+\delta p\left(x,z,t\right)\right),
\]
where we denote base-state quantities by a subscript zero.
Since the flow is turbulent, and since the perturbations take the form of
waves, they must satisfy the RANS equations for a linear wave of speed $c$, $\partial_t=-c\partial_x$:
\begin{subequations}
\begin{equation}
\rho\left[\left(U_0-c\right)\frac{\partial}{\partial x} \delta u+\frac{\mathd{U_0}}{\mathd{z}}\delta{w}\right]=
-\frac{\partial}{\partial x}\left(\delta p-\rho\delta\sigma_z\right)+\mu\left(\frac{\partial^2}{\partial
x^2}+\frac{\partial^2}{\partial z^2}\right)\delta u+\rho\frac{\partial}{\partial
x}\delta\sigma+\rho\frac{\partial}{\partial
z}\delta\tau,
\end{equation}
\begin{equation}
\rho\left(U_0-c\right)\frac{\partial}{\partial x}\delta w=-\frac{\partial}{\partial
z}\left(\delta p-\rho\delta\sigma_z\right)+\mu\left(\frac{\partial^2}{\partial x^2}+\frac{\partial^2}{\partial
z^2}\right)\delta w+\rho\frac{\partial}{\partial x}\delta\tau,
\end{equation}
\begin{equation}
\frac{\partial}{\partial x}\delta u+\frac{\partial}{\partial z}\delta w=0.
\end{equation}%
\end{subequations}%
The quantities
\[
\delta\tau=-\langle u' w'\rangle-\tau^{(0)},\,\,
\delta\sigma_x=-\langle u'^2\rangle-\sigma_x^{(0)},\,\,
\delta\sigma_z=-\langle w'^2\rangle-\sigma_z^{(0)},\,\,\delta\sigma=\delta\sigma_x-\delta\sigma_z
\]
are the perturbation stresses due to the turbulence in the perturbed state,
while the quantities with the zero-superscript are base-state stresses.
 Using the streamfunction representation $\left(\delta u,\delta w\right)=\left(\phi_z,-\phi_x\right)$,
 and the normal-mode decomposition $\partial_x=i\alpha$, the perturbed
 RANS equations reduce to a single equation:
%
%
%\begin{framed}
\begin{equation}
i\alpha\rho\left[\left(U_0-c\right)\Diff^2\phi-\frac{\mathd^2U_0}{\mathd z^2}\phi\right]=\mu\left(\Diff^2-\alpha^2\right)^2\phi+\rho
i\alpha\Diff\delta\sigma+\left(\Diff^2+\alpha^2\right)\delta\tau,
\label{eq:os_turb}
\end{equation}
%\end{framed}
%
%
\noindent where $\Diff=\mathd/\mathd z$.
\subsection{Separation of domains}
To model the turbulent stresses in detail, we need to understand
the features of the turbulence.  Because the instabilities we observe arise
due to conditions in the gas, we pay particular attention to the gas layer.
If the liquid layer is deep, then the standard Orr--Sommerfeld equation gives
an appropriate description of the flow there, provided an accurate base-state
profile is supplied~\cite{Boomkamp1996,Zeisel2007}, while if the liquid resides in a thin layer, the Orr--Sommerfeld
equation
gives an exact description of the flow.  Thus, in both cases, we specify the
standard Orr--Sommerfeld equation in the liquid, without the additional stresses.
The turbulence in the gas is characterised by two timescales:
the eddy turnover timescale, and the advection timescale.
Roughly
speaking, the eddy turnover or turbulent timescale is the time required for
a typical turbulent eddy to interact with the surrounding fluid and come
into equilibrium (where the production equals dissipation).  An estimate
of this timescale, at a distance $z$ from the interface, is $T_\mathrm{t}=\kappa
z/U_{*\mathrm{i}}$, where $\kappa$ is the Von K\'arm\'an constant, $\kappa=0.41$,
and $U_{*\mathrm{i}}$ is the friction velocity at the interface.  The advection timescale is the time needed for the flow
to advect an eddy over a wave crest.  This flow distorts the turbulence and
moves it away from equilibrium.  An estimate of this timescale is  $T_\mathrm{a}=\left[\alpha\left|U_0\left(z\right)-c\right|\right]^{-1}$,
where $\alpha$ is the wavenumber and $c$ is the complex the wave speed.
  These eddy-turnover and advection effects compete: near the interface, $T_\mathrm{t}$
  is small compared with $T_{\mathrm{a}}$, that is, the frequency of turbulent
  interactions is large compared with the frequency of advection events.
   This is the region of near-equilibrium where by definition eddy viscosity
   and one-equation
   turbulent closures are expected to be appropriate.  Far away from the interface, $T_\mathrm{t}$
   is large compared with the advection timescale (at least for the large,
   most energetic turbulent structures), or equivalently, the frequency of
   turbulent interactions is small compared with the frequency of advection
   events.  This is the region where rapid-distortion theory is expected
   to apply~\cite{Belcher1993, Belcher1994,Belcher1998}.
    In an obvious way, we call the region where eddy viscosity and one-equation
    formalisms are appropriate the \textit{near-interface region}, while
    the region where the rapid-distortion calculations work is called the
    \textit{far-field region}.  Crossover occurs at $z=z_\mathrm{t}$, for
    which $T_\mathrm{t}\left(z_\mathrm{t}\right)=T_\mathrm{a}\left(z_\mathrm{t}\right)$.
     For further discussion of this separation of
    domains, see~\cite{Cohen1998, Belcher1998}.

\subsection{Turbulence modelling in the distinct domains}
\label{subsec:turb_domains}
In modelling the perturbation turbulent stresses (PTS), we are particularly
interested in  the anisotropy tensor
\begin{equation}
\ntensor_{ij}=\frac{\langle u'_i u'_j\rangle}{\langle u'_ku'_k\rangle }
=\frac{\rtensor_{ij}}{q}
%-\delta_{ij}d^{-1}
=-\left(\begin{array}{cc}\frac{\sigma_x}{q}&\frac{\tau}{q}\\
\frac{\tau}{q}&\frac{\sigma_z}{q}\end{array}\right),
\label{eq:nij}
\end{equation}
where $q=2k=\mathrm{tr}\left(\rtensor_{ij}\right)$, and $k$ is the
turbulent kinetic energy.
The exact equations for the fluctuating component of the velocity in the RANS equations
are
\begin{equation}
\left(\frac{\partial}{\partial t}+\bm{U}\cdot\nabla\right)u'_i=\frac{\partial\rtensor_{ik}}{\partial
x_k}-u'_k\frac{\partial U_i}{\partial x_k}-u_k'\frac{\partial u_i'}{\partial x_k}+\nu\nabla^2u_i'-\frac{1}{\rho}\frac{\partial p'}{\partial
x_i},\qquad \nu=\frac{\mu}{\rho}.
\end{equation}
In the far field, the mean strain produced by the wave is large.  We therefore
study the equations of motion in the limit of strong mean strain.  In that
case the terms involving gradients of fluctuating quantities drop out:
\begin{equation}
\left(\frac{\partial}{\partial t}+\bm{U}\cdot\nabla\right)u'_i=-u'_k\frac{\partial
U_i}{\partial x_k}-\frac{1}{\rho}\frac{\partial p^{(r)}}{\partial
x_i},
\label{eq:rapid_evolution}
\end{equation}
where
\[
\frac{1}{\rho}\nabla^2 p^{(r)}=-2\frac{\partial U_i}{\partial
x_j}\frac{\partial u'_j}{\partial x_i},
\]
with $p^{(r)}$ being the so-called rapid pressure.
By a straightforward manipulation of Eq.~\eqref{eq:rapid_evolution},
%
%
%\begin{equation}
%\left(\frac{\partial}{\partial t}+\langle\bm{U}\rangle\cdot\nabla\right)\rtensor_{ij}+\frac{\partial}{\partial
%x_k}\mathcal{T}_{kij}=\mathcal{P}_{ij}+\mathcal{R}_{ij}^{(r)},
%\label{eq:rapid_tensor0}
%\end{equation}
%
%
%or, by incompressibility,
%
%
\begin{equation}
\frac{\partial\rtensor_{ij}}{\partial t}+\frac{\partial}{\partial
x_k}\left( U_k\rtensor_{ij}+\mathcal{T}_{kij}\right)=\mathcal{P}_{ij}+\mathcal{R}_{ij}^{(r)},
\label{eq:rapid_tensor0}
\end{equation}
in the limit of strong mean strain, where the tensor quantities have the
following meaning:
\begin{itemize}
\item The production of stress by the mean flow:
\[
\mathcal{P}_{ij}=-\rtensor_{ik}\frac{\partial U_j}{\partial x_k}-\rtensor_{jk}\frac{\partial U_i}{\partial x_k};
\]
\item The transport of stress by the rapid pressure:
\[
\mathcal{T}_{kij}=-\frac{1}{\rho}\langle u'_i p^{(r)}\rangle\delta_{jk}-\frac{1}{\rho}\langle u'_j p^{(r)}\rangle\delta_{ik}
\]
\item The pressure rate-of-strain term:
\[
\mathcal{R}_{ij}^{(r)}=\bigg\langle\frac{p^{(r)}}{\rho}\left(\frac{\partial u'_i}{\partial x_j}+\frac{\partial u'_j}{\partial x_i}\right)\bigg\rangle
\]
\end{itemize}
For waves, the shear rate $\alpha\left(U_0-c\right)$ enters into
the advection term in Eq.~\eqref{eq:rapid_tensor0}.  In the rapid-distortion
domain, this rate is large, and thus advective transport dominates
over pressure-driven transport.
 We therefore omit the pressure transport term in~\eqref{eq:rapid_tensor0},
 which now simplifies further:
\begin{equation}
\left(\frac{\partial}{\partial t}+\bm{U}\cdot\nabla\right)\rtensor_{ij}=\mathcal{P}_{ij}+\mathcal{R}_{ij}^{(r)},
\label{eq:rapid_tensor}
\end{equation}
Equation~\eqref{eq:rapid_tensor} contains only one term that is not available
in closed form: the rapid pressure-rate of strain tensor $\mathcal{R}_{ij}^{(r)}$.
 This term has been modelled accurately by
 %Naot, Shavit and Wolfshtein,and by
 Launder, Reece and Rodi~\cite{LRR1975} (see also~\cite{TurbulencePope}).  They use the following form:
\begin{equation}
\mathcal{R}_{ij}^{(\mathrm{r})}=-C_{\mathrm{I}}\left(\mathcal{P}_{ij}-\tfrac{1}{2}\mathcal{P}_{kk}\delta_{ij}\right),
\label{eq:model_r}
\end{equation}
where $C_{\mathrm{I}}=\tfrac{3}{5}$.  This model also
has
the property of linearity in the tensor $\rtensor_{ij}$, which is desirable
in any rapid-distortion theory~\cite{TurbulencePope}.
The advection
equation~\eqref{eq:rapid_tensor} now becomes
\begin{equation}
\frac{\partial \rtensor_{ij}}{\partial t}+\bm{U}\cdot\nabla\rtensor_{ij}=
\left(1-C_{\mathrm{I}}\right)\mathcal{P}_{ij}+\tfrac{1}{2}C_{\mathrm{I}}\mathcal{P}_{kk}\delta_{ij}.
\label{eq:rapid_tensor1}
\end{equation}
Using the definition~\eqref{eq:nij} and Eq.~\eqref{eq:rapid_tensor1}, the
$\ntensor_{ij}$'s satisfy the relation
\[
\frac{\partial \ntensor_{ij}}{\partial t}+\bm{U}\cdot\nabla\ntensor_{ij}=-\left(1-C_{\mathrm{I}}\right)\ntensor_{ik}\frac{\partial
U_j}{\partial x_k}-\left(1-C_{\mathrm{I}}\right)\ntensor_{jk}\frac{\partial U_i}{\partial
x_k}-2\ntensor_{k\ell}\frac{\partial U_\ell}{\partial x_k}\left(\tfrac{1}{2}C_{\mathrm{I}}\delta_{ij}-\ntensor_{ij}\right).
\]
We decompose the tensor $\ntensor_{ij}$ into a flat-interface component and
a perturbation component: $\ntensor_{ij}=\ntensor_{ij}^{(0)}+\delta\ntensor_{ij}$,
where $\ntensor_{ij}^{(0)}$ is the equilibrium anisotropy tensor, and is
thus constant in time.  Hence,
\begin{eqnarray*}
\ntensor_{12}&=&\ntensor_{12}^{(0)}-\frac{1}{q_0}\left(\delta\tau+\ntensor_{12}^{(0)}\delta
q\right),\\
\ntensor_{11}&=&\ntensor_{11}^{(0)}-\frac{1}{q_0}\left(\delta\sigma_x+\ntensor_{11}^{(0)}\delta
q\right),\\
\ntensor_{22}&=&\ntensor_{22}^{(0)}-\frac{1}{q_0}\left(\delta\sigma_z+\ntensor_{22}^{(0)}\delta
q\right),
\end{eqnarray*}
where $q_0\left(z\right)$ represents twice the turbulent kinetic energy associated
with the base state and $\delta q$ a small-amplitude perturbation about this
state.
%
% From the rapid distortion of stress equation~\eqref{eq:stress_rapid},
% %
% \[
% \frac{\partial\mathrm{n}_{ij}}{\partial t}+\bm{U}\cdot\nabla\mathrm{n}_{ij}=
% -\mathrm{n}_{ik}\frac{\partial U_k}{\partial x_j}-\mathrm{n}_{jk}\frac{\partial
% U_k}{\partial x_i}+2\mathrm{n}_{ij}\mathrm{n}_{\ell k}\frac{\partial U_k}{\partial
% x_\ell}+\frac{\mathcal{R}^{(\mathrm{r})}_{ij}}{q}
% \]
% %
% %
% %
%
%
%
%
%
%
%
%
We therefore have perturbation equations for $\delta\ntensor_{12}$ and $\delta\ntensor:=\delta\ntensor_{11}-\delta\ntensor_{22}$:
\begin{eqnarray}
\left(\frac{\partial}{\partial t}+\bm{U}\cdot\nabla\right)\delta\ntensor_{12}&=&
\alpha_1\frac{\partial}{\partial z}\delta{u}+\alpha_2\frac{\partial}{\partial{x}}\delta{w}+\alpha_3\frac{\partial}{\partial{x}}\delta{u}+\alpha_4\delta\ntensor_{12}+\alpha_5\delta\ntensor,\nonumber\\
\left(\frac{\partial}{\partial t}+\bm{U}\cdot\nabla\right)\delta\ntensor&=&
\beta_1\frac{\partial}{\partial z}\delta{u}+\beta_2\frac{\partial}{\partial{x}}\delta{w}+\beta_3\frac{\partial}{\partial{x}}\delta{u}+\beta_4\delta\ntensor_{12}+\beta_5\delta\ntensor.
\label{eq:advection_n}
\end{eqnarray}
Given the base-state shear rate $\Sigma=\mathd U_0/\mathd z$, the coefficients
in Eqs.~\eqref{eq:advection_n} are defined as follows:
\begin{eqnarray*}
\alpha_1&=&2\left(\ntensor^{(0)}_{12}\right)^2-\left(1-C_{\mathrm{I}}\right)\ntensor^{(0)}_{11},\nonumber\\
\alpha_2&=&2\left(\ntensor^{(0)}_{12}\right)^2-\left(1-C_{\mathrm{I}}\right)\ntensor^{(0)}_{22},\nonumber\\
\alpha_3&=&2\ntensor^{(0)}_{12}\left(\ntensor^{(0)}_{11}-\ntensor^{(0)}_{22}\right),\nonumber\\
\alpha_4&=&4\Sigma\ntensor^{(0)}_{12},\nonumber\\
\alpha_5&=&\tfrac{1}{2}\Sigma\left(1-C_{\mathrm{I}}\right),
\end{eqnarray*}
\begin{eqnarray*}
\beta_1&=&2\ntensor^{(0)}_{12}\left(\ntensor^{(0)}_{11}-\ntensor^{(0)}_{22}\right)-2\left(1-C_{\mathrm{I}}\right)\ntensor^{(0)}_{12},\nonumber\\
\beta_2&=&2\ntensor^{(0)}_{12}\left(\ntensor^{(0)}_{11}-\ntensor^{(0)}_{22}\right)+2\left(1-C_{\mathrm{I}}\right)\ntensor^{(0)}_{12},\nonumber\\
\beta_3&=&2\left(\ntensor^{(0)}_{11}-\ntensor^{(0)}_{22}\right)^2-2\left(1-C_{\mathrm{I}}\right)\left(\ntensor^{(0)}_{11}+\ntensor^{(0)}_{22}\right),\nonumber\\
\beta_4&=&2\Sigma\ntensor^{(0)}_{12},\nonumber\\
\beta_5&=&2\Sigma\left[\left(\ntensor^{(0)}_{11}-\ntensor^{(0)}_{22}\right)-2\left(1-C_{\mathrm{I}}\right)\right].
\label{eq:coeff}
\end{eqnarray*}
Far from the interface, the
effects of rapid distortion due to the base flow are negligible, since the
shear rate $\Sigma\left(z\right)$ is small there; thus $\alpha_4$, $\alpha_5$,
$\beta_4$, and $\beta_5$ are small and are neglected in this formalism.
In Secs.~\ref{sec:thif} and~\ref{sec:deep_water}, we estimate the $\ntensor_{ij}^{(0)}$'s
by constant values obtained from the properties of log layer in turbulent
shear flow.  Then, Eq.~\eqref{eq:advection_n}, in normal-mode form, reduces
to a set of equations that are algebraic in $\delta\ntensor_{12}$ and $\delta\ntensor$:
\begin{eqnarray}
i\alpha\left(U_0-c\right)\delta\ntensor_{12}&=&
\alpha_1\frac{\partial}{\partial z}\delta{u}+\alpha_2\frac{\partial}{\partial{x}}\delta{w}+\alpha_3\frac{\partial}{\partial{x}}\delta{u}
,\nonumber\\
i\alpha\left(U_0-c\right)\delta\ntensor&=&
\beta_1\frac{\partial}{\partial z}\delta{u}+\beta_2\frac{\partial}{\partial{x}}\delta{w}+\beta_3\frac{\partial}{\partial{x}}\delta{u}.
\label{eq:advection_n_algebraic}
\end{eqnarray}
The right-hand side of both these equations can easily be recast in terms of the streamfunction, using $\left(\partial/\partial z\right)\delta u=\Diff^2\phi$ etc.  However, to highlight the effect of the distortion of the Reynolds stresses by the perturbation velocity, we leave Eq.~\eqref{eq:advection_n_algebraic} in its present form throughout the paper.

The derivation of a near-field model is less involved.  Near the interface,
the turbulence tends quickly to equilibrium, and the $\ntensor_{ij}$'s take
their constant value.  Thus,
\begin{equation}
\delta\ntensor_{ij}=0.
\label{eq:n_ij_near}
\end{equation}
Next, we develop a strategy to join up these domains and the models~\eqref{eq:advection_n}
and~\eqref{eq:n_ij_near} into a single formalism.  This requires a detailed picture
of the turbulent kinetic energy.

\subsection{Turbulent kinetic energy}

The governing equation for the perturbation turbulent kinetic energy
$\delta k=\tfrac{1}{2}\langle u'^2+w'^2\rangle-k_0$ can be written as:
\[
\frac{\partial}{\partial t}\delta k+\bm{U}_0\cdot\nabla\delta k+\delta\bm{U}\cdot\nabla
k=-\nabla\cdot\left(\delta\bm{J}\right)+\delta\mathcal{P}-\delta\varepsilon,
\]
where the terms are explained in detail in what follows.
Using the properties of the basic state, this equation reduces to
\[
\frac{\partial}{\partial t}\delta k+U_0\frac{\partial}{\partial x}\delta k+\delta w\frac{\mathd
k_0}{\mathd z}=-\nabla\cdot\left(\delta\bm{J}\right)+\delta\mathcal{P}-\delta\varepsilon.
\]
We now turn to the problem of modelling the terms on the right-hand side
of this equation.
\begin{itemize}
\item\textit{The energy flux:}  In Townsend~\cite{Townsend1972,Townsend1979} and Ierley and Miles~\cite{Miles2001}, the energy flux $\delta\bm{J}$ is ignored, since turbulent
transport is negligible in the bulk gas flow.  It is important, however, near
the interface, where molecular viscosity dominates.  We therefore take $\delta\bm{J}=-\nu\nabla\delta
k$.
\item\textit{Dissipation:}  In previous works,~\cite{Townsend1972,Townsend1979,Miles2001}, the dissipation function $\delta\varepsilon$ in the bulk gas
region was constructed
as a cumbersome function of $k_0$ and the mixing length.  In this work,
we make use of the simple model
\[
\delta\varepsilon=\frac{1}{\chi}\delta k,
\]
where $\chi$ is a timescale constructed from dimensional analysis:
\[
\chi=\frac{\nu}{U_{*\mathrm{i}}^2},
\]
We have verified that the choice of form for this relation makes little difference to the results of the stability
calculation, and we therefore settle on the linear form, which has the added advantage that $\chi^{-1}$ does not diverge at $z=0$.

% since then although it is desirable from a mathematical point of view,
% since the decay function $\chi^{-1}$ does not diverge near the interface.
%
%
%
\item\textit{Production:}  The production of turbulent kinetic energy is
half the trace of $\mathcal{P}_{ij}$,
%
%$=-\rtensor_{ik}\left(\partial U_j/\partial x_k\right)-\rtensor_{jk}\left(\partial{U_i}/\partial{x_k}\right)$
%
and thus the perturbed production rate has the form
\[
\delta\mathcal{P}=-\delta\rtensor_{ij}\frac{\partial U_i^{(0)}}{\partial x_j}-\rtensor_{ij}^{(0)}\frac{\partial}{\partial
x_j}\delta U_i,
\]
which can be re-expressed as
\[
\delta\mathcal{P}=\delta\tau\frac{\mathd U_0}{\mathd z}+\tau^{(0)}\left(\Diff^2+\alpha^2\right)\phi+i\alpha\Diff\phi\left(\sigma_x^{(0)}-\sigma_z^{(0)}\right),
\]
where $\delta\tau=-\langle u' w'\rangle-\tau^{(0)}$ is the perturbed turbulent
shear stress, and $\tau^{(0)}$, $\sigma_x^{(0)}$, and $\sigma_z^{(0)}$ represent
the turbulent stresses in the flat-interface state.  Thus, we obtain the
following kinetic-energy equation:
\begin{multline*}
i\alpha\left(U_0-c\right)\delta k+\chi^{-1}\delta k\\=\nu\left(\Diff^2-\alpha^2\right)\delta
k-\delta\rtensor_{12}\frac{\mathd U_0}{\mathd z}-\rtensor_{12}^{(0)}\left(\Diff^2+\alpha^2\right)\phi-i\alpha\rtensor^{(0)}\Diff\phi+i\alpha\frac{\mathd
k_0}{\mathd z}\phi,
\end{multline*}
where $\rtensor^{(0)}=-\left(\sigma_x^{(0)}-\sigma_z^{(0)}\right)$ and $\rtensor_{12}^{(0)}=-\tau^{(0)}$.
\end{itemize}

\subsection{Interpolation between domains}

The algebraic model for the tensor $\delta\ntensor_{ij}$ is given by Eq.~\eqref{eq:advection_n_algebraic}
in the far field, and by Eq.~\eqref{eq:n_ij_near} in the near field.  We can interpolate
between these domains by using a hybrid stress model:
\begin{equation*}
\left[i\alpha\left(U_0-c\right)\mathcal{I}\left(z\right)+\left(1-\mathcal{I}\left(z\right)\right)\right]\delta\ntensor_{12}=
%\left(1-\mathcal{I}\left(z\right)\right)\frac{\left(c^2/2\right)^{1/2}\ell_{\mathrm{m}}}{q^{1/2}}\delta{S}_{12}\\
\mathcal{I}\left(z\right)\left(\alpha_1\frac{\partial}{\partial z}\delta{u}+\alpha_2\frac{\partial}{\partial{x}}\delta{w}+\alpha_3\frac{\partial}{\partial{x}}\delta{u}\right),
\end{equation*}
\begin{equation*}
\left[i\alpha\left(U_0-c\right)\mathcal{I}\left(z\right)+\left(1-\mathcal{I}\left(z\right)\right)\right]\delta\ntensor=
%-\left(1-\mathcal{I}\left(z\right)\right)\frac{\left(c^2/2\right)^{1/2}\ell_{\mathrm{m}}}{q^{1/2}}\left(\delta{S}_{11}-\delta{S}_{22}\right)\\
\mathcal{I}\left(z\right)\left(\beta_1\frac{\partial}{\partial z}\delta{u}+\beta_2\frac{\partial}{\partial{x}}\delta{w}+\beta_3\frac{\partial}{\partial{x}}\delta{u}\right),
\end{equation*}
\noindent where $\mathcal{I}\left(z\right)$ is an interpolating function
that is zero at $z=0$ and asymptotes to $\mathcal{I}=1$, with a characteristic
lengthscale $z_{\mathrm{t}}$.

We now have all the components to assemble the model, which we present below
in non-dimensional form.  We
non-dimensionalize on as-yet undefined length and velocity scales $h$ and
$V$ respectively, and the gas viscosity,
which gives rise to the gas Reynolds number $Re=V h/\nu_G$.
In the liquid, a single equation describes the flow, which in non-dimensional
form is simply
\begin{subequations}
%\begin{framed}
\begin{equation}
i\alpha r\left[\left(U_L-c\right)\left(\Diff^2-\alpha^2\right)\phi_L-\frac{\mathd^2U_L}{\mathd z^2}\phi_L\right]=\frac{m}{Re}\left(\Diff^2-\alpha^2\right)^2\phi_L,\qquad
z\leq 0,
\end{equation}
%\end{framed}
%
%
where the subscripts `L' and `G' denote evaluation in the gas and liquid
phases respectively: $U_L$ and $U_G$ are the base-state velocities in the
liquid and gas phases, and $r=\rho_L/\rho_G$ and $m=\mu_L/\mu_G$ are the
density and viscosity ratios.
In the turbulent gas, we must solve the following coupled
set of equations:
%
%
%
%\begin{framed}
\begin{multline}
i\alpha\left[\left(U_G-c\right)\left(\Diff^2-\alpha^2\right)\phi_G-\frac{\mathd^2U_G}{\mathd z^2}\phi_G\right]=
\frac{1}{Re}\left(\Diff^2-\alpha^2\right)^2\phi_G
-i\alpha\Diff\delta\rtensor-\left(\Diff^2+\alpha^2\right)\delta\rtensor_{12},
\end{multline}
\begin{multline}
\left[i\alpha\left(U_G-c\right)+\frac{Re_*^2}{Re}\right]\delta k
\\=\frac{1}{Re}\left(\Diff^2-\alpha^2\right)\delta
k-\delta\rtensor_{12}\frac{\mathd U_G}{\mathd z}-\rtensor_{12}^{(0)}\left(\Diff^2+\alpha^2\right)\phi_G-i\alpha\rtensor^{(0)}\Diff\phi+i\alpha\frac{\mathd
k_0}{\mathd z}\phi_G,
\end{multline}
where $Re_*=U_{*\mathrm{i}}h/\nu_G$, the Reynolds number based on the interfacial friction velocity $U_{*\mathrm{i}}$, and where
\begin{multline}
\left[i\alpha\left(U_G-c\right)\mathcal{I}\left(z\right)+\left(1-\mathcal{I}\left(z\right)\right)\right]\delta\rtensor_{12}
-\frac{\rtensor_{12}^{(0)}}{q_0}\left[i\alpha\left(U_G-c\right)\mathcal{I}\left(z\right)+\left(1-\mathcal{I}\left(z\right)\right)\right]\delta{q}\\=
%
%
%-\left(1-\mathcal{I}\left(z\right)\right)\frac{\left(2c^2\right)^{1/2}\ell_{\mathrm{m}}}{q^{1/2}}\delta{S}_{12}
%
%
q_0\left(z\right)\mathcal{I}\left(z\right)\left(\alpha_1\frac{\partial}{\partial z}\delta{u}+\alpha_2\frac{\partial}{\partial{x}}\delta{w}+\alpha_3\frac{\partial}{\partial{x}}\delta{u}\right),
\label{eq:rs_interp1}
\end{multline}
\begin{multline}
\left[i\alpha\left(U_G-c\right)\mathcal{I}\left(z\right)+\left(1-\mathcal{I}\left(z\right)\right)\right]\delta\rtensor
-\frac{\rtensor^{(0)}}{q_0}\left[i\alpha\left(U_G-c\right)\mathcal{I}\left(z\right)+\left(1-\mathcal{I}\left(z\right)\right)\right]\delta{q}\\=
%
%
%
%
%-\left(1-\mathcal{I}\left(z\right)\right)\frac{\left(2c^2\right)^{1/2}\ell_{\mathrm{m}}}{q^{1/2}}\left(\delta{S}_{11}-\delta{S}_{22}\right)
%
%
%
%
q_0\left(z\right)\mathcal{I}\left(z\right)\left(\beta_1\frac{\partial}{\partial z}\delta{u}+\beta_2\frac{\partial}{\partial{x}}\delta{w}+\beta_3\frac{\partial}{\partial{x}}\delta{u}\right).
\label{eq:rs_interp2}
\end{multline}
%
%
%\end{framed}
\label{eq:model_eqns}
\end{subequations}
\noindent At the wall $z=-d_L/h$, we have the no-slip conditions on the streamfunction
$\phi_L$,
\[
\phi_L\left(-d_L/h\right)=\Diff\phi_L\left(-d_L/h\right)=0.
\]
 The equations are also matched across the undisturbed interface $z=0$, where we have
 the following conditions:
\begin{subequations}
\begin{align}
\phi_{L}&=\phi_{G},\\
\Diff\phi_{L}&=\Diff\phi_{G}+\frac{\phi_L}{c-U_L}\left(\frac{\mathd U_{G}}{\mathd{z}}-\frac{\mathd
U_{L}}{\mathd{z}}\right),\\
% \end{align}
%
%
%
%
%
m\left(\Diff^2+\alpha^2\right)\phi_L&=\left(\Diff^2+\alpha^2\right)\phi_{G}+Re\delta\tau_{G},
\end{align}
\vskip -0.3in
\begin{multline}
m\left(\Diff^3\phi_L-3\alpha^2\Diff\phi_L\right)+\mathrm{i}\alpha rRe\left(c-U_L\right)\Diff\phi_L+i\alpha
rRe  \frac{\mathd U_L}{\mathd z}\phi_L-
\frac{\mathrm{i}\alpha r Re}{c-U_L}\left(Fr+\alpha^2 S\right)\phi_L
\\
=\left(\Diff^3\phi_{G}-3\alpha^2\Diff\phi_{G}\right)+\mathrm{i}\alpha  Re\left(c-U_L\right)\Diff\phi_{G}+i\alpha Re\frac{\mathd U_{G}}{\mathd z}\phi_{G}+
Re\Diff\delta\tau+i\alpha Re\delta\sigma,
\label{eq:ic_normal}%
\end{multline}%
\label{eq:ic_interface}%
\end{subequations}%
where we have introduced the inverse Froude and inverse Weber numbers, respectively
\begin{equation}
Fr=\frac{g\left(\rho_L-\rho_G\right)h}{\rho_GV^2},\qquad S=\frac{\sigma}{\rho_GV^2h}.
\label{eq:Fr_def}
\end{equation}
Here $g$ and $\sigma$ are the gravity and surface tension constants.   For
clarity, we shall refer to the inverse Froude number as the gravity number,
being the ratio of gravitational to inertial forces, likewise we refer to
the inverse Weber number as the surface-tension number.
%
%, with
%units of $\text{[Length][Time]}^{-2}$ and $\text{[Energy]}\text{[Length]}^{-1}$
%(in two dimensions) respectively.
%
%

%In addition, we must take account of the equation for the turbulent kinetic
%energy.  This is a second-order equation, and therefore requires only two
%boundary conditions.
%
The turbulent kinetic-energy equation is second-order, and therefore requires only two boundary conditions.  We apply the conditions $k_0+\delta k=0$ on $z=\eta$
and $z=d_G/h$, where $d_G/h$ is the non-dimensional vertical extent of the gas layer.  For boundary-layer flow, $d_G=\infty$, while for channel flow $d_G$ is finite.
Upon linearization these kinetic energy conditions are
\begin{equation}
\delta k=0\qquad\text{at }z=0,\,\,\text{and at } z=d_G/h.
\end{equation}
Furthermore, if we take
\[
\mathcal{I}\left(0\right)=\frac{\mathd\mathcal{I}\left(0\right)}{\mathd z}=0,
\]
then, by inspection of Eqs.~\eqref{eq:rs_interp1} and~\eqref{eq:rs_interp2}, $\delta\rtensor_{12}=\left(\rtensor_{12}/q\right)\delta q=0$
at $z=0$, and similarly for $\delta\rtensor$ and $\Diff\delta\rtensor$.  (The justification for this choice of $\mathcal{I}$ is given at the end of Sec.~\ref{subsec:base_state_thif}.)
 Consequently, Eqs.~\eqref{eq:ic_interface} reduce to the standard interfacial
 equations: %
\begin{subequations}
\begin{align}
\phi_{L}&=\phi_{G},\\
\Diff\phi_{L}&=\Diff\phi_{G}+\frac{\phi_L}{c-U_L}\left(\frac{\mathd U_{G}}{\mathd{z}}-\frac{\mathd
U_{L}}{\mathd{z}}\right),\\
% \end{align}
%
%
%
%
%
m\left(\Diff^2+\alpha^2\right)\phi_L&=\left(\Diff^2+\alpha^2\right)\phi_{G},
\end{align}
\vskip -0.3in
\begin{multline}
m\left(\Diff^3\phi_L-3\alpha^2\Diff\phi_L\right)+\mathrm{i}\alpha rRe\left(c-U_L\right)\Diff\phi_L+i\alpha
rRe  \frac{\mathd U_L}{\mathd z}\phi_L-
\frac{\mathrm{i}\alpha r Re}{c-U_L}\left(Fr+\alpha^2 S_{\mathrm{cap}}\right)\phi_L
\\
=\left(\Diff^3\phi_{G}-3\alpha^2\Diff\phi_{G}\right)+\mathrm{i}\alpha  Re\left(c-U_L\right)\Diff\phi_{G}+i\alpha Re\frac{\mathd U_{G}}{\mathd z}\phi_{G}.
\end{multline}%
\label{eq:ic_interface1}%
\end{subequations}%
Finally, at the top of the gas domain $z=d_G/h$, we have the requirement that
the perturbation velocities should vanish:
\[
\phi_G\left(d_G/h\right)=\Diff\phi_G\left(d_G/h\right)=0.
\]
Thus, we obtain a total of $10$ conditions, which provides sufficient information
to close the system of equations $\left(\phi_L,\phi_G,\delta k,\delta\rtensor_{12},\delta\rtensor\right)$.
 No conditions are required on
 $\delta\rtensor$, and $\delta\rtensor_{12}$, since these variables appear  only in an algebraic way in Eqs.~\eqref{eq:rs_interp1} and~\eqref{eq:rs_interp2}.  We now turn to the numerical
 solution of this system of equations
%
%
%
% Putting it all together, we have an eigenvalue problem to solve: schematically
% this is
% %
% %
% \[
% \Ltensor\left(\begin{array}{c}\phi_L\\\phi_G\\\delta k\\\delta\rtensor_{12}\\\delta\rtensor\end{array}\right)=\lambda
% \Mtensor\left(\begin{array}{c}\phi_L\\\phi_G\\\delta k\\\delta\rtensor_{12}\\\delta\rtensor\end{array}\right),\qquad
% \lambda=-i\alpha c.
% \]
% %
% %
% %
% We now turn to the solution of this
% system of equations.

\section{Numerical method}
\label{sec:numerics}

In the liquid and turbulent gas domains, we propose
an approximate solution to the Orr--Sommerfeld (OS) equations~\eqref{eq:model_eqns}
\begin{figure}
\begin{center}
\includegraphics[width=0.65\textwidth]{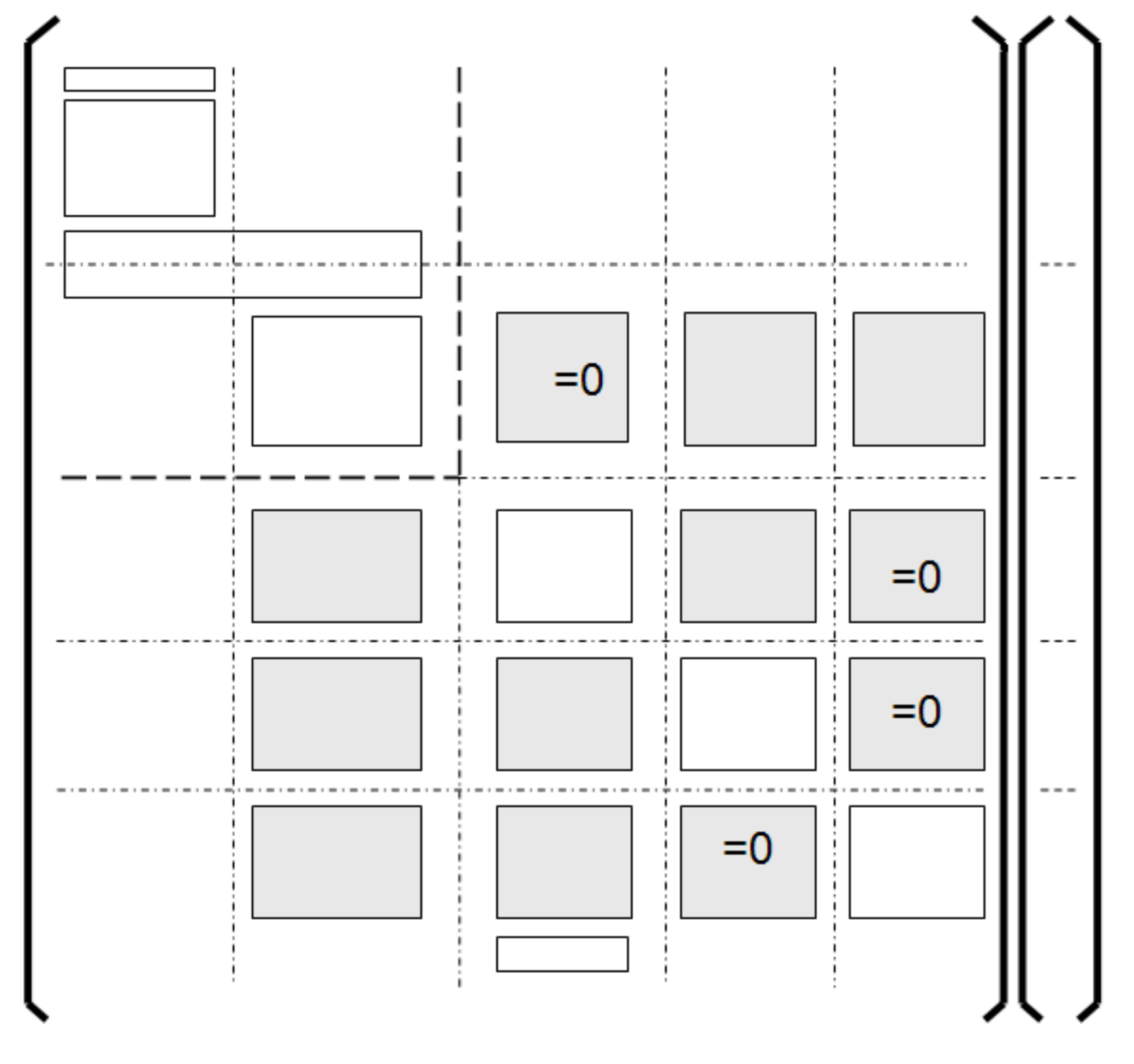}
\end{center}
\caption{A schematic description of the left-hand side of Eq.~\eqref{eq:eig_eqn}.
 The shaded blocks represent the linear interaction between the turbulence
 variables.  The upper right-hand block enclosed by the thick dotted lines
 is the matrix for the eigenvalue problem in the absence of the turbulent
 stress and energy variables.}
\label{fig:eig_eqn}
\end{figure}
in terms of Chebyshev polynomials, such that
\begin{subequations}
\begin{equation}
\phi_L\left(z\right)\approx\sum_{n=0}^{N_L}a_n T_n\left(\eta_L\right),\qquad
\phi_{G}\left(z\right)\approx\sum_{n=0}^{N_{G}}b_{n}T_n\left(\eta_{G}\right),
%\label{eq:cheby_ansatz}
\end{equation}
\begin{equation}
\delta k=\sum_{n=0}^{N_{G}-2} \zeta_n T_n\left(\eta_{G,1}\right),\qquad
\delta\rtensor_{12}=\sum_{n=0}^{N_{G}-4} \theta_n T_n\left(\eta_{G}\right),\qquad
\delta\rtensor=\sum_{n=0}^{N_{G}-4} \xi_n T_n\left(\eta_{G}\right),\qquad
\end{equation}
where we have mapped
the domains of the various layers into the interval $[-1,1]$. For our two-layer
system, with $z\in\left[-d_L,0\right]$ for the liquid and $z\in\left[0,d_G\right]$
for the gas turbulent layer, the appropriate coordinate transformations are
the following:
\begin{align*}
\eta_L=\frac{2z}{d_L}+1,\,\,\, -d_L\leq z\leq0&;\qquad
\eta_{G}=\frac{2z}{d_G}-1,\,\, 0\leq z\leq d_G.
\end{align*}
Thus, we assume that the boundary layer is of finite depth.  This setup has
the advantage that the gas domain is finite in extent, and thus there are
no virtual interfaces in the problem.  In our experience, these virtual interfaces
can interfere
with the stability properties of the system, and are therefore undesirable.  Although there are differences
between the stability behaviour of the finite- and infinite-domain frameworks,
these are apparent only at small $\alpha$ (large-wavelength disturbances).
 We shall also show that this is far from the most dangerous mode,
 and these differences are therefore unimportant.
The derivatives of our approximate solution are obtained through the expressions
\begin{equation}
\frac{\mathd^r\phi_L\left(z\right)}{\mathd z^r}=\sum_{n=0}^{N_L}a_n \left(\frac{\mathd\eta_L}{\mathd
z}\right)^rT_n^{\left(r\right)}\left(\eta_L\right),\qquad
%\end{equation}
%
%
%\begin{equation}
\frac{\mathd^r\phi_{G}\left(z\right)}{\mathd z^r}=\sum_{n=0}^{N_{G}}b_{n}\left(\frac{\mathd\eta_{G}}{\mathd
z}\right)^rT_n^{\left(r\right)}\left(\eta_{G}\right),
\end{equation}%
\label{eq:cheby_ansatz}%
\end{subequations}%
where the derivatives of the Chebyshev polynomial $T_n\left(\eta\right)$
are obtained through the appropriate recurrence relation.

To reduce the problem~\eqref{eq:model_eqns} to a matrix form, we substitute
the approximate solution~\eqref{eq:cheby_ansatz} into the OS equations~\eqref{eq:model_eqns}
and evaluate the resulting identity at the collocation points
\begin{eqnarray*}
\eta_L&=&\cos\left(\frac{p\pi}{N_L-2}\right),\qquad p=1,...,N_L-3,\\
\eta_{G}&=&\cos\left(\frac{p\pi}{N_{G}-2}\right),\qquad p=1,...,N_{G}-3.
\end{eqnarray*}
In the two-layer application that we have in mind, this gives
\[
\left(N_L-3\right)+4\left(N_{G,1}-3\right)=N_L+4N_{G,1}-15
\]
equations in $\left(N_L+1\right)+\left(N_{G,1}+1\right)+\left(N_{G,1}-1\right)+2\left(N_{G,1}-3\right)$
$=N_L+4N_{G,1}-5$ unknowns. We
must therefore find
an additional $10$ equations to close the system. These are provided by
the $10$ boundary and interfacial conditions.
Thus, we are reduced to solving for the
wave speed $c$, given $N_L+4N_{G,1}-3$ linear equations in as many unknowns.
In symbols, we therefore have a generalized eigenvalue problem
\begin{equation}
\Ltensor\bm{x}=\lambda\Mtensor\bm{x},\qquad \lambda=-i\alpha c,
\label{eq:eig_eqn}
\end{equation}
where positivity of the real part of $\lambda$ ($\lambda_\mathrm{r}>0$) indicates
instability.
A schematic description of the left-hand side of this system is shown in
Fig.~\ref{fig:eig_eqn}.  The solution of~\eqref{eq:eig_eqn} is best found
using a linear algebra
package (as in~\cite{TrefethenBook}). Then, a built-in eigenvalue
solver automatically balances the
matrices $\Ltensor$ and $\Mtensor$, an important issue here
since these matrices can be very badly conditioned, owing to the high-order
derivatives of the Chebyshev polynomials
that appear in the expansion of the streamfunction (as in the work by Boomkamp
\textit{et al.}~\cite{Boomkamp1997}).
Furthermore,
a specialized eigenvalue solver takes into account
the appearance of zero rows (hence infinite eigenvalues) in the matrix $\Mtensor$,
and thus gives an accurate answer. Following standard practice, we modify
the number of collocation points until convergence is achieved (typically, $N_L=200$, $N_G=150$ for convergence to four significant figures).

\section{Results for a thin film}
\label{sec:thif}

In this section, we investigate the effects of turbulence on the interfacial
stability of a thin laminar liquid layer exposed to fully-developed turbulent
shear flow in an overlying gas layer.  The parameter regime we study involves
$Re$, $Fr$, $S$, and the ratio
\[
\delta=\frac{d_L}{d_G}=\frac{\text{liquid-film thickness}}{\text{gas-layer thickness}},
\]
and is chosen to mimic the conditions in pipe flow in oil-gas transport problems~\cite{Miesen1995}.
 The liquid Reynolds number is such that the liquid flow is laminar.
We use a base state that mimics flow in a boundary layer.  In this case the
flow is confined by a flat plat at $z=d_G$, a large distance from the interface.
 This plate moves at velocity $U_0$ relative to the interface.  It is thus
 appropriate to re-scale on the velocity $U_0$, and on the gas-layer depth
 $d_G$ (thus $h=d_G$ and $V=U_0$, in the notation of Sec.~\ref{sec:model}).  This setup
 has the advantage that the gas domain is finite in extent, while having
 little effect on the results (see Fig.~\ref{fig:growth_rate_thif} for a comparison between
 finite- and infinite-domain growth rates).
Using this framework, we constitute the base-state velocity.

\subsection{Base-state determination}
\label{subsec:base_state_thif}
\begin{figure}
\begin{center}
\subfigure[]{
\includegraphics[width=0.32\textwidth]{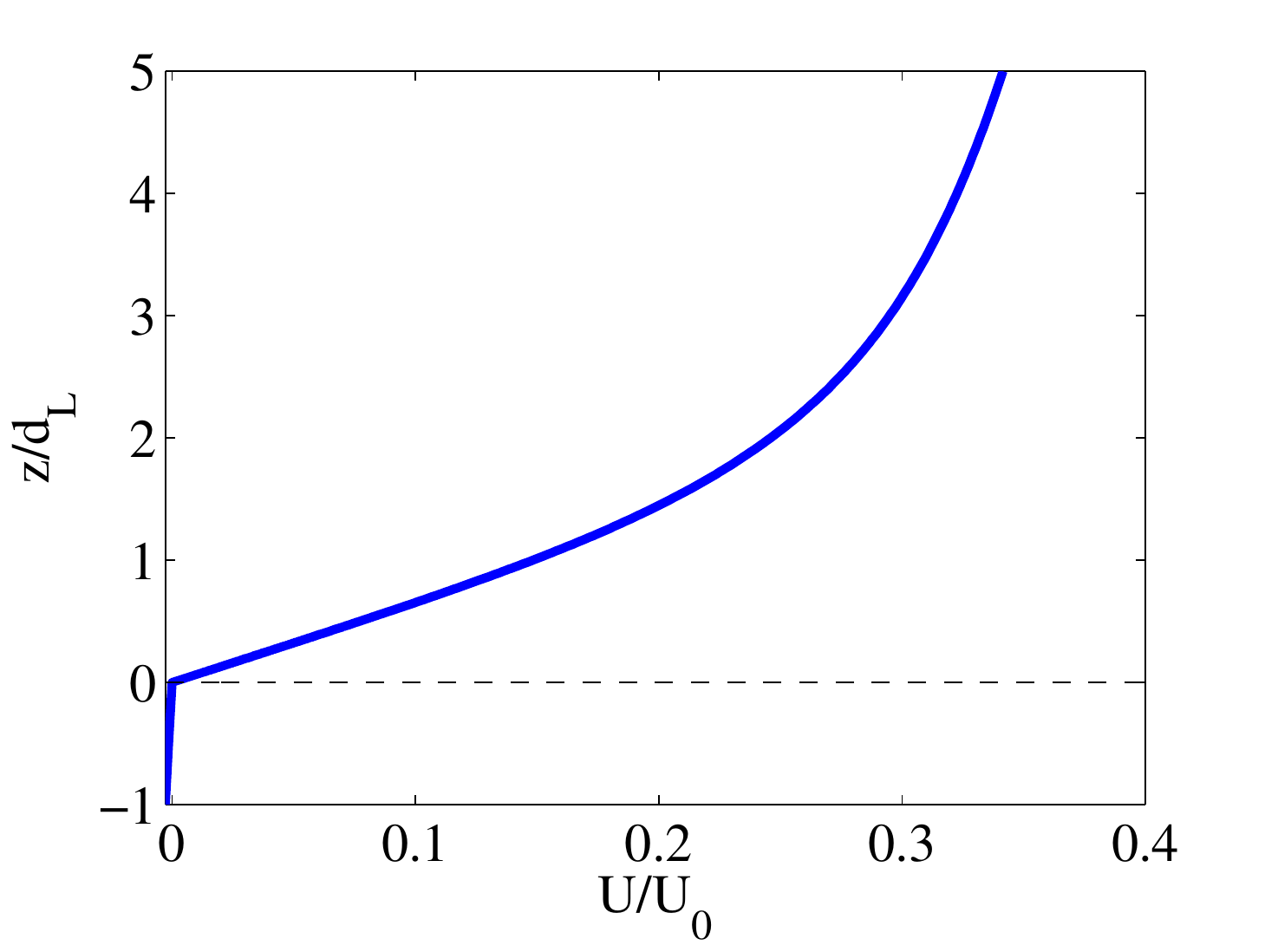}}
\subfigure[]{
\includegraphics[width=0.32\textwidth]{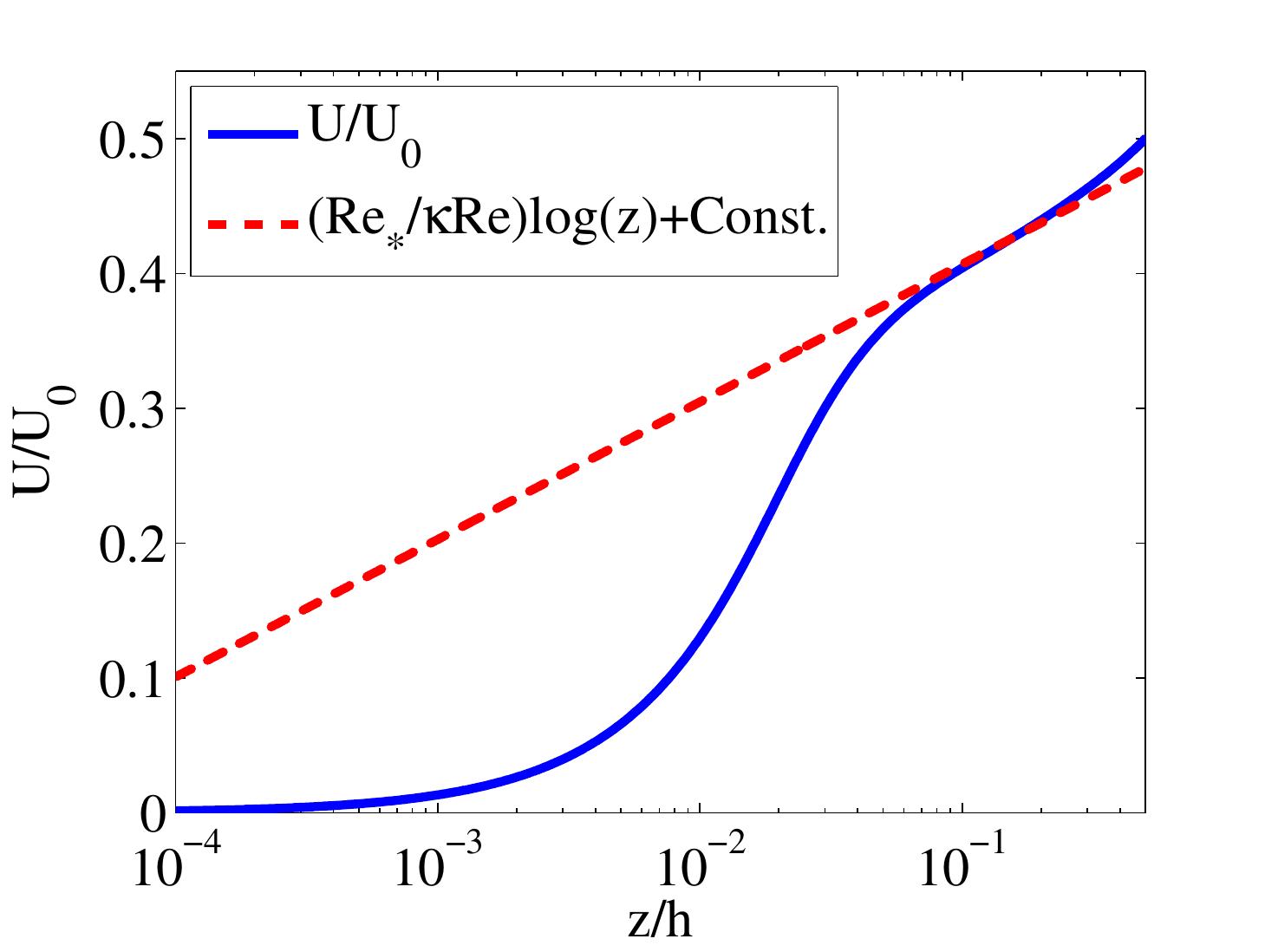}}
\subfigure[]{
\includegraphics[width=0.32\textwidth]{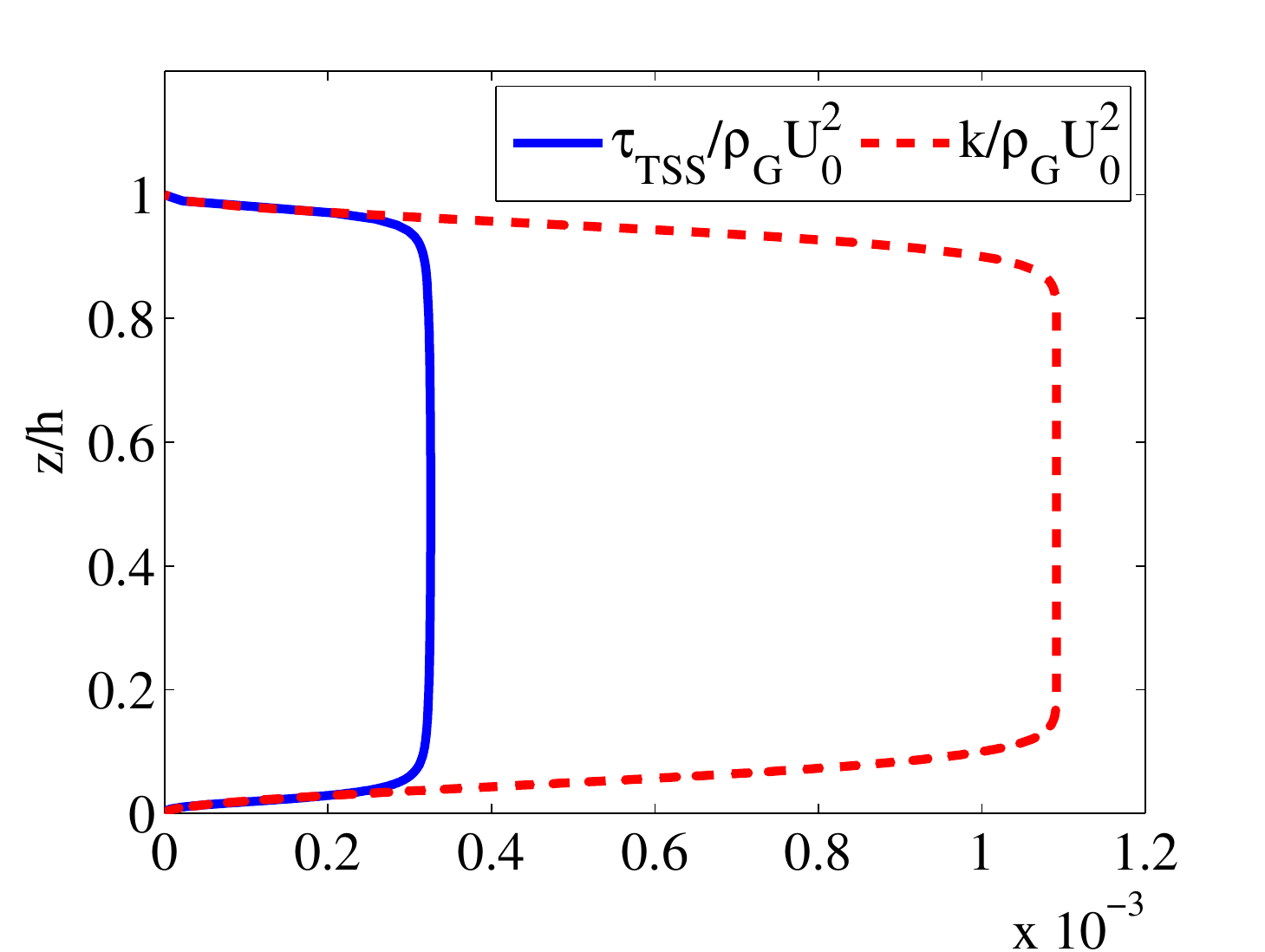}}
\end{center}
\caption{The base-flow variables for $Re=8000$, $m=55$, $r=1000$, and $\delta=1/50$.  Subfigure (a) shows the
velocity profile in the liquid and the gas, with distances normalized on
the liquid-film thickness; subfigure (b) shows the development
of a logarithmic profile near the centreline of the gas domain.  Subfigure
(c) shows the spatial dependence of the turbulent shear stress and kinetic
energy.}
\label{fig:Ubase}
\end{figure}

\paragraph*{The gas layer:}  In the gas, we have the following momentum balance
for the base state:
\begin{subequations}
\begin{equation}
\mu_G\frac{\mathd U_G}{\mathd z}+\tau_{TSS}=\tau_{\mathrm{i}},
\end{equation}
where $\tau_{\mathrm{i}}$ is the interfacial shear stress, and $\tau_{TSS}$
is the turbulent shear stress $-\rho_G\langle u' w'\rangle$.  To close this
term, we make use
of the interpolation function
\begin{equation}
G\left(\tilde{z}\right)=\tilde{z}\left(1-\tilde{z}\right),\qquad
\tilde{z}=\frac{z}{d_G},
\label{eq:G}
\end{equation}
where $\tilde{z}$ is the non-dimensional vertical co-ordinate (the tilde decoration has been introduced to designate dimensionless quantities).  Now the turbulent
shear stress has a characteristic Taylor expansion near the walls, and for the interpolation function~\eqref{eq:G} to agree with this expansion, we must modify
Eq.~\eqref{eq:G} near the upper wall and the interface, such that
\begin{equation}
\tau_{TSS}=\kappa\rho_G d_G U_{*\mathrm{i}}G\left(\tilde{z}\right)\psi\left(\tilde{z}\right)\psi\left(1-\tilde{z}\right)\frac{\mathd
U_G}{\mathd z},\qquad
\psi\left(\tilde{z}\right)=1-e^{-\tilde{z}^n/A},
\label{eq:wall_fn}
\end{equation}
where $n$ and $A$ (which depends on $Re$) are model parameters fixed below with reference to Eq.~\eqref{eq:wall_taylor}, and $U_{*\mathrm{i}}=\sqrt{\tau_{\mathrm{i}}/\rho_G}$
is the interfacial friction velocity.  Hence,
\begin{equation*}
U_G=\tau_{\mathrm{i}}d_G\int_0^{z/d_G}\frac{\mathd s}{\mu_G+\kappa\rho_G d_G U_{*\mathrm{i}} G\left(s\right)\psi\left(s\right)\psi\left(1-s\right)},
\end{equation*}
where we have set the frame of reference to move with the interface.  We
non-dimensionalize on the plate velocity $U_0$, on the gas height $d_G$, and
on the gas viscosity $\mu_G$,
which gives a non-dimensional velocity
\begin{equation}
\tilde{U}_G=\frac{Re_*^2}{Re}\int_0^{\tilde{z}}\frac{\mathd s}{1+\kappa Re_* G\left(s\right)\psi\left(s\right)\psi\left(1-s\right)},
\label{eq:u_nondim}
\end{equation}
where $Re_*=\rho_G U_{*\mathrm{i}} d_G/\mu_G$, in contrast to the Reynolds number set by
the non-dimensionalization, $Re=\rho_G U_0 d_G/\mu_G$.  The Reynolds number
$Re_*$ is determined as the root of the equation
\begin{equation}
\frac{Re_*^2}{Re}\int_0^{1}\frac{\mathd s}{1+\kappa Re_* G\left(s\right)\psi\left(s\right)\psi\left(1-s\right)}=1.
\end{equation}
\paragraph*{The liquid layer:}  The momentum balance in the laminar liquid is simply
\begin{equation}
\mu_L\frac{\mathd U_L}{\mathd z}=\tau_{\mathrm{i}}.
\end{equation}
Integrating and non-dimensionalizing, this is
\begin{equation}
\tilde{U}_L=\frac{Re_*^2}{mRe}\tilde{z}.
\end{equation}

\paragraph*{The turbulence-related variables:} From Eq.~\eqref{eq:wall_fn},
we obtain the following non-dimensional form for the gas Reynolds stress:
\begin{equation}
\tilde{\tau}_{TSS}=\frac{\tau_{TSS}}{\rho_GU_0^2}=\frac{Re_*^2}{Re^2}\frac{\kappa Re_*G\left(\tilde{z}\right)\psi\left(\tilde{z}\right)\psi\left(1-\tilde{z}\right)}{1+\kappa Re_* G\left(\tilde{z}\right)\psi\left(\tilde{z}\right)\psi\left(1-\tilde{z}\right)}.
\end{equation}
Moreover, we constitute the turbulent kinetic energy as
\begin{equation}
\tilde{k}=\frac{k}{\rho_GU_0^2}=\frac{1}{C^2}\frac{Re_*^2}{Re^2}\psi\left(\tilde{z}\right)\psi\left(1-\tilde{z}\right),
\label{eq:k_model}
\end{equation}
where $C$ is another constant.  We take $C=0.55$, which is the value appropriate for the logarithmic region of the mean velocity in a boundary layer.
%
%This value is chosen
%so as to agree with shear-driven equilibrium turbulence in the log layer.
%
The form~\eqref{eq:k_model} takes
into account the constancy of the turbulent kinetic energy in the core, while
damping the energy to zero at the interface and at the wall.
Well into the bulk of the gas flow, this model gives the ratio
\begin{equation*}
-\ntensor_{12}^{(0)}=\frac{\tau}{2k}\approx\frac{C^2}{2}.
\end{equation*}
To model the quantity $\rtensor^{(0)}=\rtensor_{11}^{(0)}-\rtensor_{12}^{(0)}=\langle u'^2\rangle-\langle
w'^2\rangle$ (where this average is over base-state variables), we make the approximation
\begin{equation}
\rtensor_{11}^{(0)}=a_{11}k,\qquad \rtensor_{22}^{(0)}=a_{22}k,
\label{eq:rtensor_k}
\end{equation}
where $a_{11}$ and $a_{22}$ are constants, such that $a_{11}+a_{22}=2$.  For linear algebraic stress models,
the convention is to take $a_{11}=a_{22}=1$, which is physically incorrect:
the turbulent kinetic energy is not equipartitioned in the streamwise and normal
directions.  One fix is to use a non-linear algebraic stress model~\cite{Speziale1987},
which gives better predictions of the partitioning of energy.  For simplicity, however, we use typical
values for the partition constants from DNS data in boundary-layer flow~\cite{Spalart1988}.
Note finally that this linear relationship, while true in the bulk flow,
breaks down near the interface and near the wall.  We have, however, conducted
tests to verify the sensitivity of the stability analysis to this detail
and have found that changing the form of the law~\eqref{eq:rtensor_k} has
little effect on the results.  Thus, we take $a_{11}=\tfrac{3}{2}$ throughout
the flow, and hence, $a_{22}=\tfrac{1}{2}$.
\begin{equation}
\rtensor^{(0)}=k,\qquad \ntensor_{11}^{(0)}=\tfrac{3}{4},\qquad
\ntensor_{22}^{(0)}=\tfrac{1}{4}
\end{equation}

Close to the interface $\tilde{z}=0$, we have the following asymptotic expressions
for the model:
\begin{equation*}
\tilde{\tau}_{TSS}\sim\frac{\kappa Re_*}{A}\tilde{z}^{1+n},\qquad
\tilde{k}\sim\left(\frac{\kappa Re_*}{
Re}\right)^2\frac{\tilde{z}^n}{C^2A},\qquad
\tilde{U}_G\sim \tilde{z}-\frac{\kappa Re_*}{\left(n+2\right)A}\tilde{z}^{n+2}.
\end{equation*}%
\label{eq:model_base}%
\end{subequations}%
The asymptotic behaviour of these model variables can be made to agree with
the true behaviour of near-wall turbulence.  Although the system we study
possesses both a wall and an interface, when the density contrast between
the liquid and the gas is large, DNS results suggest~\cite{Banerjee1995} that
a comparison between wall- and interfacial-turbulence is justified.
The fluctuating velocities for wall turbulence have the following Taylor
expansions around the wall location $z=0$:
\[
u'=a_1\left(x,t\right)+b_1\left(x,t\right)z+c_1\left(x,t\right)z^2+...,\qquad
w'=a_2\left(x,t\right)+b_2\left(x,t\right)z+c_2\left(x,t\right)z^2+....
\]
Using the no-slip conditions, we obtain $a_1=a_2=0$.  Since $u'$ is zero along the
wall $z=0$, $\left(\partial u'/\partial x\right)_{z=0}=0$, hence $\left(\partial
w'/\partial z\right)_{z=0}=b_2=0$.  Thus, near the wall,
\begin{eqnarray}
k&=&\tfrac{1}{2}\rho_G\langle u'^2+w'^2\rangle\sim\tfrac{1}{2}\rho_G\langle b_1^2\rangle
z^2,\nonumber\\
%\rtensor_{11}&=&\langle u'^2\rangle\sim\langle b_1^2\rangle z^2,\nonumber\\
%\rtensor_{22}&=&\langle w'^2\rangle\sim\langle c_2^2\rangle z^4,\nonumber\\
%=\tfrac{1}{4}\bigg\langle\left(\frac{\partial b_1}{\partial x}\right)^2\bigg\rangle{z}^4,\\
\tau&=&-\rho_G\langle u' w'\rangle\sim-\rho_G\langle b_1 c_2\rangle z^3.
%\varepsilon&=&\nu\langle s'_{ij}s'_{ij}\rangle\sim \nu\langle b_1^2\rangle.
\label{eq:wall_taylor}
\end{eqnarray}
Note that this result provides for a no-flux condition on the turbulent kinetic
energy $k$ at the wall, $\mathd k/\mathd z=0$ on $z=0$.
Choosing $n=2$ in our model forces agreement between the interface model and the near-wall asymptotics.
The results of the model are shown in Fig.~\ref{fig:Ubase}.
\paragraph*{Computation of $z_{\mathrm{t}}$:}  In Sec.~\ref{sec:model},
we introduced the turbulent and advective timescales.  These timescales vary
in the normal direction.  Close to the interface, the frequency of eddy turnovers
is $T_{\mathrm{t}}^{-1}$, which is large compared to the advection frequency
$T_{\mathrm{a}}$.  Far from the interface, the magnitude of these frequencies
is reversed.  Crossover occurs at $z_{\mathrm{t}}$, which is determined as
the root of the equation
\begin{equation}
\frac{\alpha^{-1}}{\left|U\left(z_{\mathrm{t}}\right)-c\left(\alpha\right)\right|}=\frac{\kappa
Re}{Re_*}z_{\mathrm{t}},
\label{eq:zt}
\end{equation}
Typically, $c_\mathrm{r}\gg c_\mathrm{i}$, and negligible error is incurred
by replacing $c$ by $c_\mathrm{r}$ in Eq.~\eqref{eq:zt}.  Making this replacement,
we can readily find a lower bound for $z_{\mathrm{t}}$:
\begin{eqnarray*}
Re_*\alpha^{-1}&=&\kappa z_{\mathrm{t}}\,Re\left|U\left(z_{\mathrm{t}}\right)-c_\mathrm{r}\right|,\\
&\leq&\kappa z_{\mathrm{t}}\,Re\, U\left(z_{\mathrm{t}}\right),\\
&\leq& \kappa Re_*^2 z_{\mathrm{t}}^2,
\end{eqnarray*}
hence
\begin{equation}
z_{\mathrm{t}}\geq \sqrt{\frac{1}{\kappa Re_*\alpha}}.
\label{eq:bound_zt}
\end{equation}
When the wave speed  $c_{\mathrm{r}}$ is small, the bound~\eqref{eq:bound_zt}
sharpens.  Moreover, we have verified that the propagation speed $c_{\mathrm{r}}$
is not affected by the turbulence modelling (e.g., Fig.~\ref{fig:growth_rate}~(b)), and thus the computation of
$z_{\mathrm{t}}$
\begin{figure}[htb]
\centering\noindent
\includegraphics[width=0.4\textwidth]{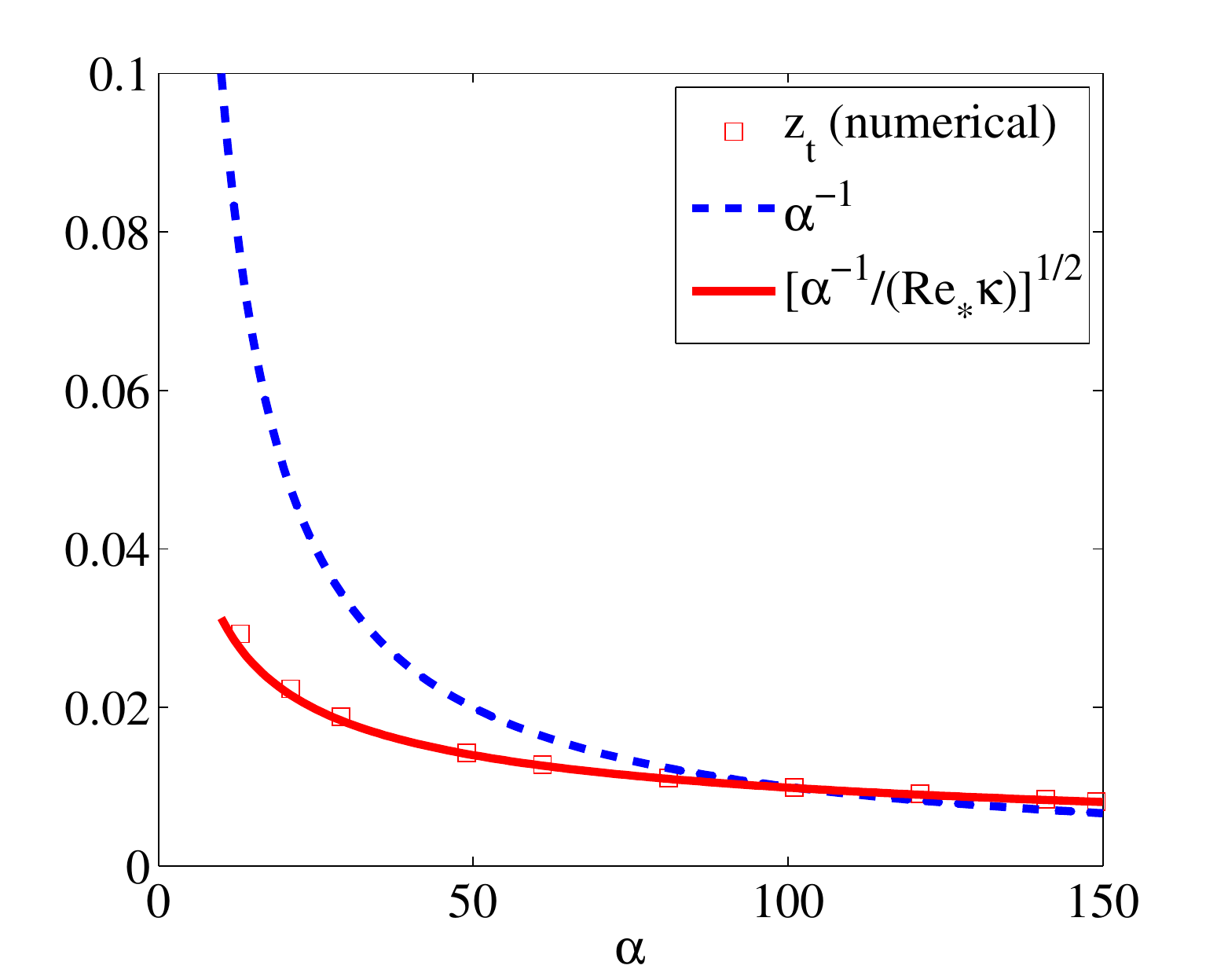}
\caption{The crossover height $z_{\mathrm{t}}$ as a function of wavenumber for $Re=8000$.
 The wavelength $\alpha^{-1}$ is shown for comparison.}
\label{fig:zt_thif}
\end{figure}
can be carried out using the model without the perturbation turbulent stresses
(PTS).
In Fig.~\ref{fig:zt_thif} we have done precisely this: the scale $z_{\mathrm{t}}$
is computed and the bound~\eqref{eq:bound_zt} is found to be sharp.  The
scale $\alpha^{-1}$, which determines the spatial extent of the streamfunction,
is shown for comparison.  For all but the largest $\alpha$-values (shortest
wavelengths), the crossover height $z_{\mathrm{t}}$ lies in a region where
the streamfunction is non-negligible.  Thus, we expect the streamfunction
to `feel' the effects of rapid distortion.
Note finally that the computation of the scale $z_{\mathrm{t}}$
enables us to develop an interpolation function $\mathcal{I}\left(z\right)$,
specifically, we take
\begin{equation}
\mathcal{I}\left(z\right)=1-e^{-\left(z/z_{\mathrm{t}}\right)^2},
\label{eq:interpolant}
\end{equation}
in the gas.  The exponent is taken to be $2$ to guarantee that the
turbulent effects vanish strongly at the interface: since the flat-interface
state has this property, it is reasonable to assign the same property to
the perturbed state.

\subsection{Linear-stability analysis}
\label{subsec:linsta_thif}

We carry out a stability analysis around the base state just constituted
for $Re=8000$, $Fr=10^{-5}$, and $S=10^{-3}$.  The density and viscosity
ratios are chosen such that the stability analysis models an air-water system
under
\begin{figure}[htb]
\centering\noindent
\subfigure[]{
\includegraphics[width=0.45\textwidth]{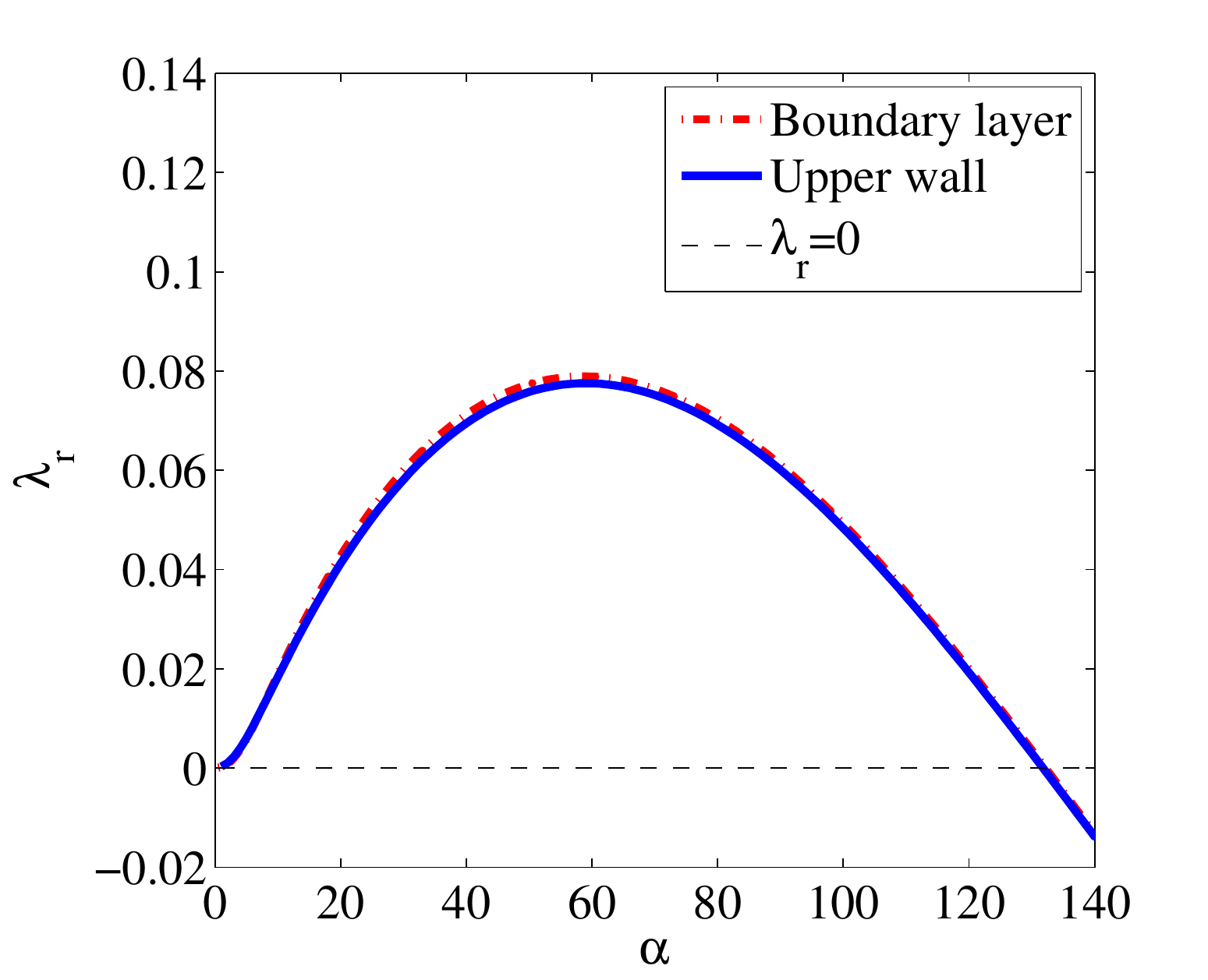}}
\subfigure[]{
\includegraphics[width=0.45\textwidth]{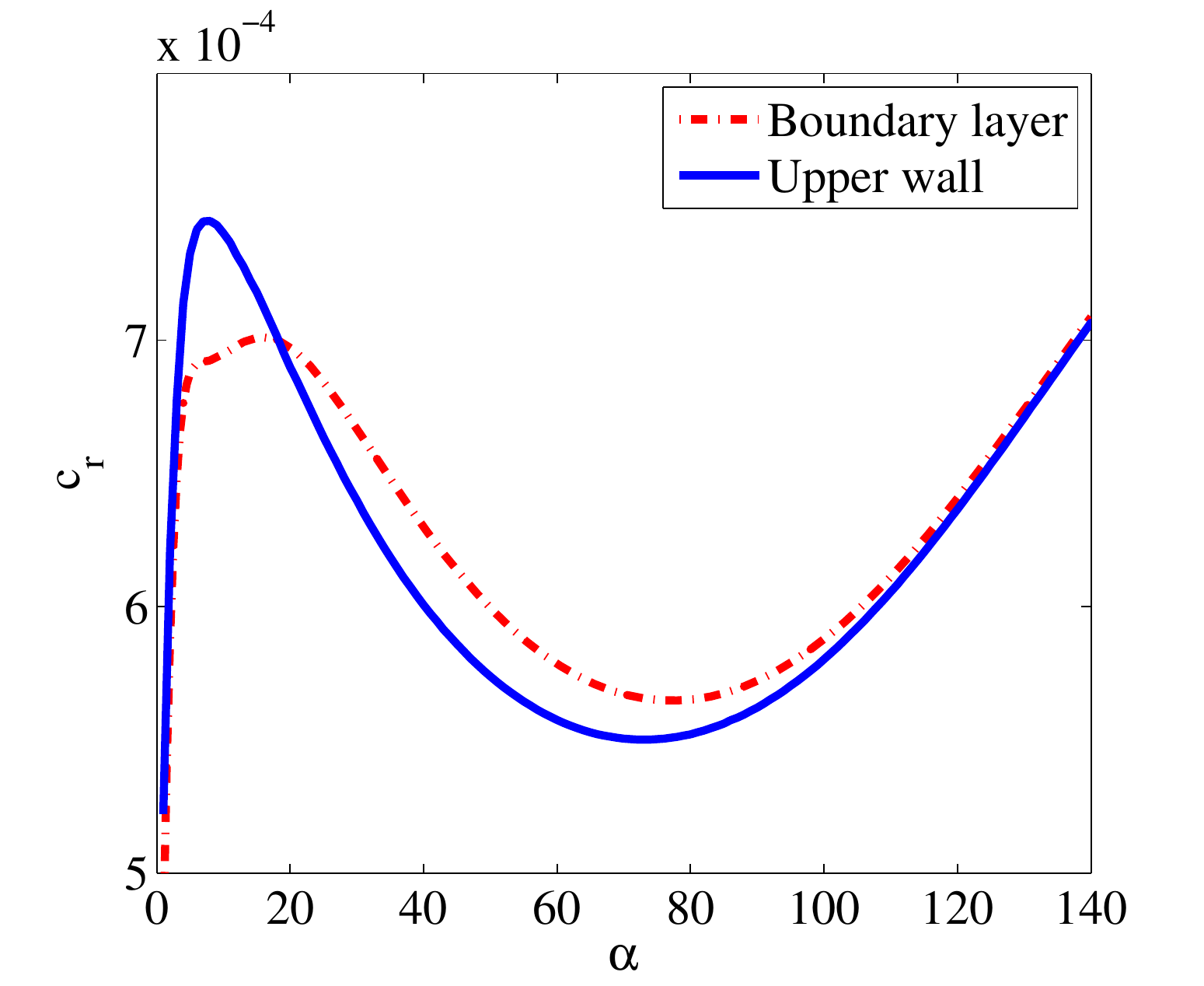}}
\caption{(a) The growth rate in the thin-film case, with $Re=8000$ (solid line).  The wavenumber
is normalized by the film thickness $d_L$, rather than the gas-domain thickness
$d_G$, and $d_G/d_L=50$.  The difference between the PTS wave and the non-PTS wave
is so small as to be negligible, and is not shown here; (b) The wave speed in the thin-film case (solid line).  A comparison with a boundary-layer base state with no upper plate is shown (dashed-dotted lines).}
\label{fig:growth_rate_thif}
\end{figure}
standard conditions $m=55$, $r=1000$.  These Reynolds, inverse Froude, and inverse Weber numbers are
chosen to correspond to typical air-water conditions for a liquid-film thickness
$d_L=d_G/50$.
We examine the growth rate of the disturbance with and without
the effects of the PTS, and the results
are shown in Fig.~\ref{fig:growth_rate_thif}, where we also plot the wave
speed.
The inclusion of the PTS has little effect on the growth rate in this case,
except at
long wavelengths.  The trend in the dispersion curve for the PTS wave shows
a tiny downward shift in the growth rate relative to the non-PTS wave.  The maximum
growth rate occurs at $\alpha\approx60$.  Since we have non-dimensionalized on the gas-layer height, this corresponds to a wavelength $\ell/d_L\approx 2\pi\delta/\alpha=2\pi\times\left(50/60\right)$,
that is, a wavelength greater than, but comparable to the liquid film thickness.
 Finally, we note that our finding that the maximum growth rate occurs at
 a wavelength comparable to the liquid-film thickness is similar to results
 obtained elsewhere for thin liquid films.  The u-shaped trough in the wave-speed
 diagram is also characteristic of thin-film flow: we have observed it by
 reproducing the work of Miesen and Boersma~\cite{Miesen1995}, and extracting
 the wave speed from the stability analysis, in addition to the growth rate.
 Our inclusion of an upper plate has little or no effect on the results:
 this can be seen by
%
%
%
%\begin{figure}[htb]
%\centering\noindent
%\subfigure[]{
%\includegraphics[width=0.45\textwidth]{growth_rate_compare_thif}}
%\subfigure[]{
%\includegraphics[width=0.45\textwidth]{wave_speed_compare_thif}}
%\caption{A comparison between the stability analysis in a turbulent channel
%with shear (solid line), and the same analysis in a turbulent boundary layer
%(dotted lines).  The channel configuration is the focus of this study.  (a) The growth rate in the thin-film case, at $Re=8000$. (b) The wave speed.}
%\label{fig:growth_rate_compare_thif}
%\end{figure}
%
%
comparing the data in Fig.~\ref{fig:growth_rate_thif} with a true boundary
layer, at the same parameter values (in particular, at the same liquid
Reynolds number).  This is shown by the broken-line curves in Fig.~\ref{fig:growth_rate_thif},
where the boundary-layer
stability analysis is conducted according to the framework of Miesen and
Boersma~\cite{Miesen1995}.

Figure~\ref{fig:growth_rate_thif} implies that the PTS have little effect on the growth rate at the Reynolds number considered ($Re=8000$).  This can be explained by studying the form of the rapid-distortion
 equations~\eqref{eq:rs_interp1} and~\eqref{eq:rs_interp2}: here the stress
 terms are damped not only by the decay function $\mathcal{I}\left(z/z_{\mathrm{t}}\right)$,
 but also by the equilibrium turbulent kinetic energy function $q\left(z\right)$,
 which is zero at the interface.  For now, let us take it on trust
 that the instability in Fig.~\ref{fig:growth_rate_thif} is due to
 the viscosity-contrast mechanism (we verify this in Tab.~\ref{tab:eb_thif}).
  Then, the instability is governed by conditions at the interface and in
  the viscous sublayer, $z\leq Re_*^{-1}$.  This is precisely the region in
  which $q\left(z\right)$ is damped to zero.  Thus, in the region $z\leq
  Re_*^{-1}$ where the growth rate is determined, rapid distortion plays
  no role, and therefore, does not affect the growth rate.  \textit{A forteriori},
  the PTS play no role in determining the growth
  rate for this particular class
  of instability.
Nevertheless, there exists a further aspect of the problem where the PTS do alter the features of the disturbance significantly, namely the \textit{shape} of the streamfunction in a zone far from the interface.  Indeed, we shall find that the far-field streamfunction has an oscillatory structure that propagates far into the gas core.  The existence of these oscillations casts doubt on whether the structure and statistics of the turbulence near non-linear interfacial waves is similar to that for a wall region.  Ultimately, this must be confirmed by accurate numerical simulation.

\begin{figure}[htb]
\centering\noindent
\subfigure[]{
\includegraphics[width=0.3\textwidth]{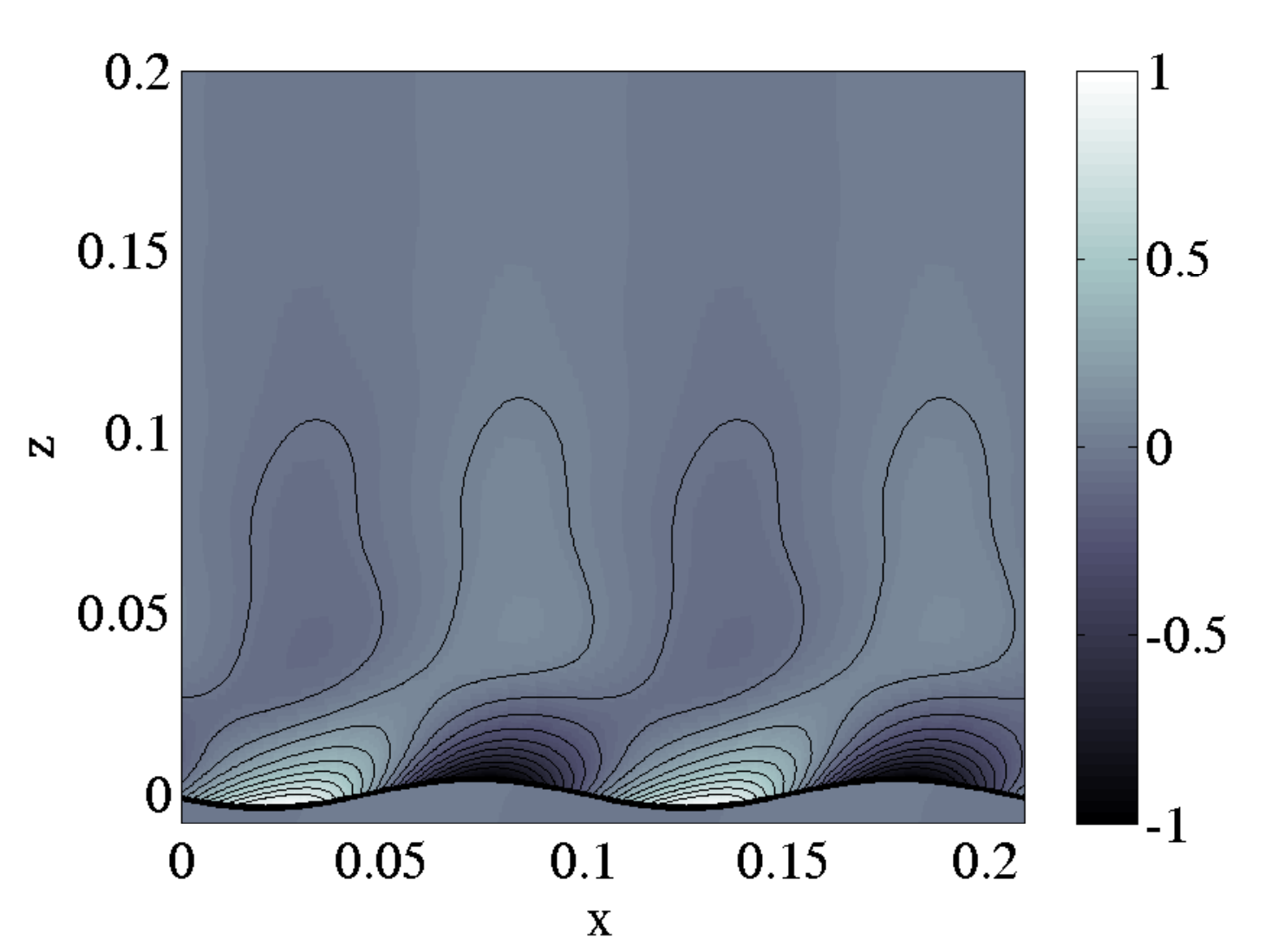}}
\subfigure[]{
\includegraphics[width=0.3\textwidth]{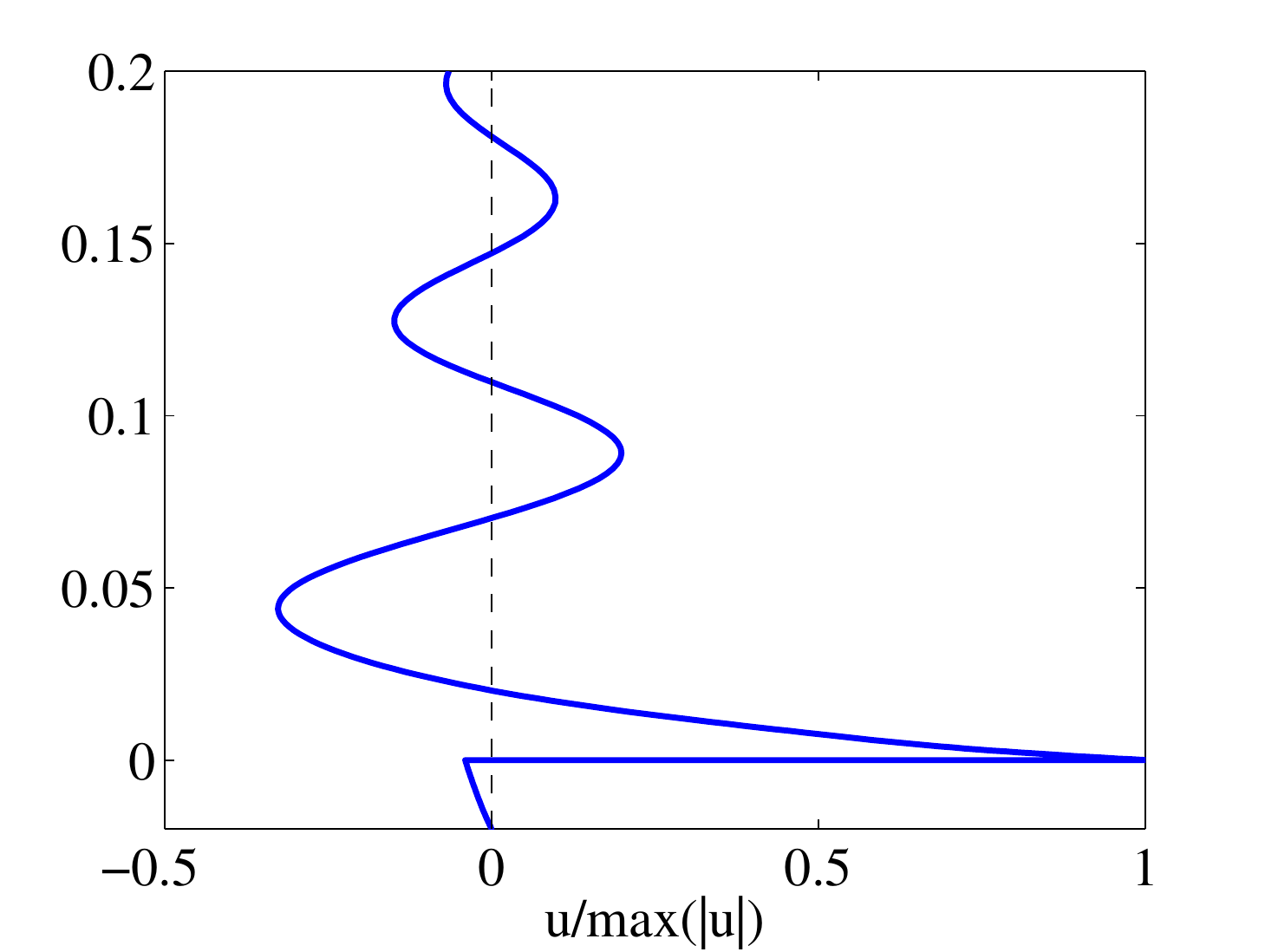}}\\
\subfigure[]{
\includegraphics[width=0.3\textwidth]{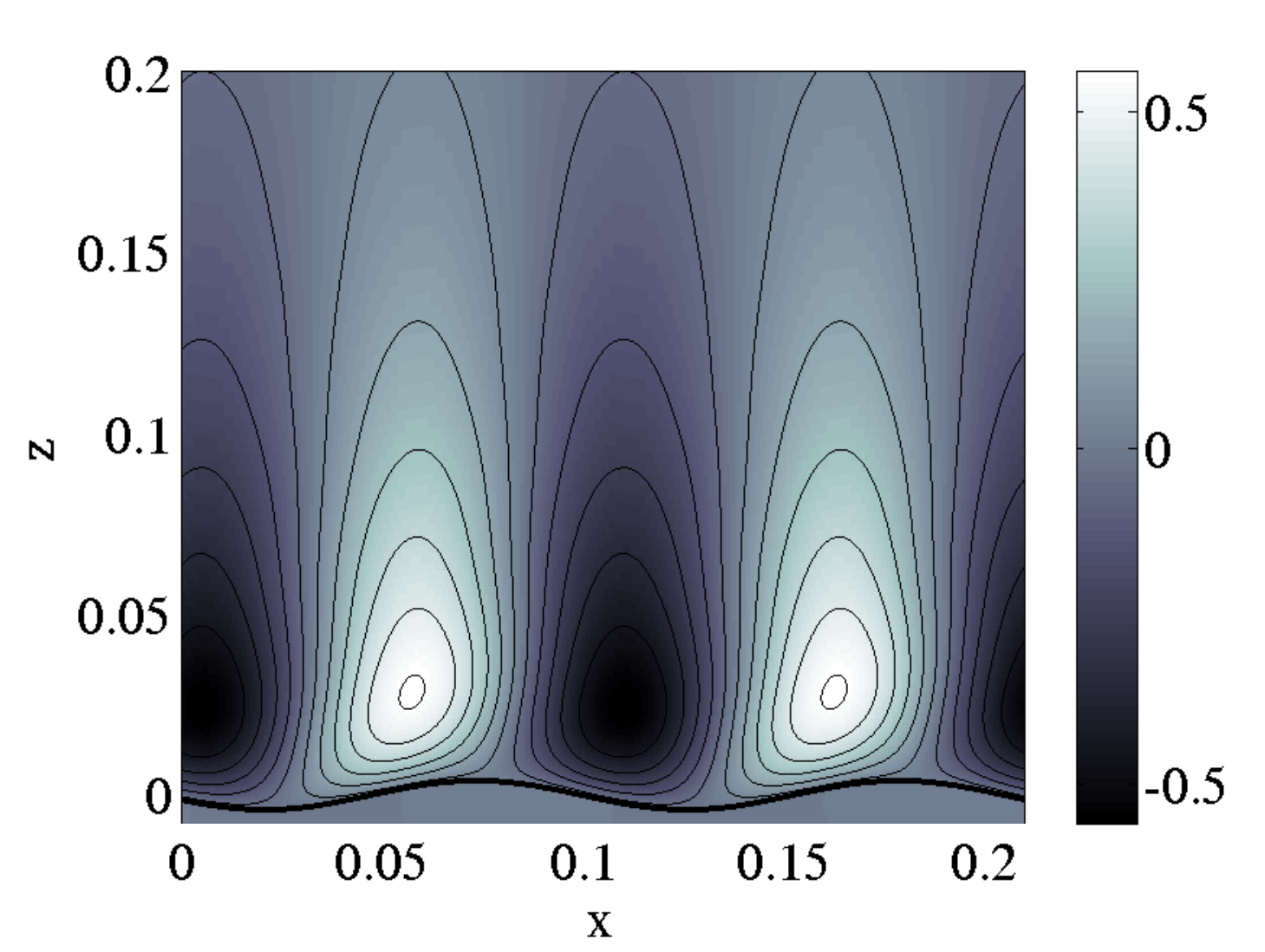}}
\subfigure[]{
\includegraphics[width=0.3\textwidth]{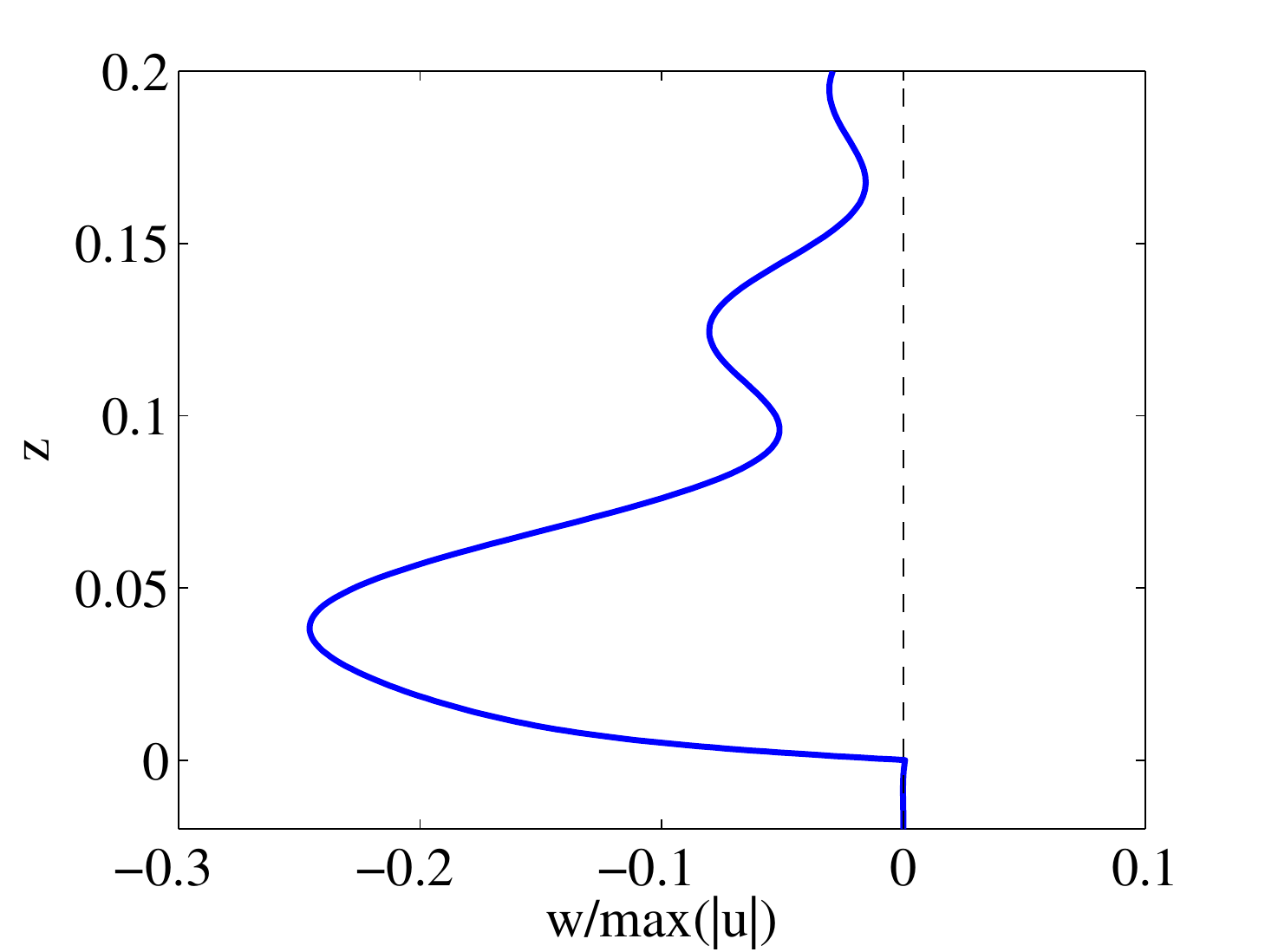}}\\
\subfigure[]{
\includegraphics[width=0.3\textwidth]{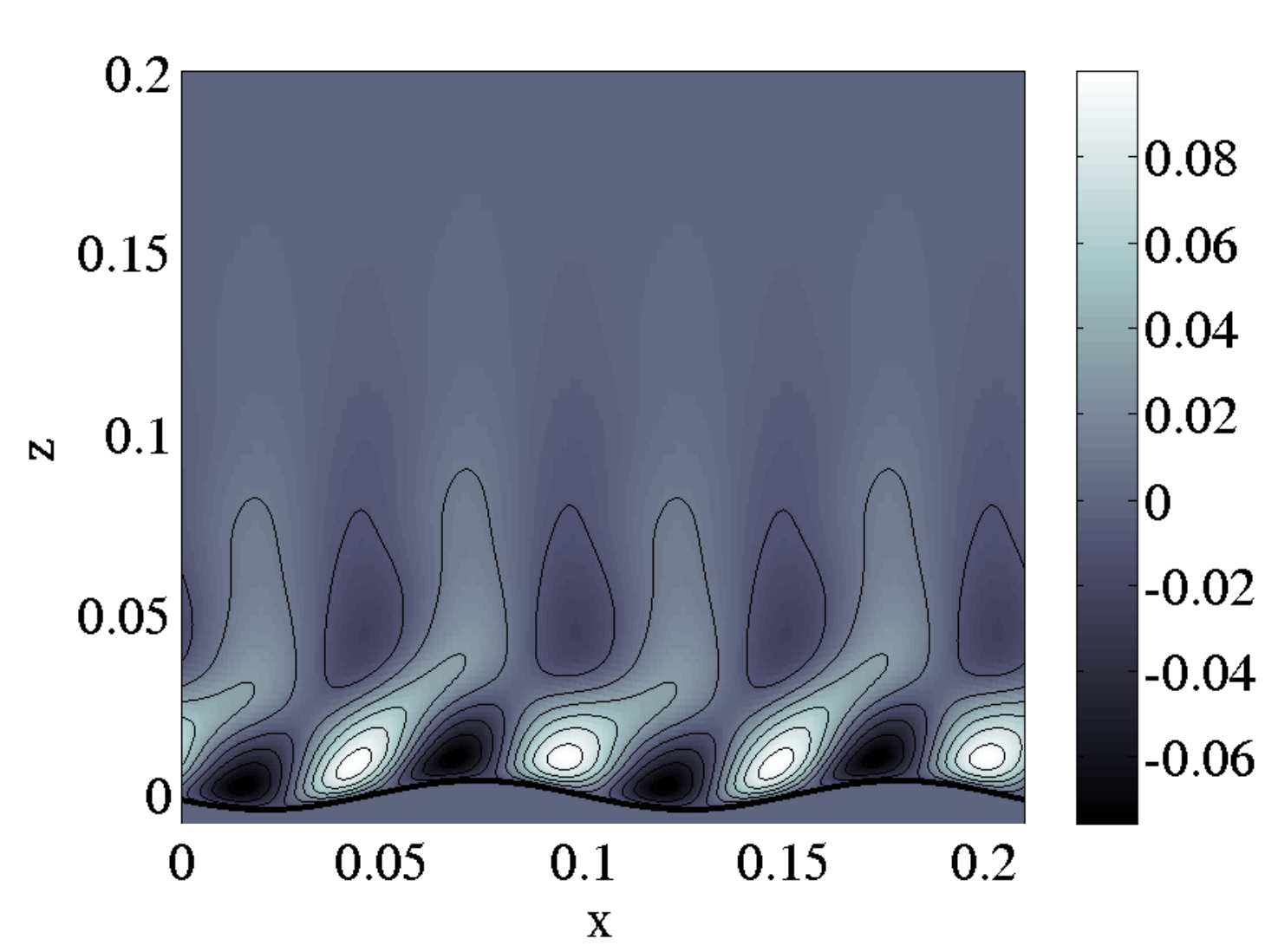}}
\subfigure[]{
\includegraphics[width=0.3\textwidth]{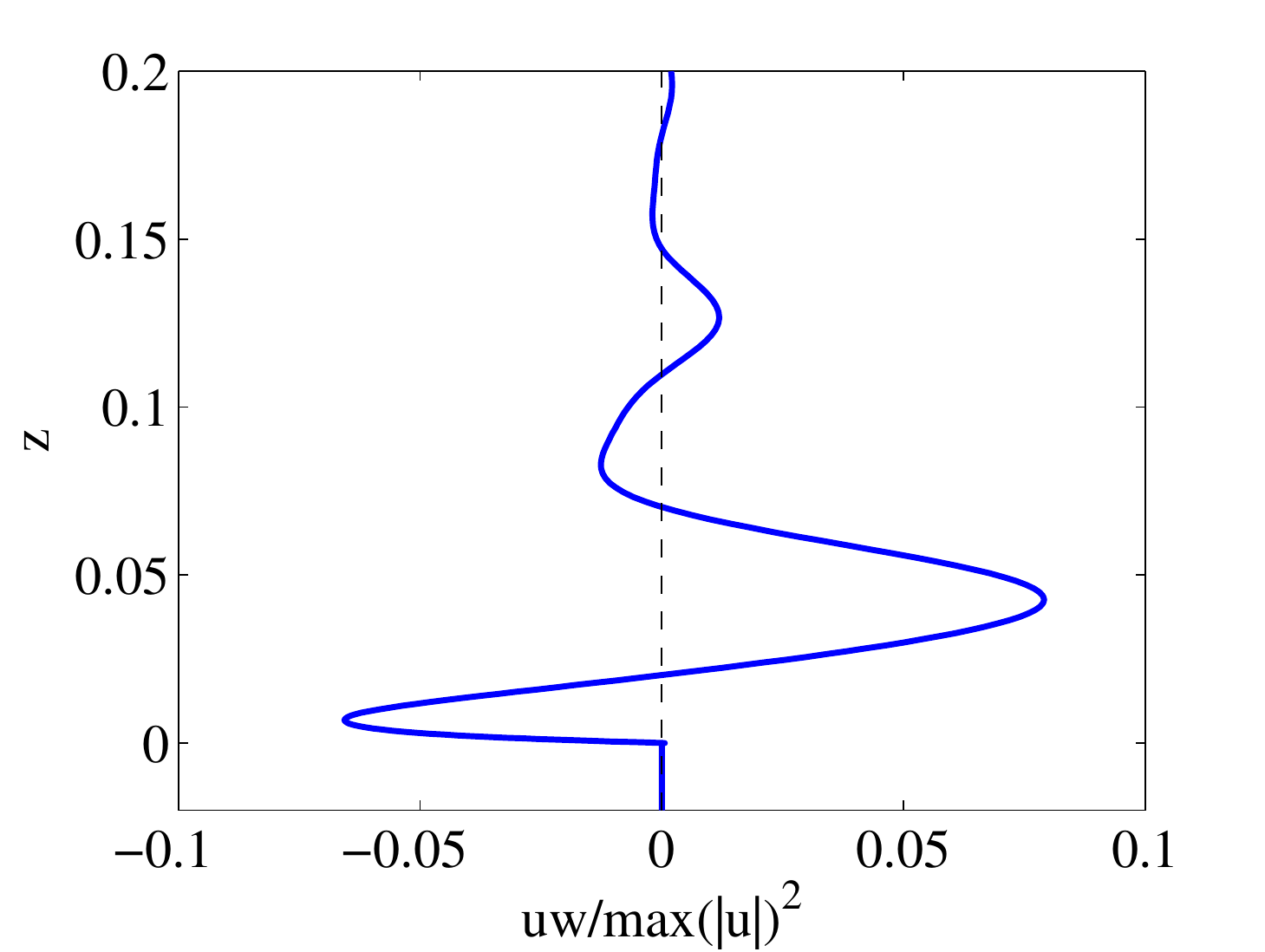}}
\caption{Flow field structure for $\alpha=15$.
 Subfigures~(a) and~(b) show the streamwise velocity field and a profile
 at $x=0$;
 subfigures~(c) and~(d) show the normal velocity field and a profile at $x=0$;
 subfigures~(e) and~(f) show the pre-averaged version of the wave Reynolds
 stress, namely the field $uw$, and its profile at $x=0$.  In each case, we have normalized
 the velocities by $\max|u|$.}
\label{fig:velocity_thif15}
\end{figure}

\begin{figure}[htb]
\centering\noindent
\subfigure[]{
\includegraphics[width=0.3\textwidth]{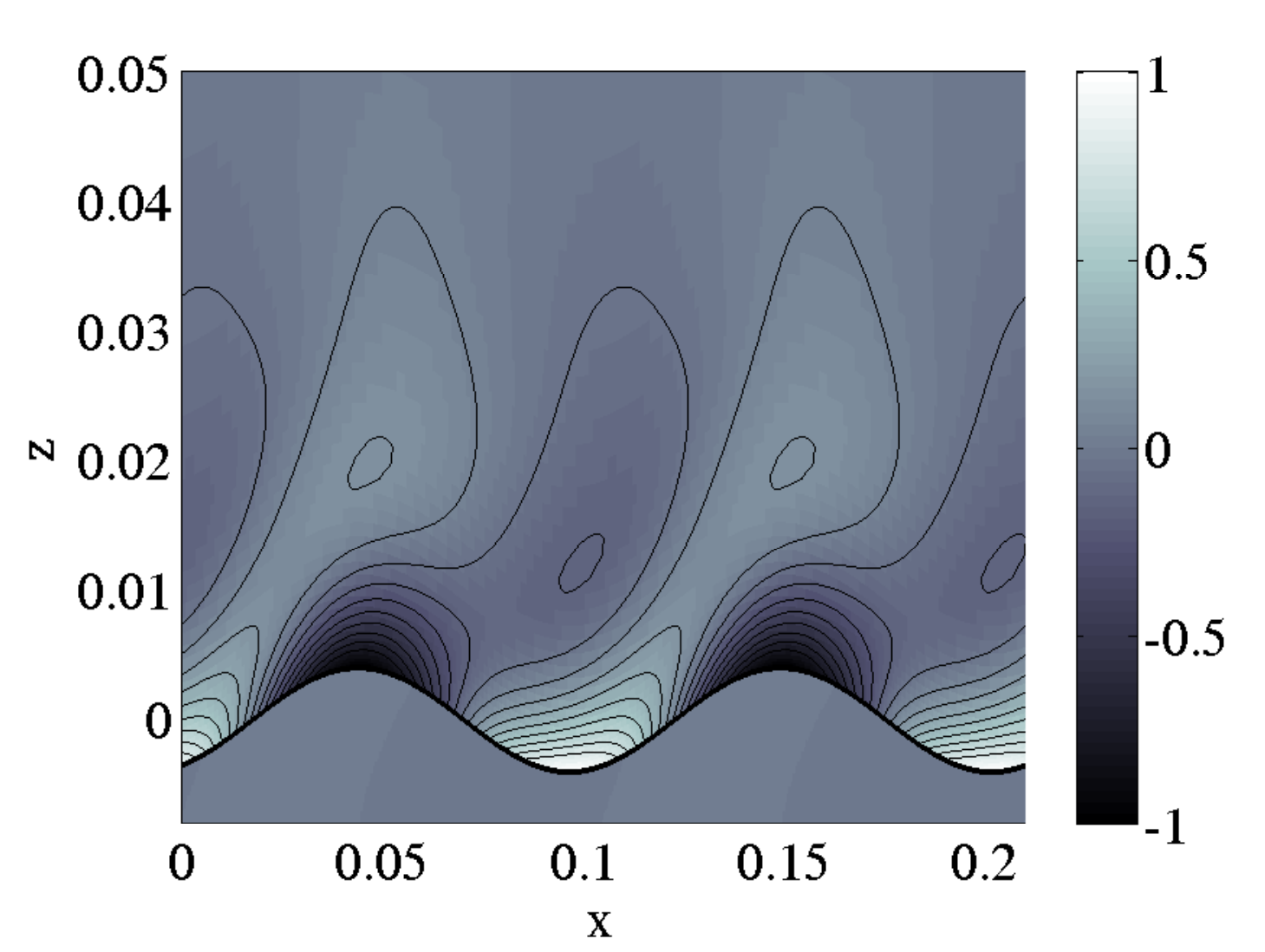}}
\subfigure[]{
\includegraphics[width=0.3\textwidth]{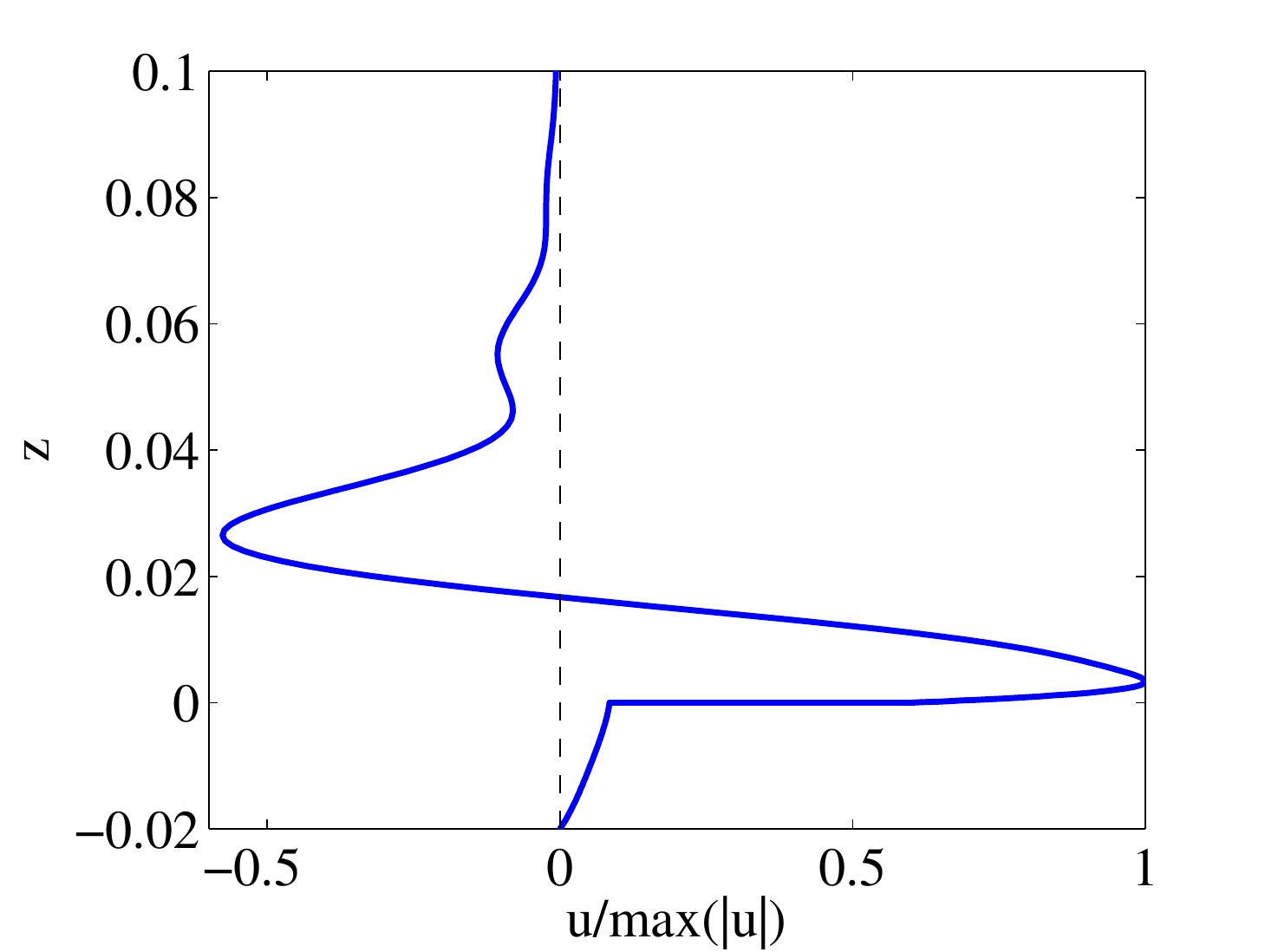}}\\
\subfigure[]{
\includegraphics[width=0.3\textwidth]{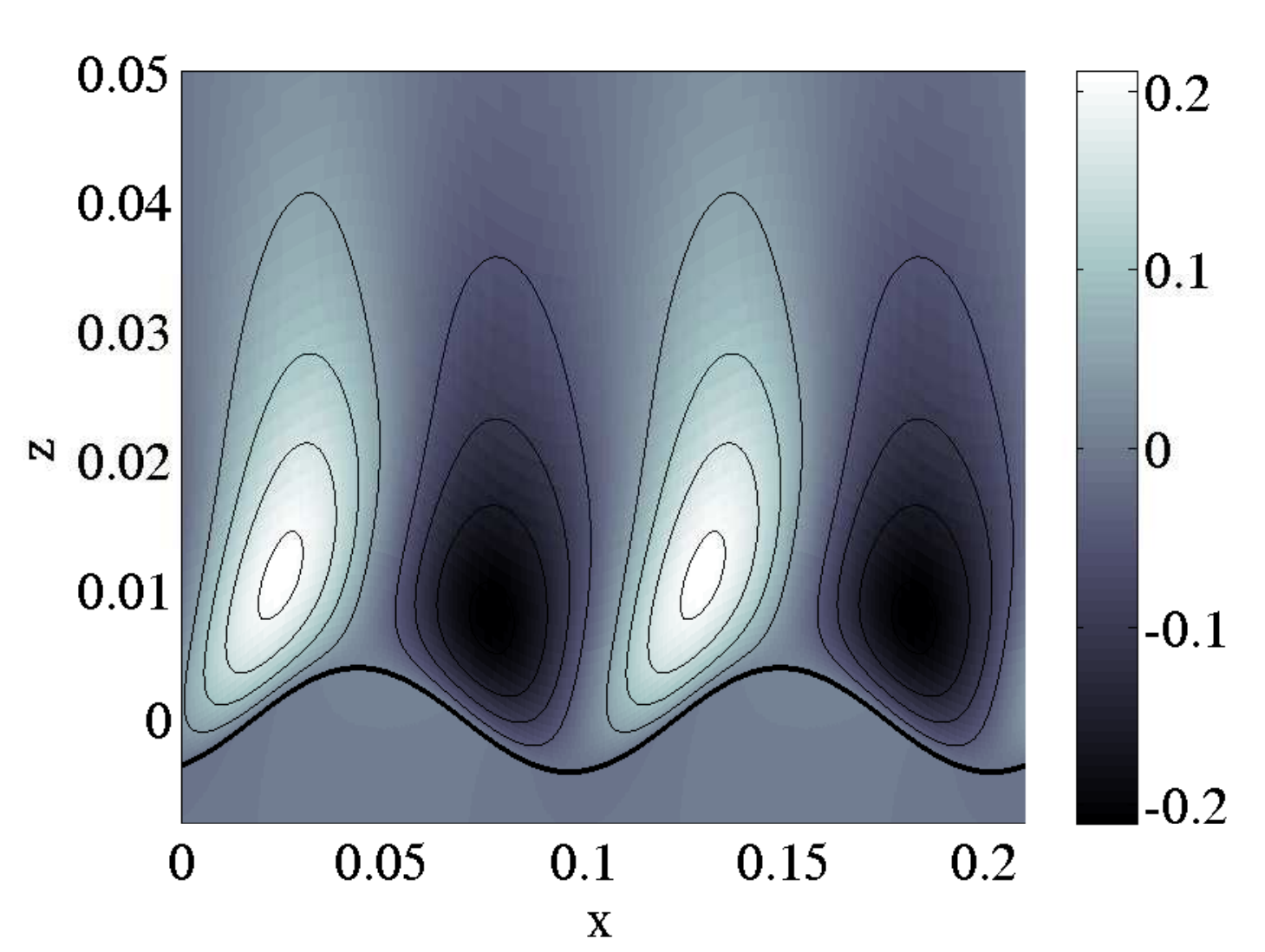}}
\subfigure[]{
\includegraphics[width=0.3\textwidth]{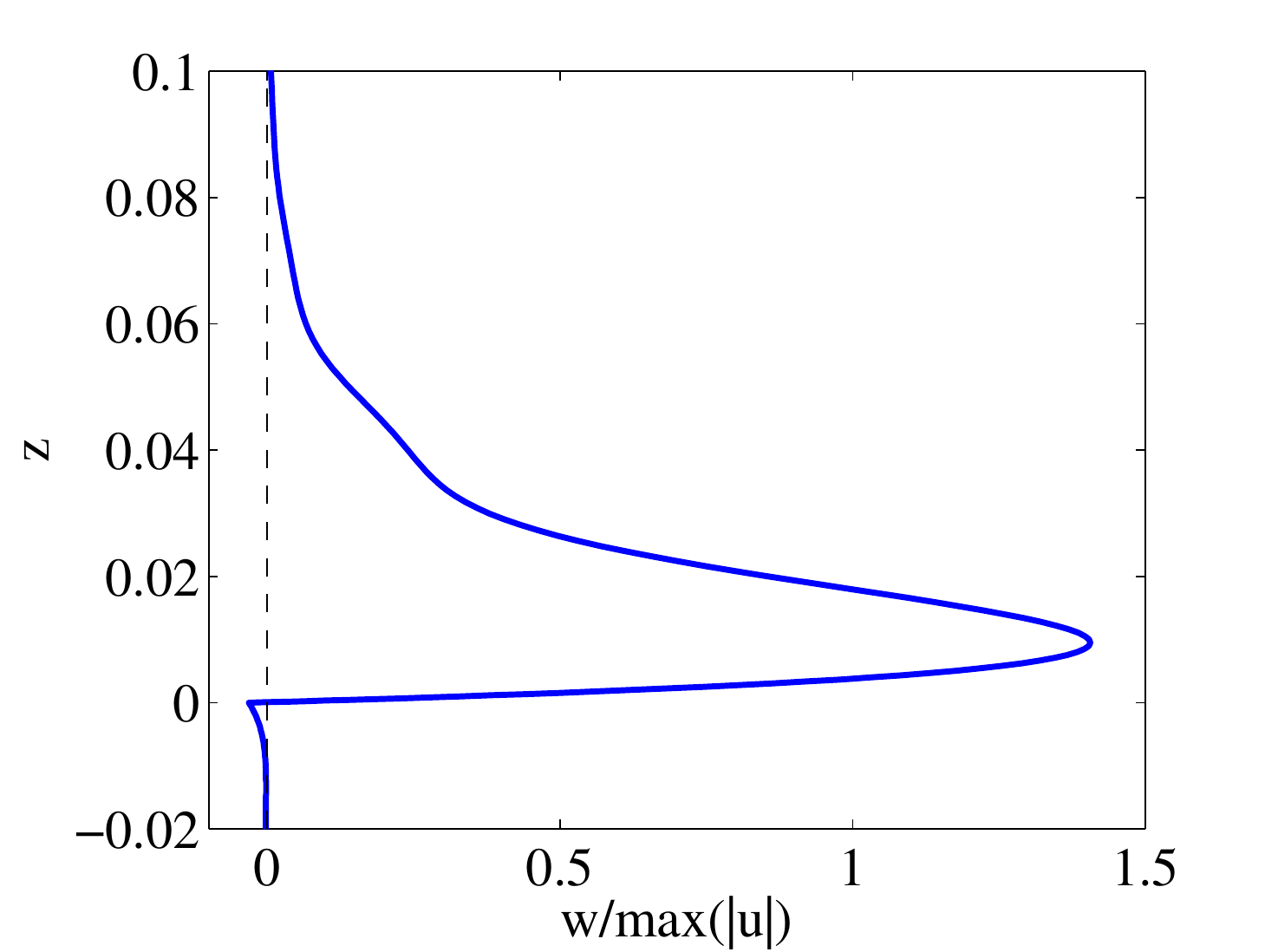}}\\
\subfigure[]{
\includegraphics[width=0.3\textwidth]{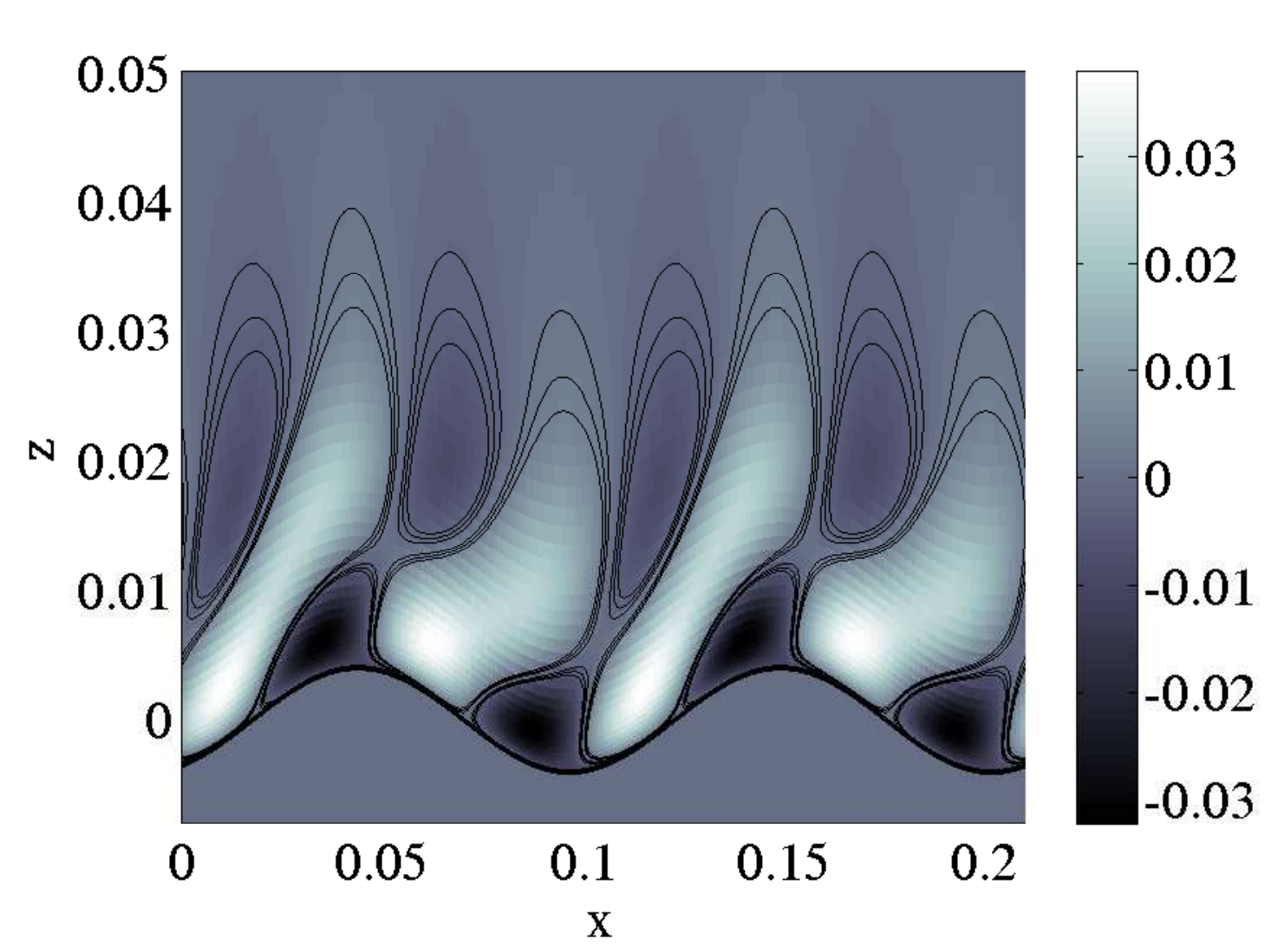}}
\subfigure[]{
\includegraphics[width=0.3\textwidth]{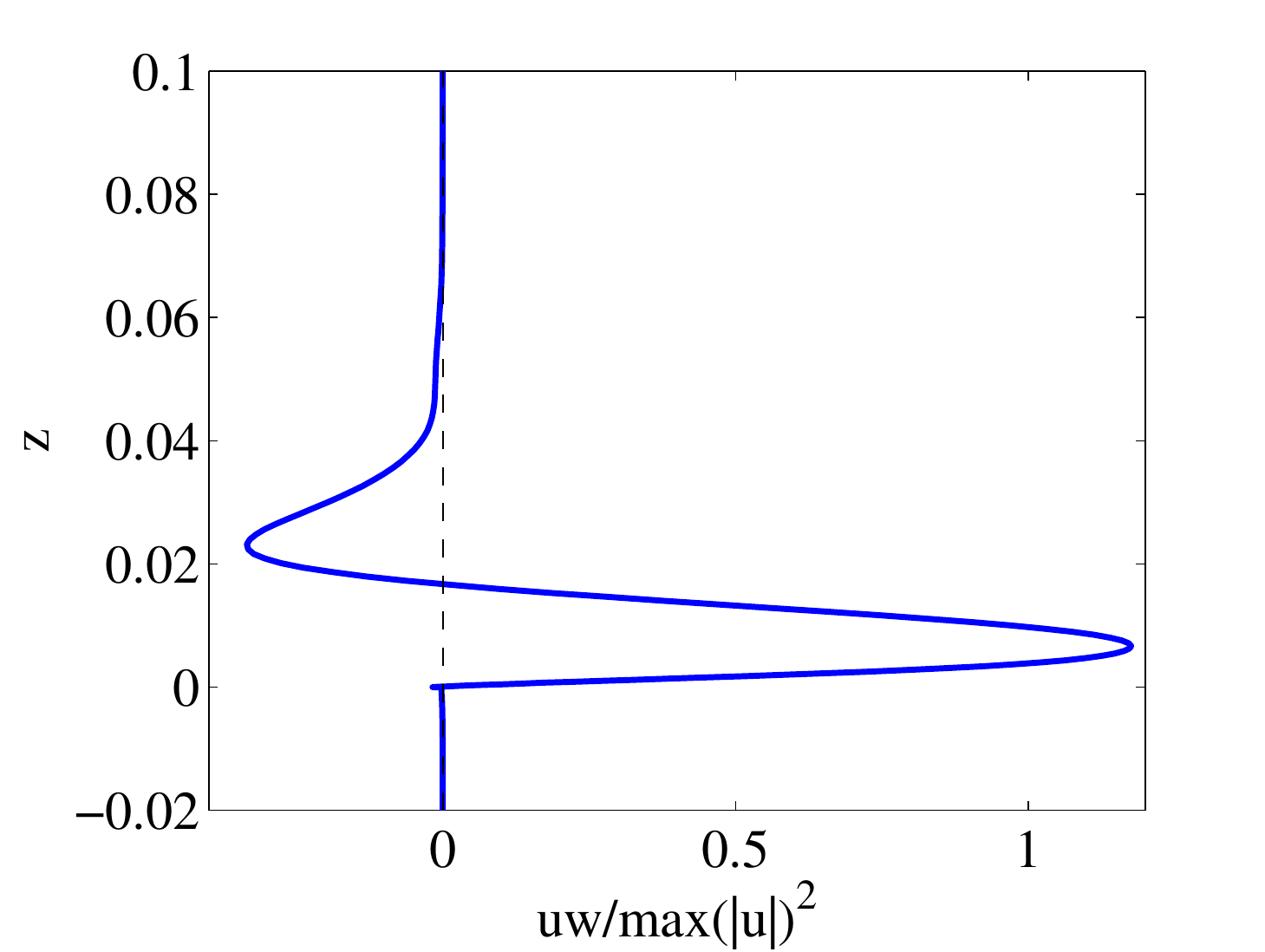}}
\caption{Flow field structure for $\alpha=60$.
 Subfigures~(a) and~(b) show the streamwise velocity field and a profile
 at $x=0$;
 subfigures~(c) and~(d) show the normal velocity field and a profile at $x=0$;
 subfigures~(e) and~(f) show the pre-averaged version of the wave Reynolds
 stress, namely the field $uw$, and its profile at $x=0$.  In each case, we have normalized the velocities
 by $\max|u|$.}
\label{fig:velocity_thif60}
\end{figure}

We investigate the spatial structure of the perturbation velocity for
wavenumbers $\alpha=15$ and $\alpha=60$; the results are shown in Figs.~\ref{fig:velocity_thif15}
and~\ref{fig:velocity_thif60}.
The effect of the rapid distortion on the flow is visible at $\alpha=15$,
and almost negligible at $\alpha=60$.
 For $\alpha=15$, the streamfunction extends into the rapid-distortion domain,
 and the wave and the turbulence interact.  This gives rise to an oscillatory
 structure in the velocity field, in the normal direction, as shown in Figs.~\ref{fig:velocity_thif15}~(b)
 and~(d).  This structure also
\begin{figure}[htb]
\centering\noindent
\subfigure[]{
\includegraphics[width=0.3\textwidth]{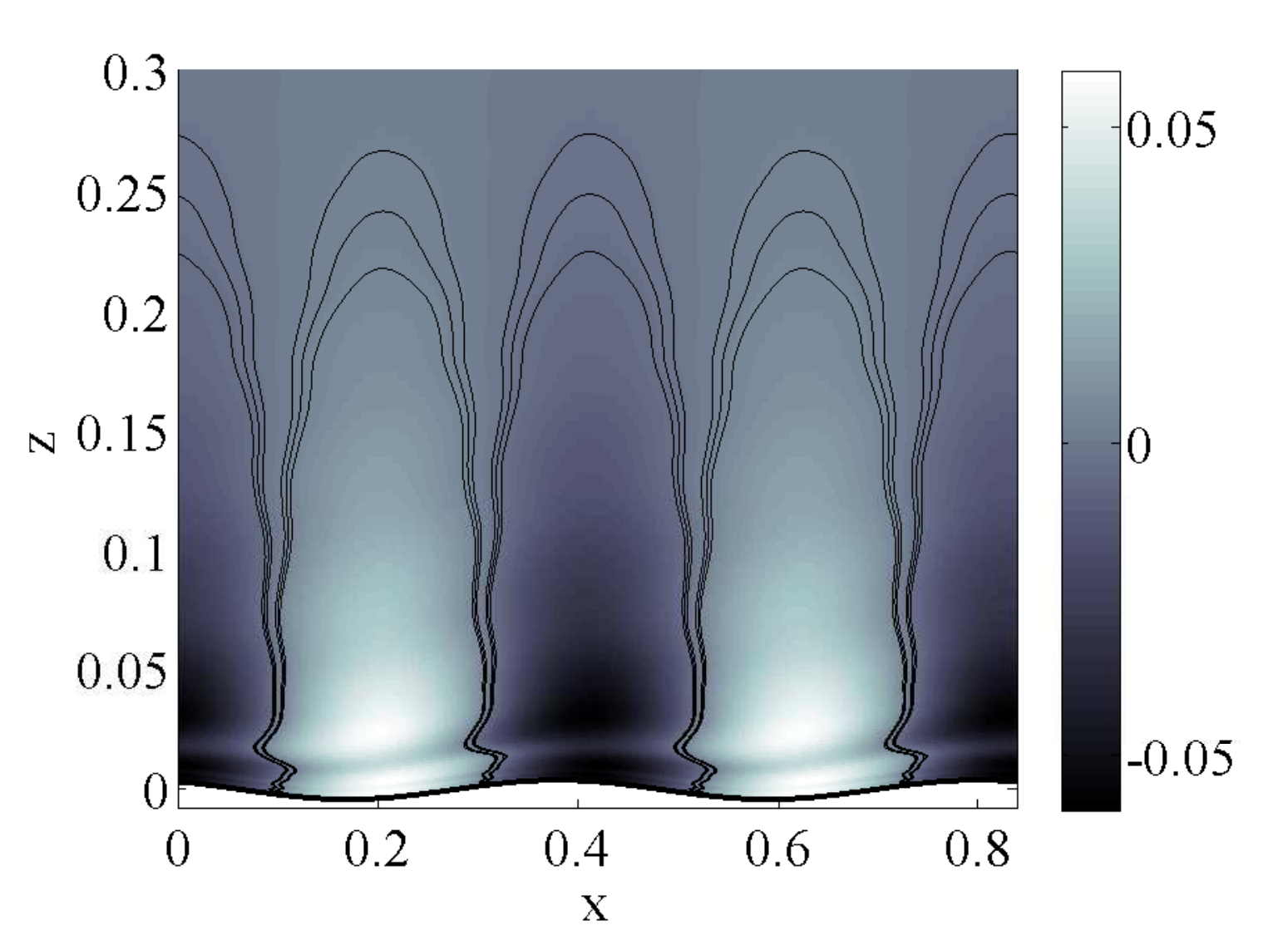}}\\
\subfigure[]{
\includegraphics[width=0.3\textwidth]{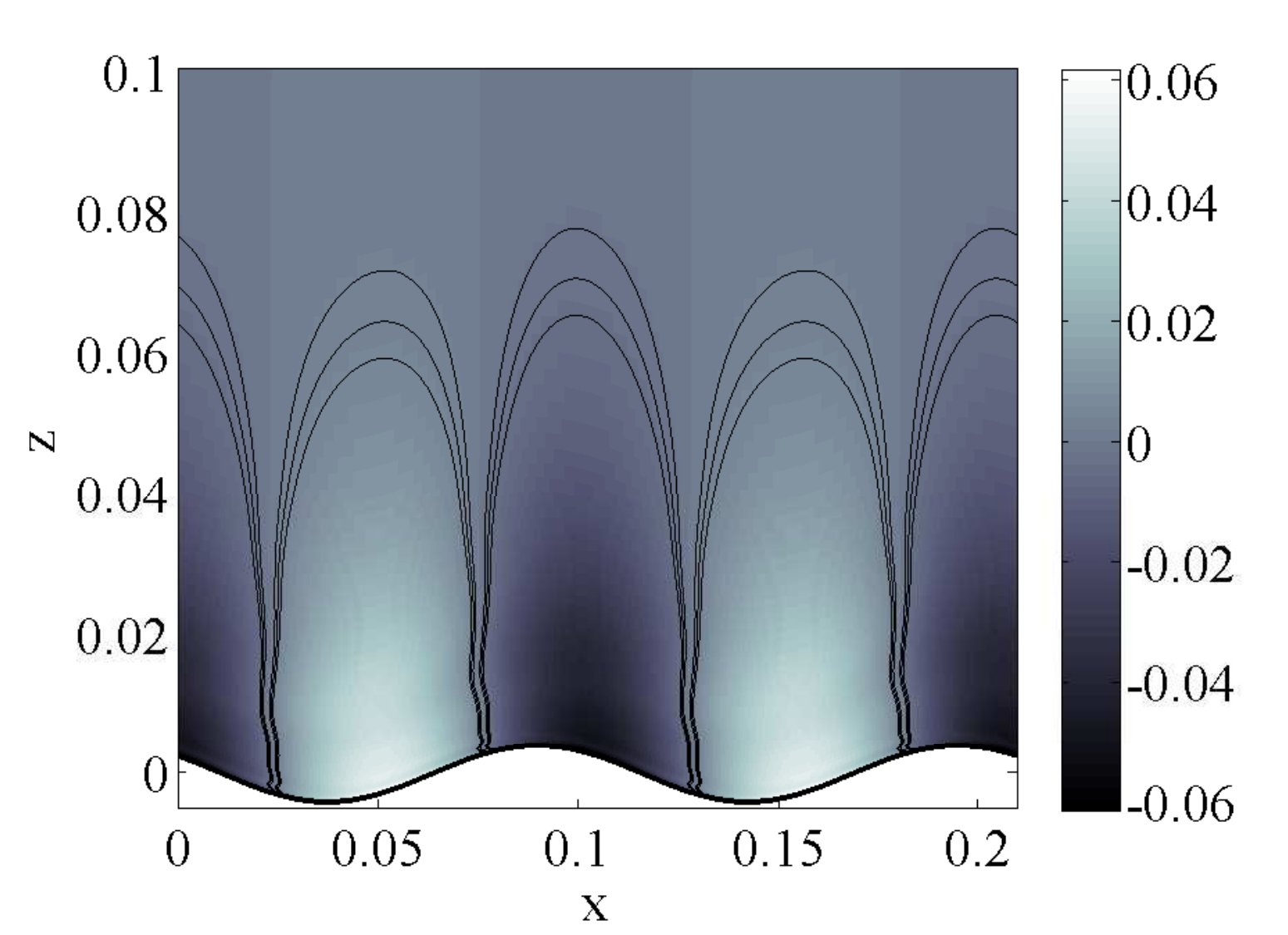}}
\caption{The pressure distribution for $\alpha=15$, (a), and $\alpha=60$, (b), normalized by $\max|u|$, the maximum streamwise perturbation velocity.}
\label{fig:pressure_thif}
\end{figure}
manifests itself as streaks, shown in Fig.~\ref{fig:velocity_thif15}~(a),
 which are inclined at an acute angle relative to the interface.  These oscillatory structures are all but invisible in the $\alpha=60$
case (Fig.~\ref{fig:velocity_thif60}).

We plot the pressure distribution
in Fig.~\ref{fig:pressure_thif}.  The minimum pressure is almost exactly
out of phase with the interfacial variation at $\alpha=15$, while at $\alpha=60$
the pressure minimum is shifted slightly downstream of the free-surface maximum.
 The co-incidence of the pressure minimum and the free-surface maximum is
 explained by Bernoulli's principle, since the gas viscosity is small.  The
 in-phase component of the pressure at $\alpha=60$ is a consequence of the  so-called quasi-separated sheltering~\cite{Benjamin1958,
 Boomkamp1997}, wherein the viscous sublayer creates a wake downstream of
 the crest, which reduces the pressure fluctuation there.

The other phase relationships, namely those between the velocity at the interface
and the interfacial
disturbance can also be understood quite readily.  The normal component of
the velocity possesses
successive extrema (maxima or minima).  The first set of extrema is located
at the surface $z=0$, and can be understood by examination
of the  kinematic condition,
which in the moving frame of reference is simply $\partial\eta/\partial t=w$
at the interface.  Upstream of a wave crest that is propagating from left
to right ($c_\mathrm{r}>0$), the interfacial height decreases in time, which implies that the
velocity $w$ is negative there, while downstream of the wave crest, the vertical
velocity component is positive.
Moreover, since $w=\Re\left[i\alpha
e^{i\alpha\left(x-ct\right)}\phi\right]$, and $u=\Re\left[e^{i\alpha\left(x-ct\right)}\left(d\phi/dz\right)\right]$,
there is a $\pi/2$ phase difference between $u$ and $w$, and thus the structure
of the streamwise velocity can be understood simply as a phase shift relative
to the normal velocity, in order to satisfy the continuity condition.

Finally, we make use of an energy-decomposition to pinpoint the source of the instability.
\begin{table}[h!b!p!]
\centering
\begin{tabular}{|c||c|c|c|c|c|c|c|c|c|}
\hline
$\alpha=15$&$KIN_L$&$KIN_G$&$REY_L$&$REY_G$&$DISS_L$&$DISS_G$&$TURB$&$NOR$&$TAN$\\
\hline
\hline
PTS&0.97&0.03&0.02&9.21&-0.20&1.35&0&0.000&1.000\\
\hline
No PTS&0.005&0.012&-0.001&-0.185&-0.019&-0.738&-0.039&0.000&1.000\\
\hline
\end{tabular}
\vskip 0.1in
\begin{tabular}{|c||c|c|c|c|c|c|c|c|c|}
\hline
$\alpha=60$&$KIN_L$&$KIN_G$&$REY_L$&$REY_G$&$DISS_L$&$DISS_G$&$TURB$&$NOR$&$TAN$\\
\hline
\hline
No PTS&0.008&0.011&-0.002&-0.089&-0.024&-0.863&0&-0.004&1.000\\
\hline
PTS&0.007&0.011&-0.002&-0.103&-0.023&-0.832&-0.018&-0.004&1.000\\
\hline
\end{tabular}
\caption{Energy budget for the thin-film flow, normalized on the tangential
term.}
\label{tab:eb_thif}
\end{table}
 This decomposition or budget is obtained from the RANS equations, and was
 introduced by Boomkamp and Miesen~\cite{Boomkamp1996}:
\begin{subequations}
\begin{equation}
r_j\left(\frac{\partial}{\partial t}\delta\bm{u}_j+\bm{U}_j\cdot\nabla\delta\bm{u}_j+\delta\bm{u}_j\cdot\nabla\bm{U}_j\right)=\nabla\cdot\delta\Ttensor^{(j)}-r_j\nabla\cdot\delta\rtensor^{(j)}
\end{equation}
\begin{equation}
\delta\Ttensor=\left(\begin{array}{cc}-\delta\pi&\mu\left(\partial_x\delta
w+\partial_z\delta u\right)\\\mu\left(\partial_x\delta
w+\partial_z\delta u\right)&-\delta\pi
\end{array}\right),\qquad
%\end{equation}
%
%
%\begin{equation}
\delta\rtensor=\left(\begin{array}{cc}-\delta\rtensor_{11}+\delta\rtensor_{22}&-\delta\rtensor_{12}\\-\delta\rtensor_{12}&0
\end{array}\right)
\end{equation}
\begin{equation}
\nabla\cdot\bm{u}_j=0,
\end{equation}%
which we multiply by the velocity $\delta\bm{u}_j$ and integrate over space.  We obtain the following
balance equation:
\begin{equation}
\sum_{j=L,G}\mathrm{KIN}_j=\sum_{j=L,G}\mathrm{REY}_j+\sum_{j=L,G}\mathrm{DISS}_j+\sum_{j=L,G}\mathrm{TURB}_j+\mathrm{INT},
\end{equation}
where
\begin{eqnarray}
\mathrm{KIN}_j&=&\tfrac{1}{2}\frac{\mathd}{\mathd t}\int dx \int dz\, r_j\delta\bm{u}_j^2,\\
\mathrm{REY}_j&=&-r_j\int dx \int dz\, \delta u_j\delta w_j\frac{\mathd U_j}{\mathd z},\\
\mathrm{DISS}_j&=&-\frac{m_j}{Re}\int dx \int dz\,\left[2\left(\frac{\partial{}}{\partial{x}}\delta
u_j\right)^2
+\left(\frac{\partial}{\partial z}\delta u_j+\frac{\partial }{\partial
x}\delta w_j\right)^2+2\left(\frac{\partial }{\partial z}\delta w_j\right)^2\right],\\
\mathrm{TURB}_j&=&
\delta_{j,G}\bigg\{r\int
dx \int dz\,\left[\delta\rtensor\frac{\partial}{\partial x}\delta u+\delta\rtensor_{12}\left(\frac{\partial}{\partial
z}\delta u+\frac{\partial}{\partial x}\delta w\right)\right]\bigg\}.
\end{eqnarray}%
\label{eq:eb}%
\end{subequations}%
Additionally,
\[
\mathrm{INT}=\int dx \,\left[\delta u_L \delta\Ttensor_{L,zx}+\delta w_L\Ttensor_{L,zz}\right]_{z=0}
-\int dx \,\left[\delta u_G \delta\Ttensor_{G,zx}+w_G\delta\Ttensor_{G,zz}\right]_{z=0},
\]
which is decomposed into normal and tangential contributions,
\[
\mathrm{INT}=\mathrm{NOR}+\mathrm{TAN},
\]
where
\[
\mathrm{NOR}=\int dx\,\left[\delta w_L\delta\Ttensor_{L,zz}-\delta w_G\delta\Ttensor_{G,zz}\right]_{z=0},
\]
and
\[
\mathrm{TAN}=\int dx \,\left[\delta u_L\delta\Ttensor_{L,zx}-\delta u_G \delta\Ttensor_{G,zx}\right]_{z=0}.
\]
The results of this study are given in Tab.~\ref{tab:eb_thif}.
We find in each case that the tangential term is the source of the instability.
 This can be re-written as
\[
\mathrm{TAN}=U_G'\left(0\right)\left(m-1\right)\int_0^\ell \delta T_{xz}\left(x,0\right)\eta\left(x\right)dx,
\]
which is positive when $m>1$, and when the viscous shear stress $\delta  T_{xz}\left(x,0\right)$ is less than $\pi/2$ out-of-phase with the interfacial
variation.  Thus, the instability is driven by the viscosity contrast $m>1$.

\subsection{A simplified calculation to elucidate the rapid-distortion effects}
\label{subsec:simple_model}

\begin{figure}[htb]
\centering\noindent
\subfigure[]{\includegraphics[width=0.35\textwidth]{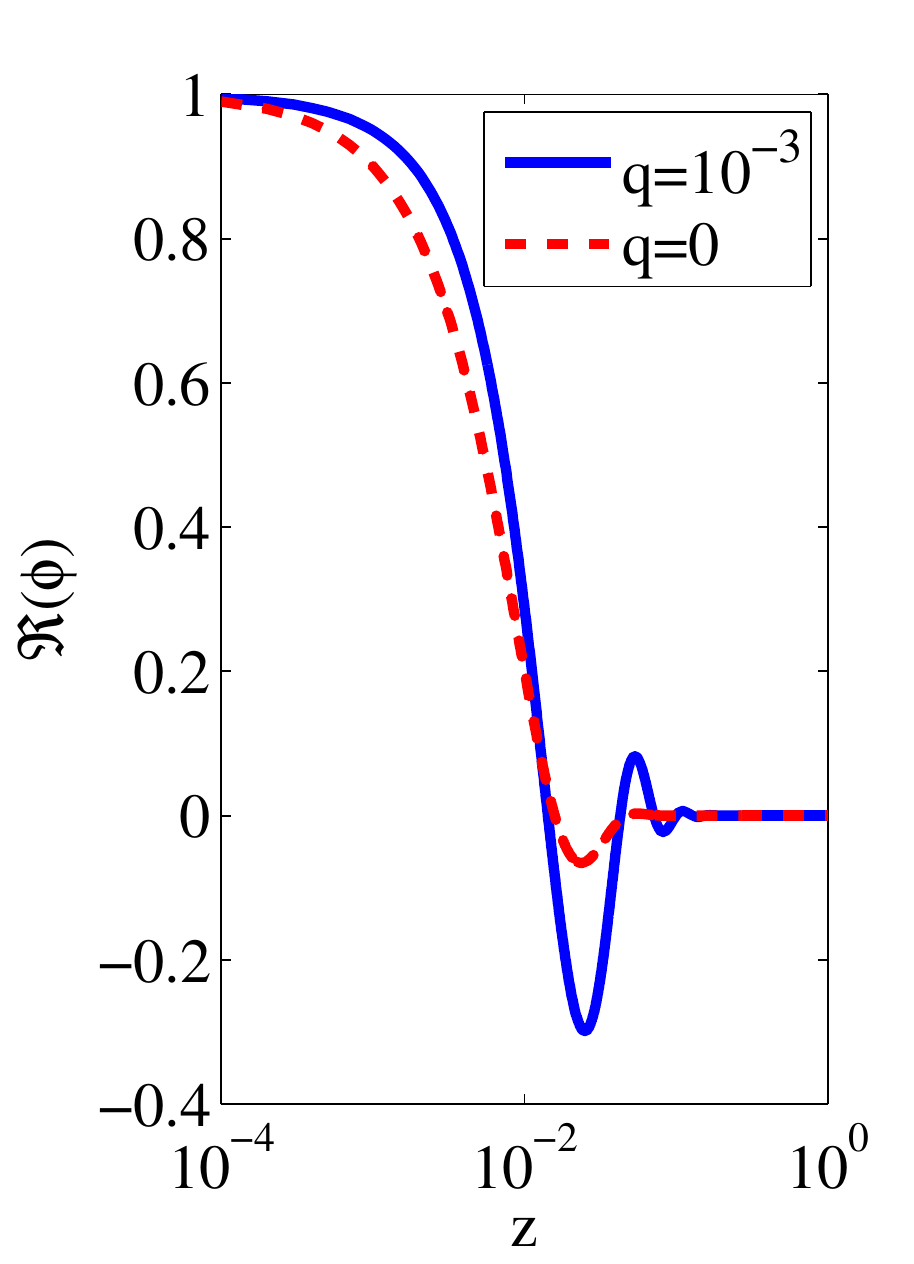}}
\subfigure[]{\includegraphics[width=0.35\textwidth]{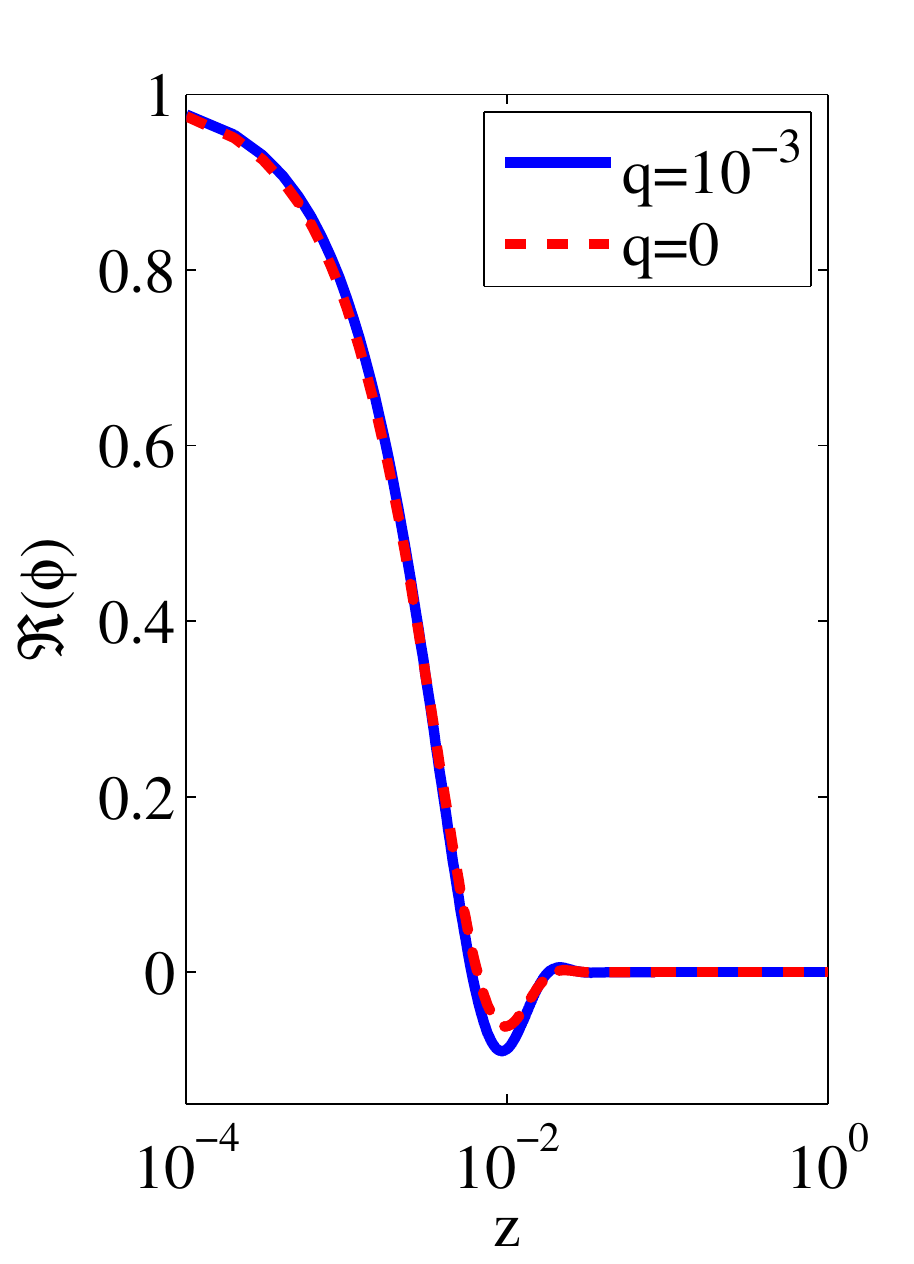}}
\caption{Far-field streamfunction for the simplified model, $Re=8000$, $b=0.25$, and $c$ taken from Fig.~\ref{fig:growth_rate_thif}.  The wavenumber is $\alpha=10$ in (a) and $\alpha=60$ in (b).  At longer wavelengths, the rapid distortion induces an oscillation in the streamfunction, seen in the comparison between the $q_\infty=0$ and $q_\infty=10^{-3}$ curves in (a).  At shorter wavelengths, this effect vanishes (see (b)).  This agrees qualitatively with the streamfunction pattern in the full model.}
\label{fig:bl_simple}%
\end{figure}

To understand the effects of rapid distortion on the \textit{structure} of
the streamfunction, rather than on the stability of the system, we resort
to a highly simplified model, where analytical progress is possible.  This
model possesses as its solution the oscillatory structures visible in Figs.~\ref{fig:velocity_thif15}
and~\ref{fig:velocity_thif60}.
 We perform a linear-stability analysis around the base state
\begin{equation}
U=\begin{cases}
mz,&-1\leq z\leq0,\\
z,&0\leq z\leq b,\\
b,&z\geq b.
\end{cases}
\end{equation}
We make use of the following Orr--Sommerfeld and PTS equations:
\begin{subequations}
\begin{equation}
i\alpha r \left(mz-c\right)\left(\Diff^2-\alpha^2\right)\phi_L=\frac{m}{Re}\left(\Diff^2-\alpha^2\right)^2\phi_L,\qquad
-1\leq z\leq0,
\end{equation}
\begin{equation}
i\alpha \left(z-c\right)\left(\Diff^2-\alpha^2\right)\phi_G=\frac{1}{Re}\left(\Diff^2-\alpha^2\right)^2\phi_G,\qquad
0\leq z\leq b,
\end{equation}
\begin{equation}
i\alpha \left(b-c\right)\left(\Diff^2-\alpha^2\right)\phi_G=\frac{1}{Re}\left(\Diff^2-\alpha^2\right)^2\phi_G+\left(\Diff^2+\alpha^2\right)\tau,\qquad
z\geq b,
\end{equation}
\begin{equation}
i\alpha\left(U_0-c\right)\tau=-q_\infty\left(\alpha_1\Diff^2\phi_G+\alpha_2\alpha^2\phi_G+i\alpha\alpha_3\Diff\phi_G\right),
\end{equation}%
\label{eq:simp_model}%
\end{subequations}%
where now $q_\infty$, the constant turbulent kinetic energy in the far field, can be thought of as parametrizing the PTS.
Given a set of boundary and interfacial conditions, it is possible to obtain
a closed-form solution to this set of equations, in terms of exponentials
and integrals of Airy functions~\cite{DrazinReidBook,abram_stegun}.  Our goal here, however, is simply to elucidate
the oscillatory nature of the solution found in the full analysis.
 To that end, we take a characteristic value of the wave speed $c$, and obtain
 a solution to the system~\eqref{eq:simp_model} in the far field $z\geq b$.
  There, the solution is $\phi_G=e^{\varsigma z}$, where $\varsigma$ solves  a fourth-order polynomial equation:
\begin{multline}
\left(1+i\alpha_1 Re_q\right)\left(\varsigma/\alpha\right)^4-\alpha_3Re_q\left(\varsigma/\alpha\right)^3
+\left[i\left(\alpha_1+\alpha_2\right)Re_q-2-iRe_b\right]\left(\varsigma/\alpha\right)^2
\\
-\alpha_3 Re_q\left(\varsigma/\alpha\right)+i Re_q\alpha_2+1+i Re_b=0,
\label{eq:poly}
\end{multline}
where
\[
Re_q=\frac{Req_\infty}{\alpha\left(b-c\right)},\qquad Re_b=\frac{Re\left(b-c\right)}{\alpha}.
\]
When $b\gg |c|$, we have the condition $|Re_b|\gg |Re_q|$, and we can treat $|Re_q|^{-1}$ as a small expansion parameter. The lowest-order solution of Eq.~\eqref{eq:poly} is then $\varsigma^2=\alpha^2$ or $\alpha^2+i\alpha Re\left(b-c\right)$, and the rapid-distortion effects appear at first order.  In fact,
%
%
%\begin{multline*}
%\sigma=
%-\alpha\left[1+\frac{q_\infty}{\left(b-c\right)^2}\left(\alpha_1+\alpha_2- % i\alpha_3\right)\right]\text{ or}\\
%\alpha\frac{q_\infty\alpha_3}{\left(b-c\right)^2}\left(\tfrac{1}{2}Re_b-i\right)-\alpha\sqrt{1+iRe_b}\times\\
%\bigg\{1-\frac{q_\infty}{\left(1+Re_b^2\right)\left(b-c\right)^2}\left[\alpha_2\left(1+\tfrac{1}{2}Re_b^2\right)+2\alpha_1Re_b^2+i\left(\alpha_1Re_b\left(\tfrac{3}{2}-Re_b\right)-\tfrac{1}{2}\alpha_2 % Re_b-\alpha_3 Re_b\right)\right]\bigg\}\\
%\end{multline*}
%
%
\begin{equation}
\varsigma=\begin{cases}
\varsigma_1=-\alpha\left[1+\frac{q_\infty}{\left(b-c\right)^2}\left(\alpha_1+\alpha_2- i\alpha_3\right)\right],\\
\varsigma_2=-\alpha\sqrt{iRe_b}\left[1+\frac{q_\infty}{\left(b-c\right)^2}\frac{\alpha_1Re_b}{2}\right],\qquad |Re_b|\gg1.
\end{cases}
\label{eq:poly_sln}
\end{equation}
 Note, however, that the rapid-distortion effects appear non-perturbatively in $\varsigma_2$ for sufficiently large values of $|Re_b| q_\infty$.  The exact condition is $\alpha_1 q_\infty Re_b/\left(b-c\right)^2=O\left(1\right)$, which equates to
\begin{equation}
%\frac{q_\infty Re |b-c|}{\alpha}=O\left(1\right),\qquad
\alpha\apprle \frac{\alpha_1q_\infty Re}{|b-c|}.
\label{eq:poly_cond}
\end{equation}
Using typical values for the thin-film waves ($Re=10^4$, $q_\infty=10^{-3}$, $|b-c|=10^{-1}$, $\alpha_1=10^{-1}$), we expect rapid distortion to affect the streamfunction at lowest order when $\alpha\apprle 10$.
This analysis is verified by the streamfunction plots in Fig.~\ref{fig:bl_simple}, using $q_\infty=10^{-3}$,
where we take $b=0.25$, and $Re=8000$.  The coefficients $\alpha_1$, $\alpha_2$, and $\alpha_3$
are taken from Sec.~\ref{subsec:turb_domains}, while the $c$-values are taken from Fig.~\ref{fig:growth_rate_thif} at $\alpha=10$ and $\alpha=60$.
In Fig.~\ref{fig:bl_simple}~(a) we compare the far-field streamfunction associated with the mode $\varsigma_2$ with and without the PTS at $\alpha=10$.  There is a clear difference between the two cases, while the difference between the two modes associated with $\varsigma_1$ is negligible (the difference is small and the plot is not shown).  This is consistent with the analysis in Eqs.~\eqref{eq:poly_sln} and~\eqref{eq:poly_cond}.  The consistency extends to the $\alpha=60$ case (Fig.~\ref{fig:bl_simple}~(b)), where there is little difference between the $\varsigma_1$ and $\varsigma_2$ streamfunctions.  The enhanced oscillations in the streamfunction visible in Fig.~\ref{fig:bl_simple}~(a) are similar to those obtained by Zaki and Saha~\cite{Zaki2009} in their study of the interaction between disturbances in the free stream and the boundary layer.  There, the authors used the continuous-spectrum Orr--Sommerfeld equation as a model, and the qualitative agreement between our findings and theirs further vindicates our results.

Note finally that there is another way for the rapid distortion to make itself felt at zeroth order in the polynomial equation~\eqref{eq:poly}.  When $b\approx c_\mathrm{r}$, the speed $b-c$ is small, and $|Re_q|\gg |Re_b|$.  By treating $|Re_b|^{-1}$ as an expansion parameter, we obtain the zeroth-order polynomial to solve:
\[
i\alpha_1\left(\varsigma/\alpha\right)^4-\alpha_3\left(\varsigma/\alpha\right)^3+i\left(\alpha_1+\alpha_2\right)\left(\varsigma/\alpha\right)^2-\alpha_3\left(\varsigma/\alpha\right)+i\alpha_2=0.
\]
In this case, the oscillatory term in the streamfunction $\phi=e^{\Re\left(\varsigma\right)z}e^{i\Im\left(\varsigma\right)z}$ is due mostly to the rapid distortion, and the contribution from the oscillation coming from the base flow is a perturbation.
This is made manifest when we write down the closed-form solution of the equation that exists when $\alpha_3=0$:
\[
\varsigma=\pm\alpha\sqrt{-\tfrac{1}{2}\left(1+\frac{\alpha_2}{\alpha_1}\right)\pm\tfrac{1}{2}\sqrt{\left(\frac{\alpha_2}{\alpha_1}\right)^2+2\left(\frac{\alpha_2}{\alpha_1}\right)-3}}.
\]
We expect this last mechanism to be important for fast waves, that is, for a case in which the equation $b-c_\mathrm{r}\left(\alpha\right)=0$ has a solution.  We therefore turn our attention to a situation in which such fast waves occur.

%  Lengthscale obtained as a solution to the
%  polynomial equation $b_4\sigma^4+b_3\sigma^3+b_2\sigma^2+b_1\sigma+b_0=0$,
%  where $\Re\left(\sigma\right)\leq0$, and where
% %
% %
% \begin{eqnarray*}
% b_4&=&\frac{1}{Re}+\frac{iq_\infty\alpha_1}{\alpha\left(b-c\right)},\\
% b_3&=&- \frac{q_\infty\alpha_3}{b-c},\\
% b_2&=&-\frac{2\alpha^2}{Re}+\frac{iq_\infty\alpha}{b-c}\left(\alpha_1+\alpha_2\right)-i\alpha\left(b-c\right),\\
% b_1&=&- \frac{q_\infty \alpha^2\alpha_3}{b-c},\\
% b_0&=&\frac{\alpha^4}{Re}+\frac{iq_\infty\alpha^3\alpha_2}{b-c}+i\alpha^3\left(b-c\right).
% \end{eqnarray*}%
%
%
%
%
%
%

\section{Results for deep-water waves}
\label{sec:deep_water}

In this section we investigate the effects of turbulence on the stability
of an interface separating a deep body of liquid from a gas layer that acts
on the interface by a fully-developed turbulent shear flow.

\subsection{Base-state determination}
\label{subsec:base_state_deep}

As before, we use a base state that mimics flow in a boundary layer, and thus, the
flow is confined by a flat plate at $z=d_G$, a large distance from the interface.
 This plate moves at velocity $U_0$ relative to the interface.
Using this framework, the basic velocity is constituted as before: the non-dimensional
velocity is given by Eq.~\eqref{eq:u_nondim}, where $Re_*=\rho_G U_{*\mathrm{i}} d_G/\mu_G$,
and $Re=\rho_G U_0/d_G$.
\begin{subequations}
The eddy-viscosity and wall functions $G$ and $\psi$ have their usual meaning,
given by Eqs.~\eqref{eq:G} and~\eqref{eq:wall_fn} respectively.
The sole difference between the profile in Sec.~\ref{sec:thif} and that used
here is in the liquid, where we make use of the following deep-water profile~\cite{Zeisel2007}:
\begin{equation}
\tilde{U}_L=a\left(e^{b\tilde{z}}-1\right),
\end{equation}
where $a$ and $b$ are constants to be determined.  There is in fact, only
one constant to determine, since the continuity of tangential stress requires
that
\begin{equation}
m\frac{\mathd\tilde{U}_L}{\mathd\tilde{z}}=\frac{\mathd\tilde{U}_G}{\mathd\tilde{z}},\qquad \tilde{z}=0.
\end{equation}
Hence, $mab=Re_*^2/Re$, and there remains a single
free constant $a$ whose value is fixed in Tab~\ref{tab:model_coeff}.  Thus,
the liquid velocity has the reduced form
\begin{equation*}
\tilde{U}_L=a\left(e^{\frac{\tilde{z}Re_*^2}{maRe}}-1\right).
\end{equation*}%
\label{eq:model_base_deep}
\end{subequations}%
\begin{table}[h!b!p!]
\centering
\begin{tabular}{|c||c|c|}
\hline
$a$&Amplitude of drift &Not determined, although a value
$a=U_{*\mathrm{i}}/(2U_0)$ \\
 & & is given in the literature~\cite{Zeisel2007}.\\
\hline
$b$&Decay scale of the&Determined from $a$ and the viscosity
contrast\\
&drift velocity&$m$ through the continuity of tangential stress\\
\hline
$C$&The kinetic energy amplitude &$C=0.55$, to agree with log-layer\\
&is given by $C^{-2}$ in wall units.&conditions in boundary-layer flow.\\
\hline
$n$&Exponent in the Van&Chosen such that the Reynolds stress and the\\
&Driest damping function&kinetic energy mimic wall turbulence as $z\rightarrow0$.\\
\hline
$A$&Length scale in the Van&Chosen such that the linear region of the base\\
&Driest damping function&flow is approximately $5$ wall units in depth.\\
\hline
\end{tabular}
\caption{Summary of the parameters used in Eqs.~\eqref{eq:model_base_deep}.}
\label{tab:model_coeff}%
\end{table}%
\paragraph*{Computation of $z_{\mathrm{t}}$:}  In carrying out the stability
analysis, we have verified that the real part of the wave speed is affected only slightly by the turbulence modelling (less than \%1), and thus the computation of $z_{\mathrm{t}}$
\begin{figure}[htb]
\centering\noindent
\includegraphics[width=0.4\textwidth]{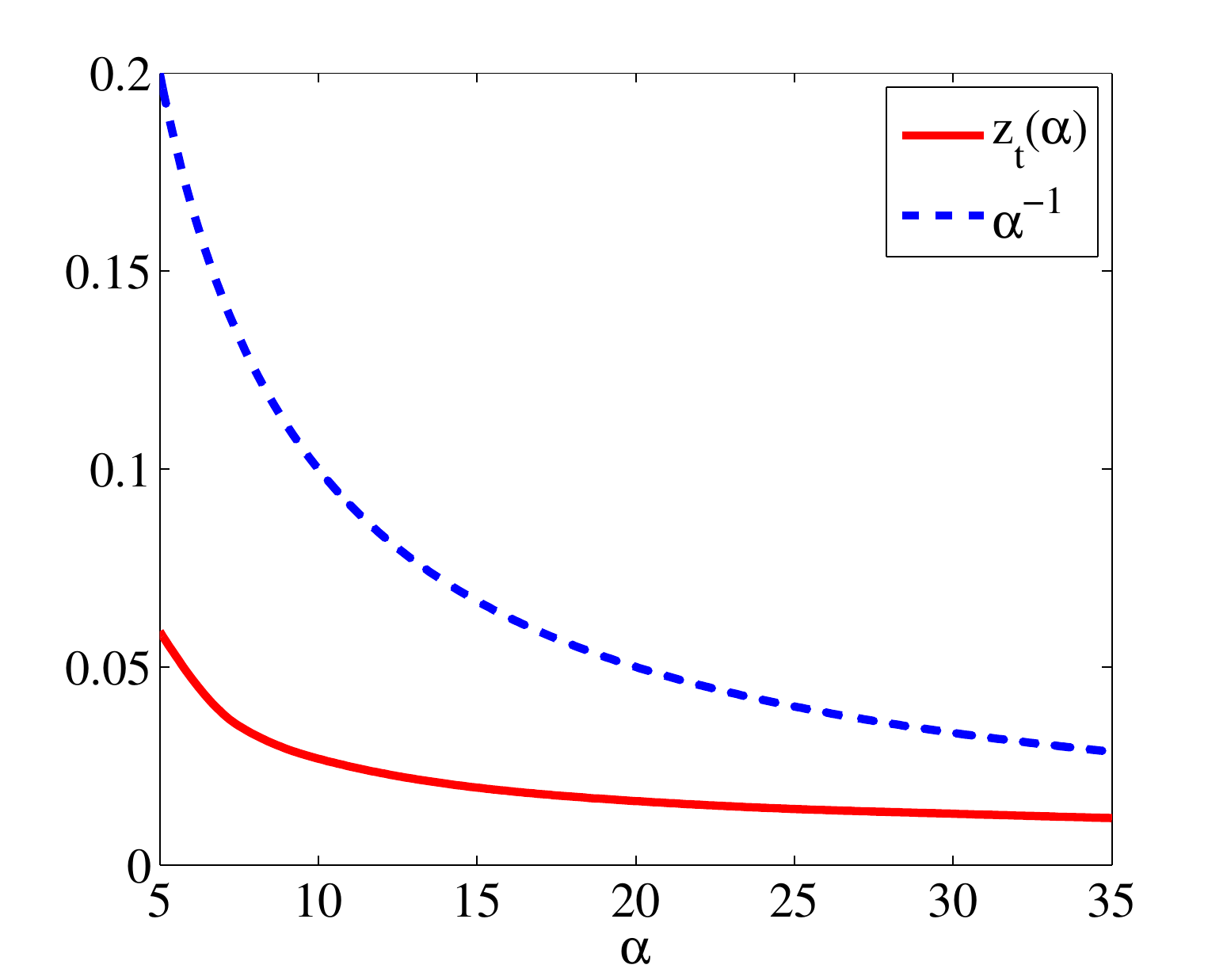}
\caption{The crossover height $z_{\mathrm{t}}$ as a function of wavenumber for $Re=10^5$, $r=1000$, and $m=55$.
 The wavelength $\alpha^{-1}$ is shown for comparison.}
\label{fig:zt}
\end{figure}
can be carried out using the model without the perturbation turbulent stresses
(PTS).  The crossover value $z_{\mathrm{t}}$ where the turbulent and advection
timescales are equal is thus determined by Eq.~\eqref{eq:zt}; the dependence
of $z_{\mathrm{t}}$ on $\alpha$ is shown in Fig.~\ref{fig:zt}, for $Re=10^5$.
 The plot is similar to Fig.~\ref{fig:zt_thif}: the streamfunction extends
 into the domain where rapid distortion is important, and we therefore expect
 to see the shape of the streamfunction adjust to take account of the wave-turbulent
 interactions.  As before, our
 computation of $z_{\mathrm{t}}$ enables us to develop an interpolation
function $\mathcal{I}\left(z\right)$ in the gas: we take $\mathcal{I}\left(z\right)=1-e^{-\left(z/z_{\mathrm{t}}\right)^2}$, and for simplicity,  we replace the $\alpha$-dependent variable $z_{\mathrm{t}}\left(\alpha\right)$ its average value, obtainable from Fig.~\ref{fig:zt}.

\subsection{Linear-stability analysis}
\label{subsec:linsta_deep}

\begin{figure}[htb]
\centering\noindent
\subfigure[]{
\includegraphics[width=0.45\textwidth]{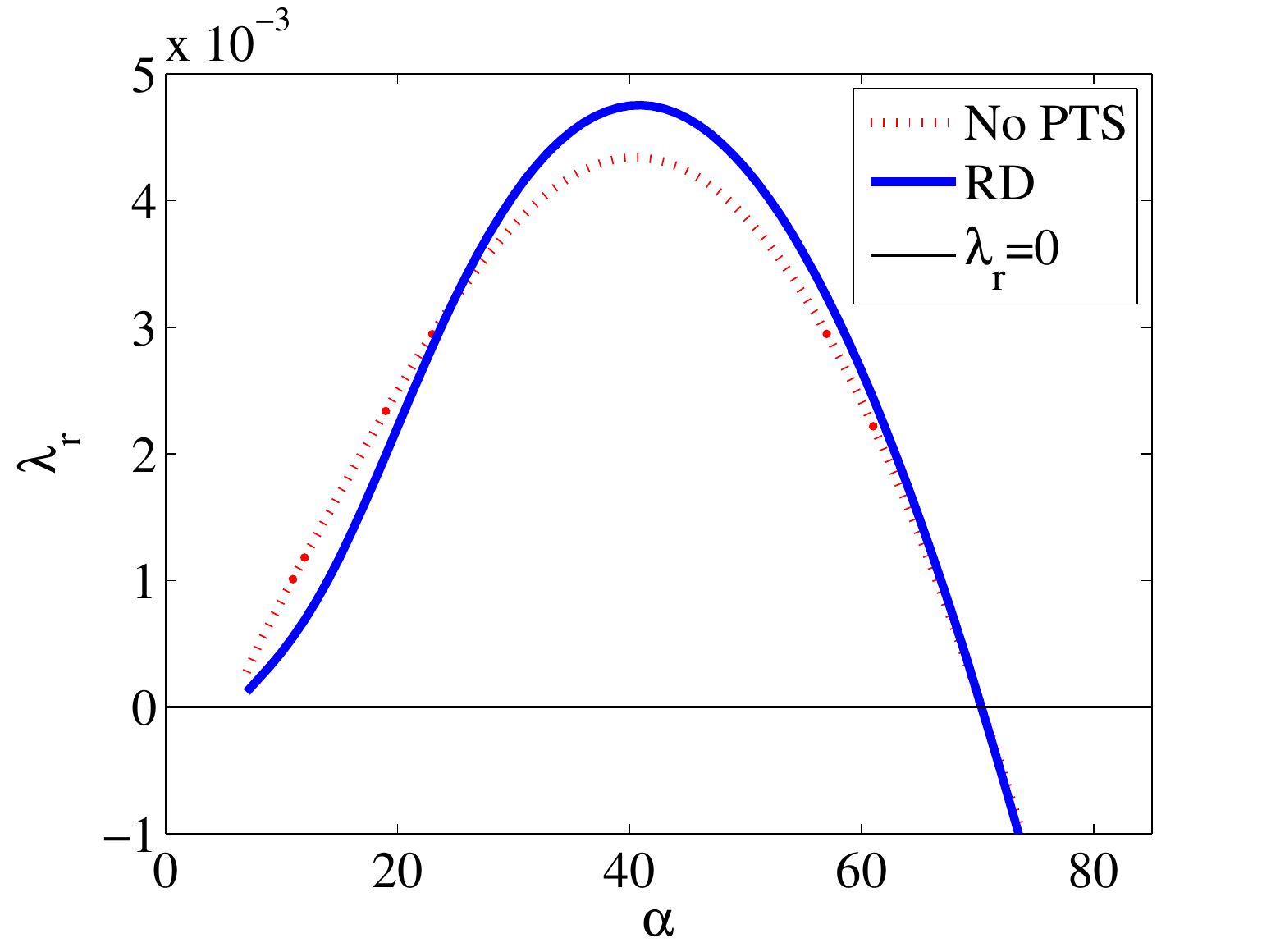}}
\subfigure[]{
\includegraphics[width=0.45\textwidth]{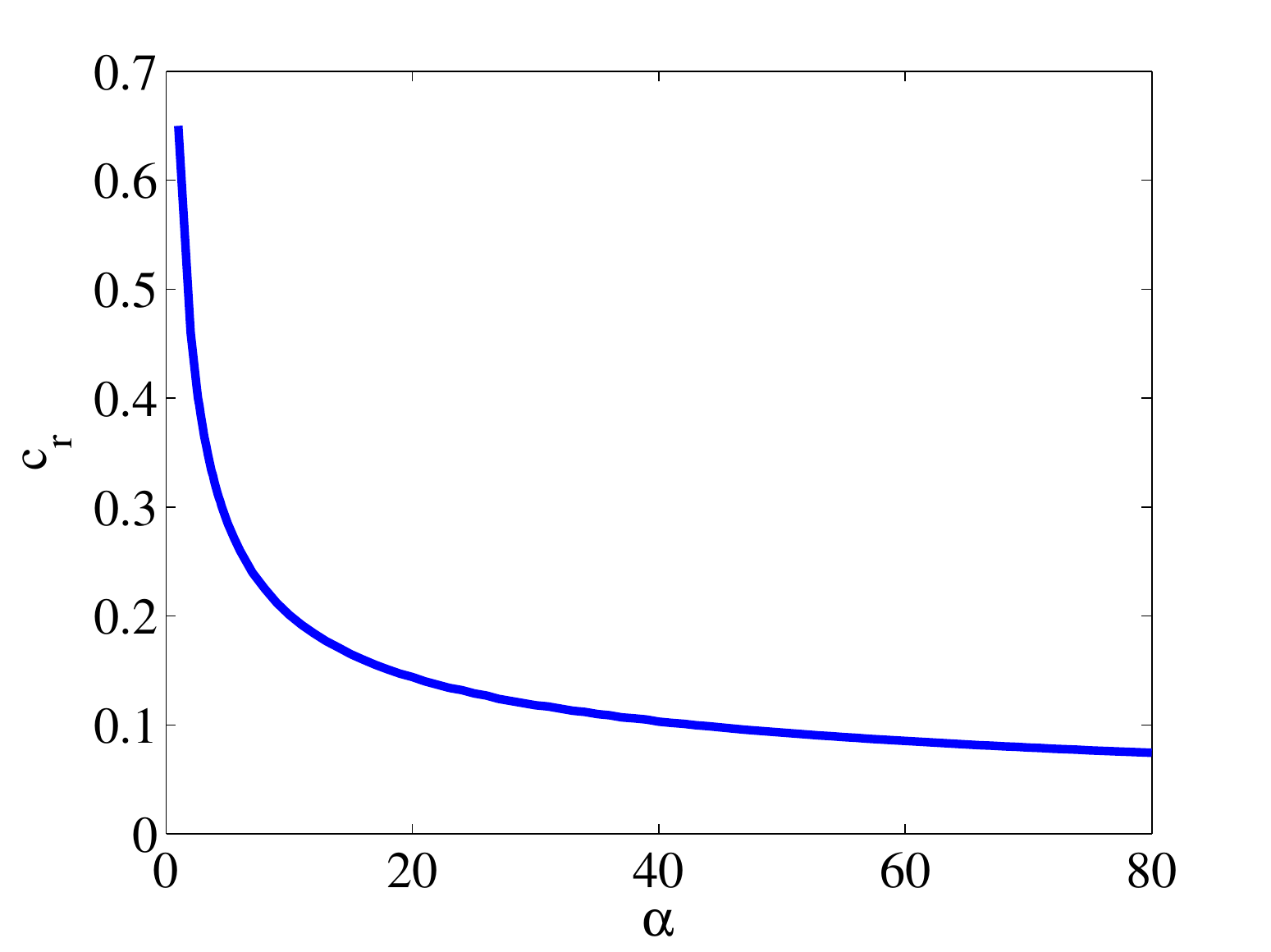}}
\caption{The growth rate and wave speed of the PTS (continuous) and non-PTS waves (dotted lines), as a function
of wavenumber $\alpha$, normalized on the gas-layer thickness.}
\label{fig:growth_rate}
\end{figure}

\begin{figure}[htb]
\centering\noindent
\includegraphics[width=0.45\textwidth]{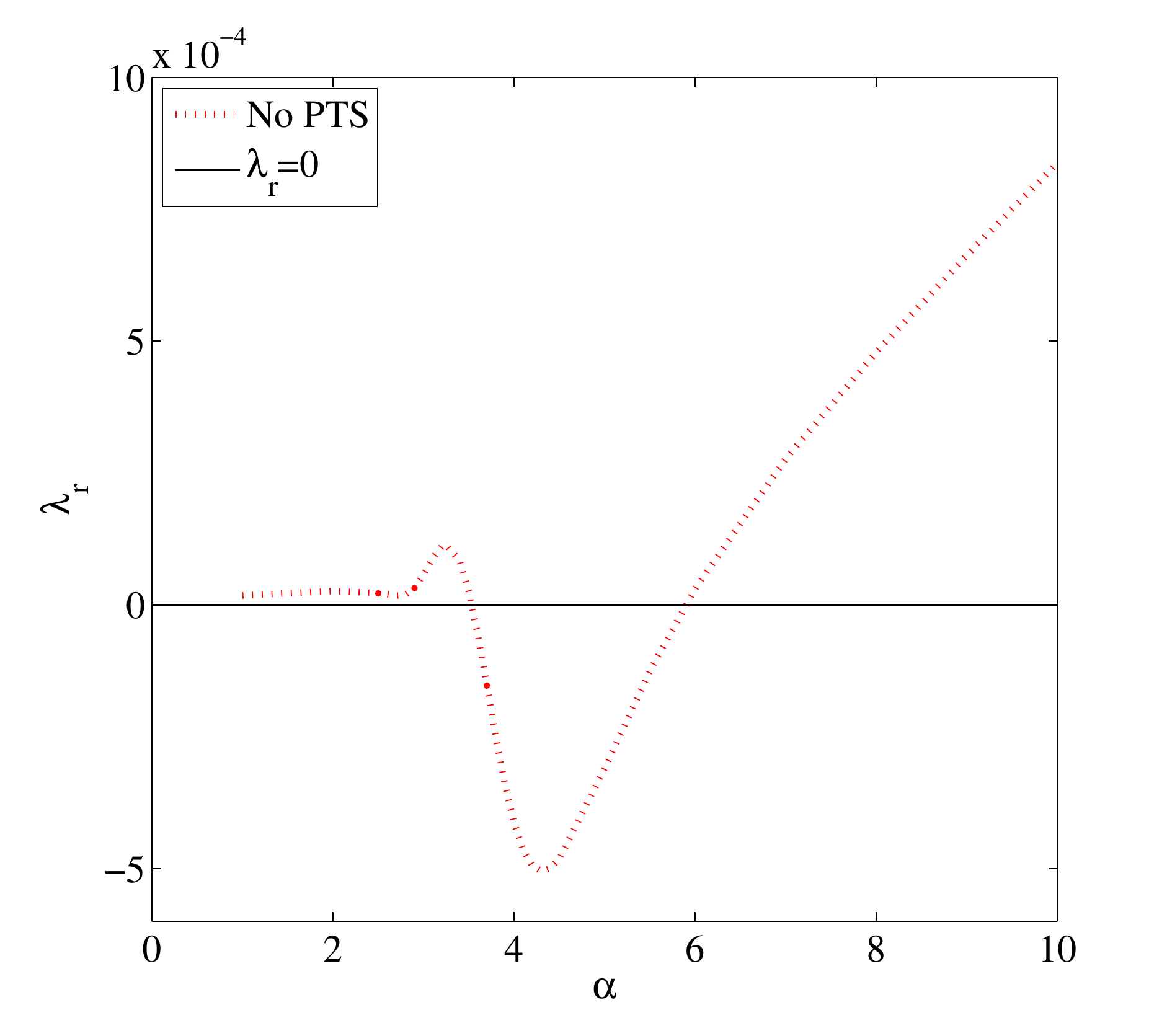}
\caption{The growth rate at long wavelengths in the non-PTS case.}
\label{fig:growth_rate_zoom}
\end{figure}

We carry out a stability analysis around the base state just constituted
at a Reynolds number $Re=10^5$, and at an inverse Froude number $Fr=500$.  The inverse Weber
number $S$ is set to zero, a realistic assumption since the Froude number is large and thus gravity dominates over capillarity.  The density and viscosity ratios
are chosen such that the stability analysis models an air-water system under
standard conditions: $r=1000$ and $m=55$ respectively.  These Reynolds and Froude numbers are chosen such that
the so-called \textit{critical-layer instability} is observed.
We examine the growth rate of the disturbance with and without the PTS.  A description
of the change in the growth rate upon adding the PTS is shown in Fig.~\ref{fig:growth_rate}.
 We also plot the wave speed in Fig.~\ref{fig:growth_rate}, which is unchanged by the PTS-modelling.
The segment of the dispersion
curve that is of interest is the short-wave limit, for which $\alpha\gg2\pi$ (where all lengths are measured relative to the gas-layer height $d_G$).
For longer waves, there is an interaction between the upper plate and the
wave streamfunction, and the model no longer describes boundary-layer phenomena.
The maximum growth rate occurs substantially above this lower cutoff
value, at a wavenumber $\alpha_\mathrm{max}\approx40$, while gravity
stabilizes the wave above an upper critical wavenumber $\alpha_{\mathrm{c}}\approx70$.
The location of the maximum growth rate is not changed by the PTS, although
its value is shifted upwards, by about $10\%$.  We do not continue the PTS dispersion curve below $\alpha=10$, since the PTS streamfunction is difficult to resolve numerically below this value.  We do, however, describe the dispersion curve for the non-PTS case below this threshold in Fig.~\ref{fig:growth_rate_zoom}.  There growth rate oscillates between positive and negative values.  It is tempting to dismiss this as an effect of the upper plate; this is not the case however, since the effect persists upon increasing the depth of the gas layer.  This anomalous region is far from the most dangerous mode, and we do not study it further.
Now the structure of the flow field also changes as a consequence of the
varying level of intensity of the rapid distortion, and it is to this variation that we now turn examining the perturbation velocity at different wavenumbers.

\paragraph*{$\alpha=15$:}

At this wavenumber, the wave parameters (growth rate and propagation speed)
are
\begin{eqnarray}
\lambda_{\mathrm{r}}&=&0.0004,\qquad c_{\mathrm{r}}=0.2011,\qquad\text{with
the PTS};\nonumber\\
\lambda_{\mathrm{r}}&=&0.0008,\qquad c_{\mathrm{r}}=0.2011,\qquad\text{with
no PTS};\nonumber\\
z_{\mathrm{c}}&=&0.0065,
\label{eq:growth_rate10}
\end{eqnarray}
where $z_{\mathrm{c}}$ is the critical height,
for which $U\left(z_{\mathrm{c}}\right)=c_{\mathrm{r}}$.
\begin{figure}[htb]
\centering\noindent
\subfigure[]{
\includegraphics[width=0.3\textwidth]{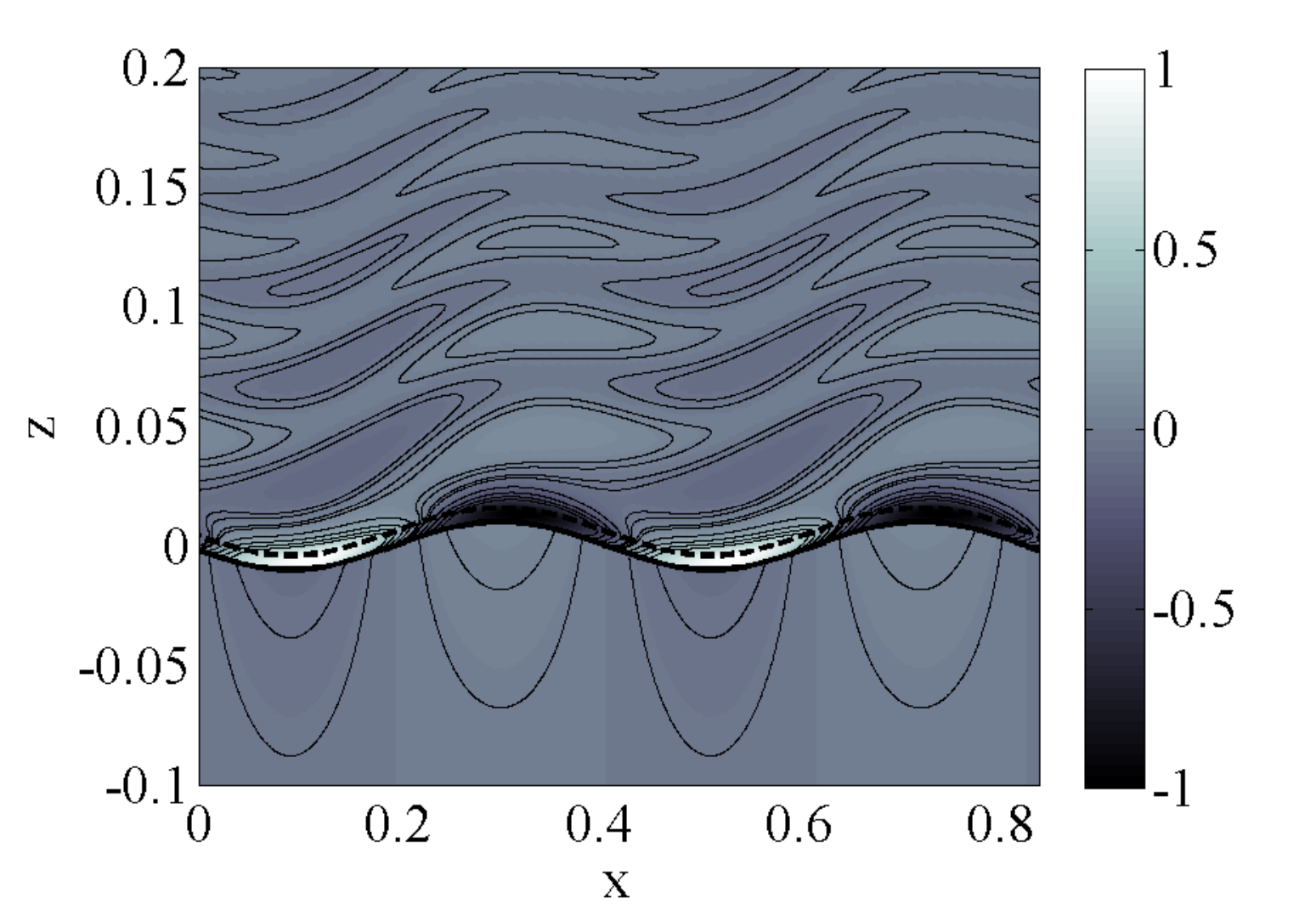}}
\subfigure[]{
\includegraphics[width=0.3\textwidth]{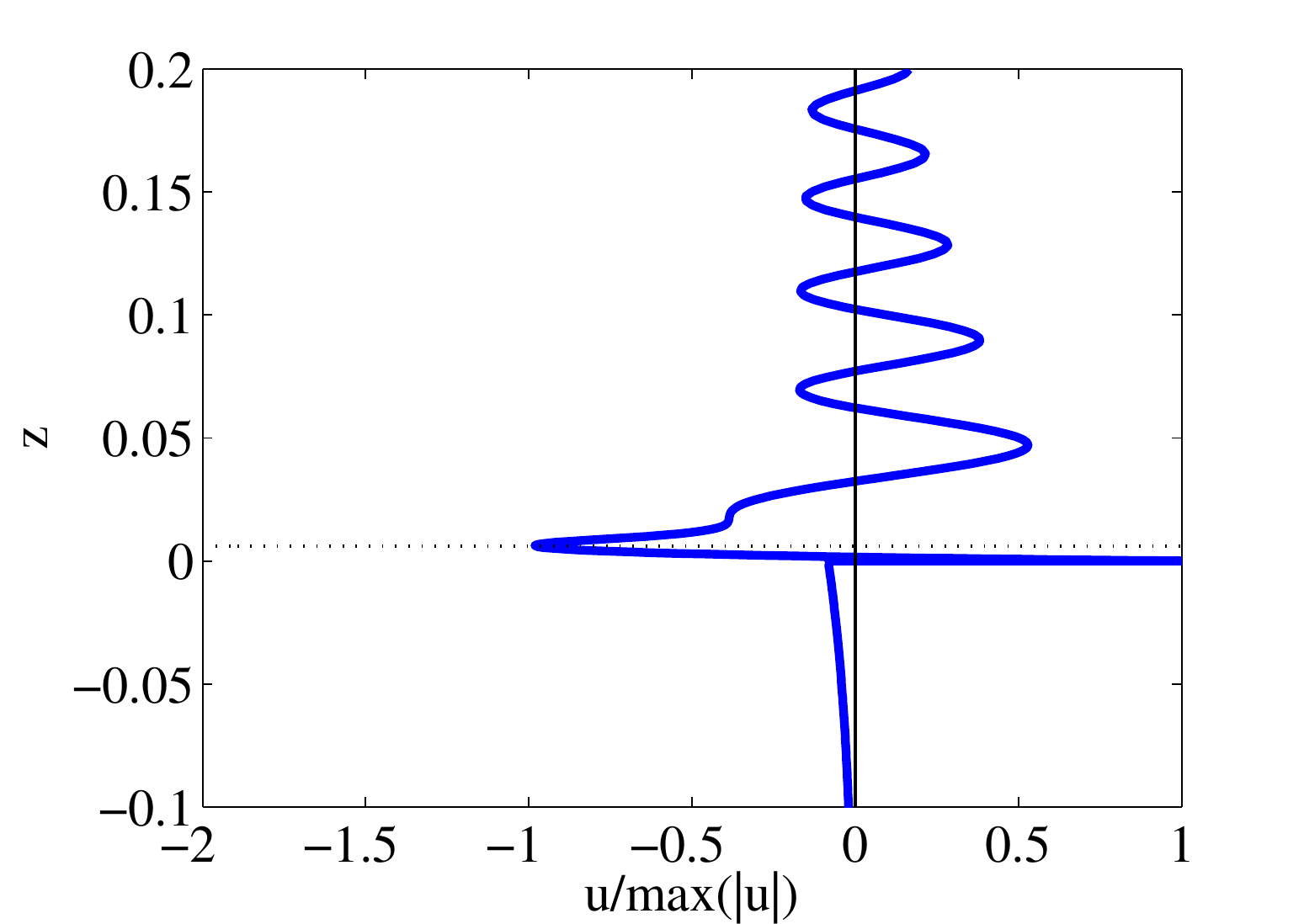}}\\
\subfigure[]{
\includegraphics[width=0.3\textwidth]{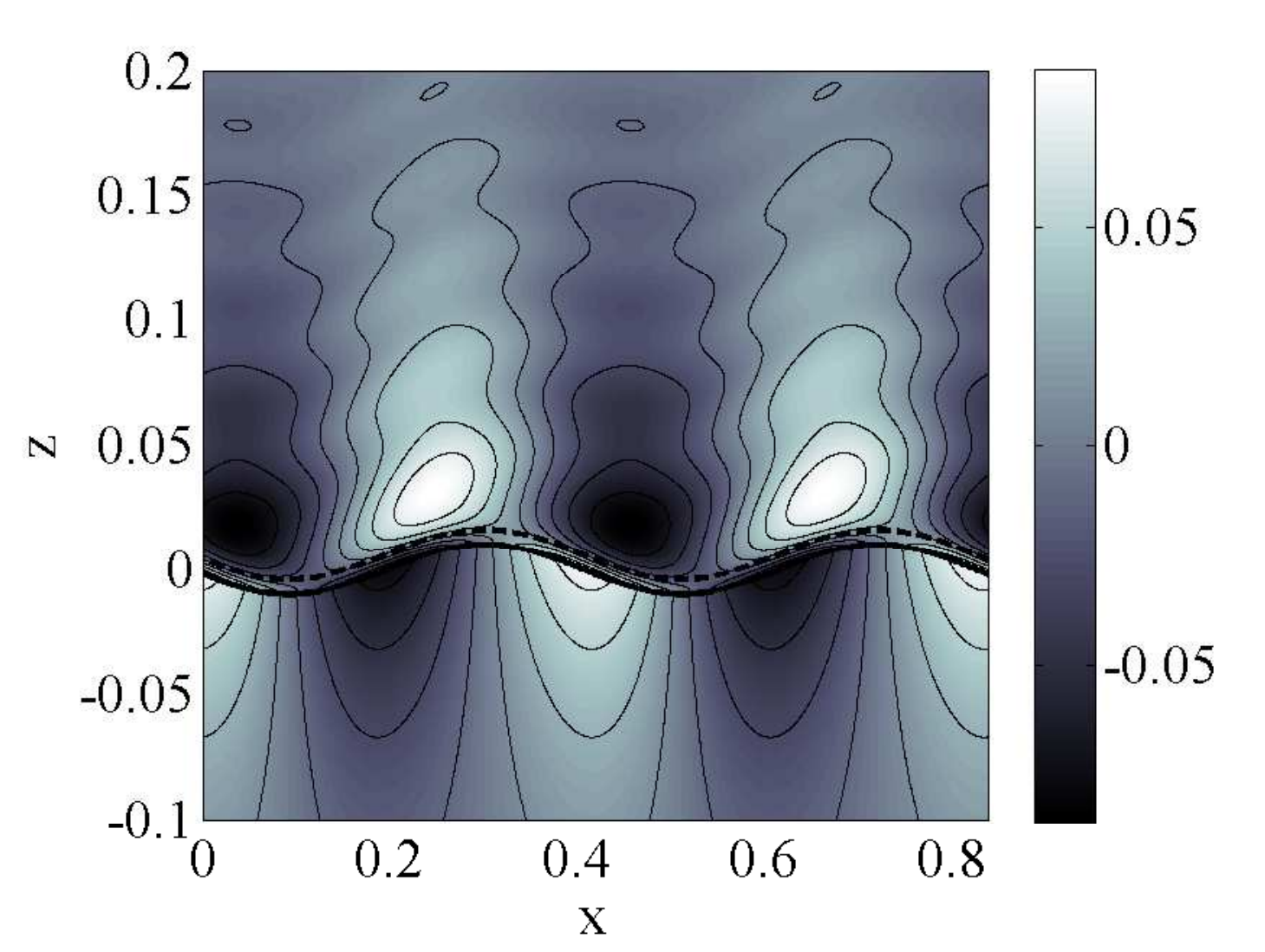}}
\subfigure[]{
\includegraphics[width=0.3\textwidth]{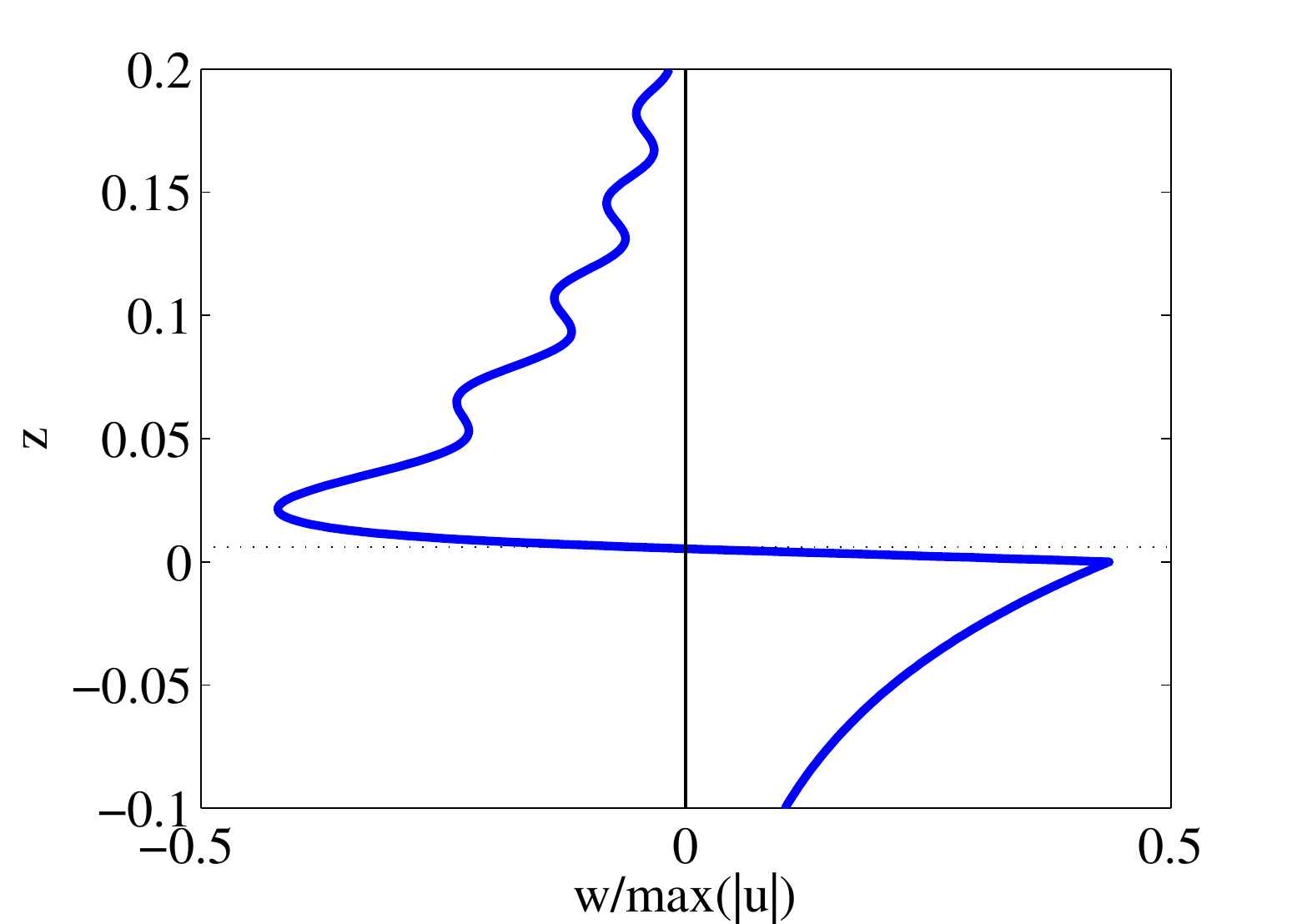}}\\
\subfigure[]{
\includegraphics[width=0.3\textwidth]{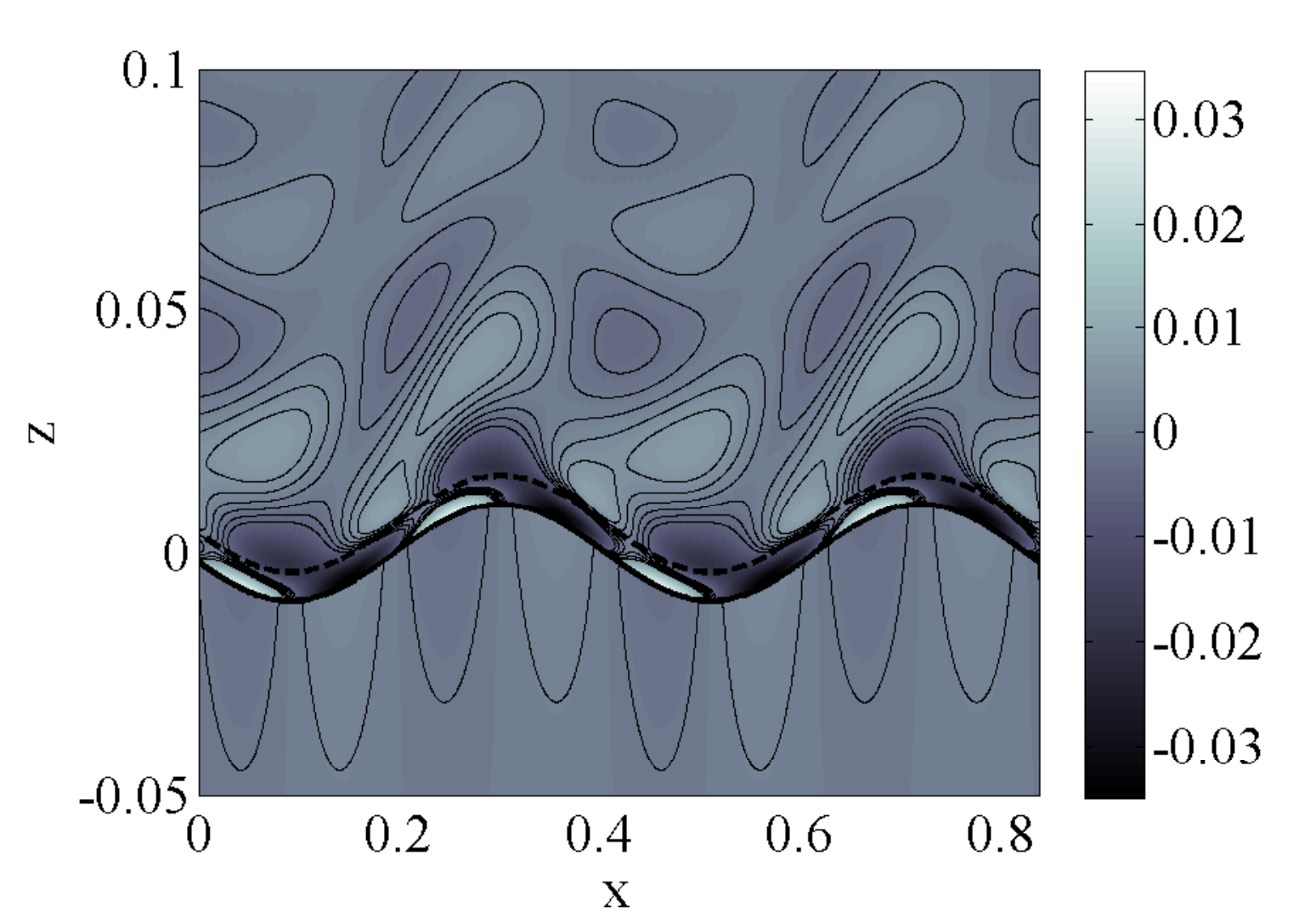}}
\subfigure[]{
\includegraphics[width=0.3\textwidth]{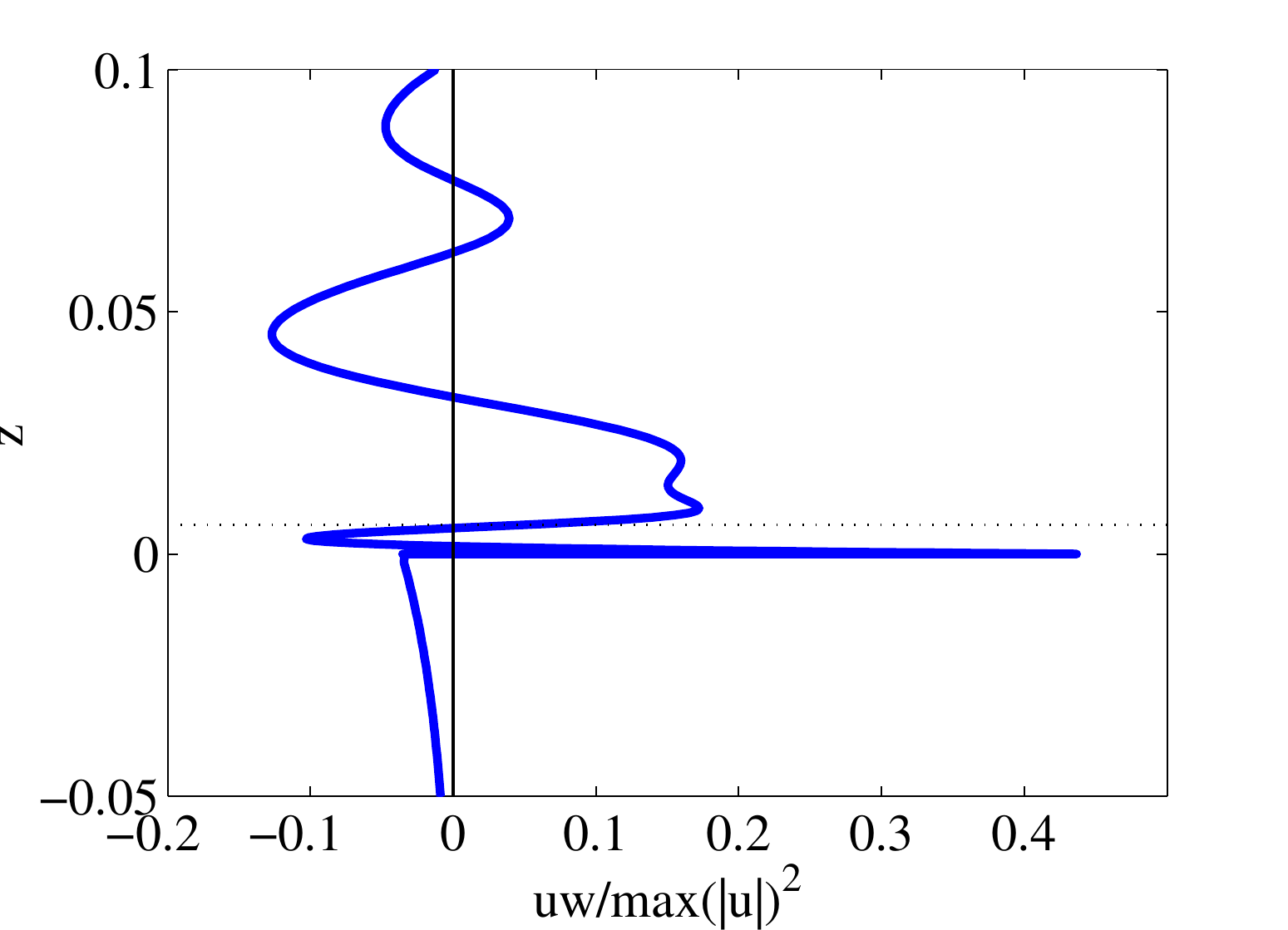}}
\caption{Flow field structure for $\alpha=15$.  The critical layer is shown as a dotted line above the interface.  The effects of the rapid
distortion are particularly visible in subfigures (a) and~(b), which display
the streamwise velocity and the streamwise velocity at $x=0$, respectively.
 Subfigures (c) and ~(d) show the normal velocity and the normal velocity
 at $x=0$.  The effects of the rapid distortion are again visible in the
 pre-averaged version of the wave Reynolds stress, namely the product $uw$,
 shown in (e) and~(f).  In each case, we have normalized the velocities
 by
 $\max|u|$.
 }
\label{fig:velocity10}
\end{figure}
% %
% %
% %
% %
% %
% which is consistent with the decrease in the $REY_G$ term in the energy
% budget,
% as shown in Tab.~\ref{tab:eb10}.
%
%
The velocity field associated with this wave is shown in Fig.~\ref{fig:velocity10}.
 These figures all possess similar features,
regardless of the PTS.  The normal component of the velocity possesses
successive extrema (maxima or minima).  The first set of extrema is located
at the surface $z=0$, and, as in Sec.~\ref{sec:thif}, can be understood
by examination of the  kinematic condition, which in the moving frame
of reference is simply $\partial h/\partial t=w$
at the interface.  Upstream of a wave for which $c_{\mathrm{r}}>0$, the interfacial
height decreases in time, implying that $w<0$ there; similarly,  $w>0$ in
the downstream region.  In contrast to Sec.~\ref{sec:thif}, there is another
set of extrema located above the critical layer; this is discussed below
in the context of the $\alpha=40$ wave.
In addition, since $w=\Re\left[i\alpha
e^{i\alpha\left(x-ct\right)}\phi\right]$, and $u=\Re\left[e^{i\alpha\left(x-ct\right)}\left(d\phi/dz\right)\right]$,
there is a $\pi/2$ phase difference between $u$ and $w$, and thus the structure
of the streamwise velocity can be understood simply as a phase shift relative
to the normal velocity.  The phase relationships
 discussed combine to give a distinctive phase relationship for the pressure
 field, shown in Fig.~\ref{fig:pressure_deep}.  In particular, at maximum
 growth, the pressure and the interfacial wave are approximately $\pi/4$
 out of phase.
Furthermore, the far-field structure of the PTS and non-PTS
waves differ by the presence in the former case of successive velocity extrema,
succeeding the one associated with the critical layer.  These are due to
 the effects of the rapid distortion on the wave-induced motion, and are
 visible as velocity `streaks' in the streamwise direction (see Fig.~\ref{fig:velocity10}~(a)); these streaks are tilted forwards in the direction of the base flow.
%These streaks are oriented at an angle of approximately $30^{\mathrm{o}}$
%
%
%
%
%
\begin{figure}[htb]
\centering\noindent
\subfigure[]{
\includegraphics[width=0.3\textwidth]{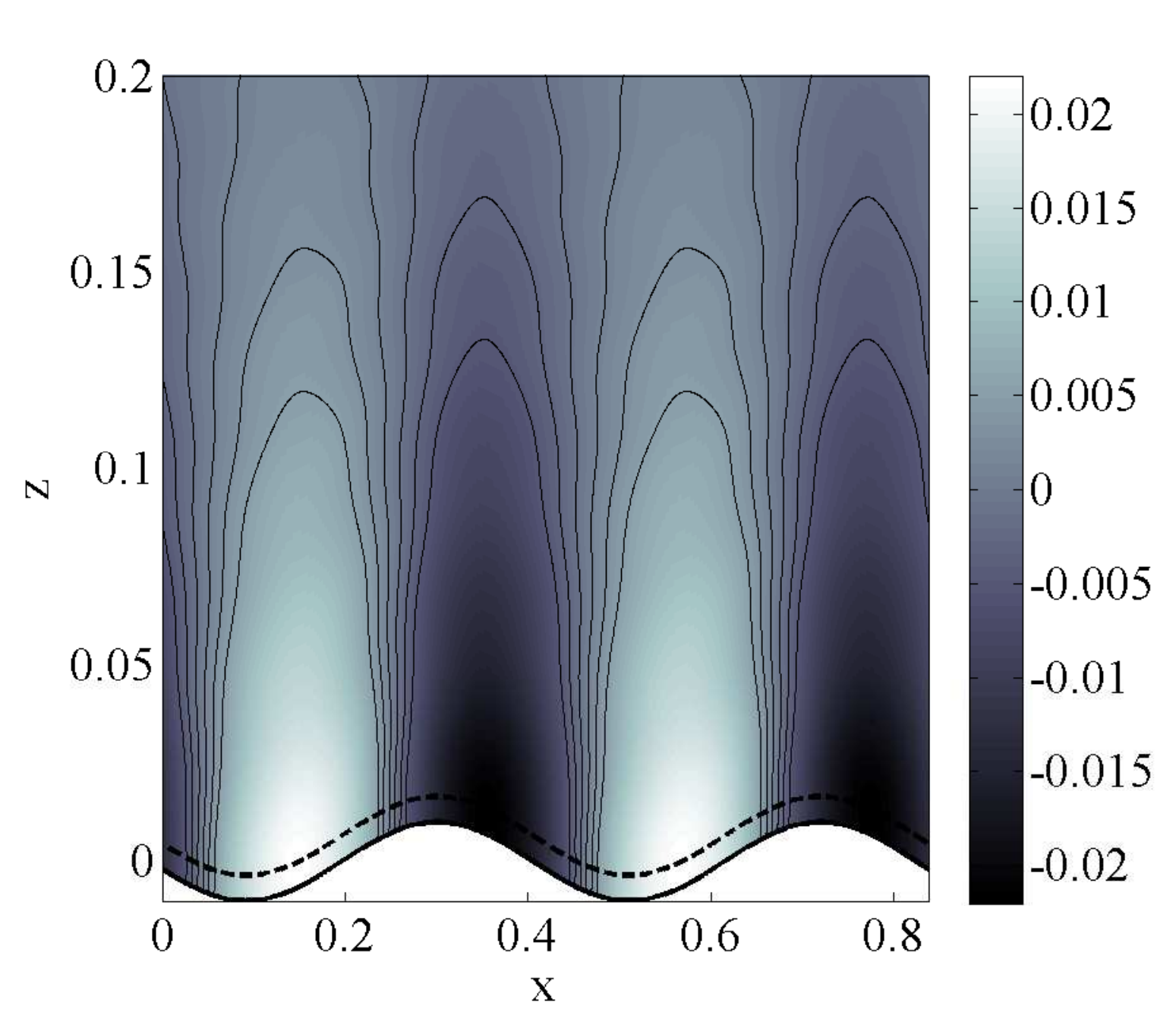}}\\
\subfigure[]{
\includegraphics[width=0.3\textwidth]{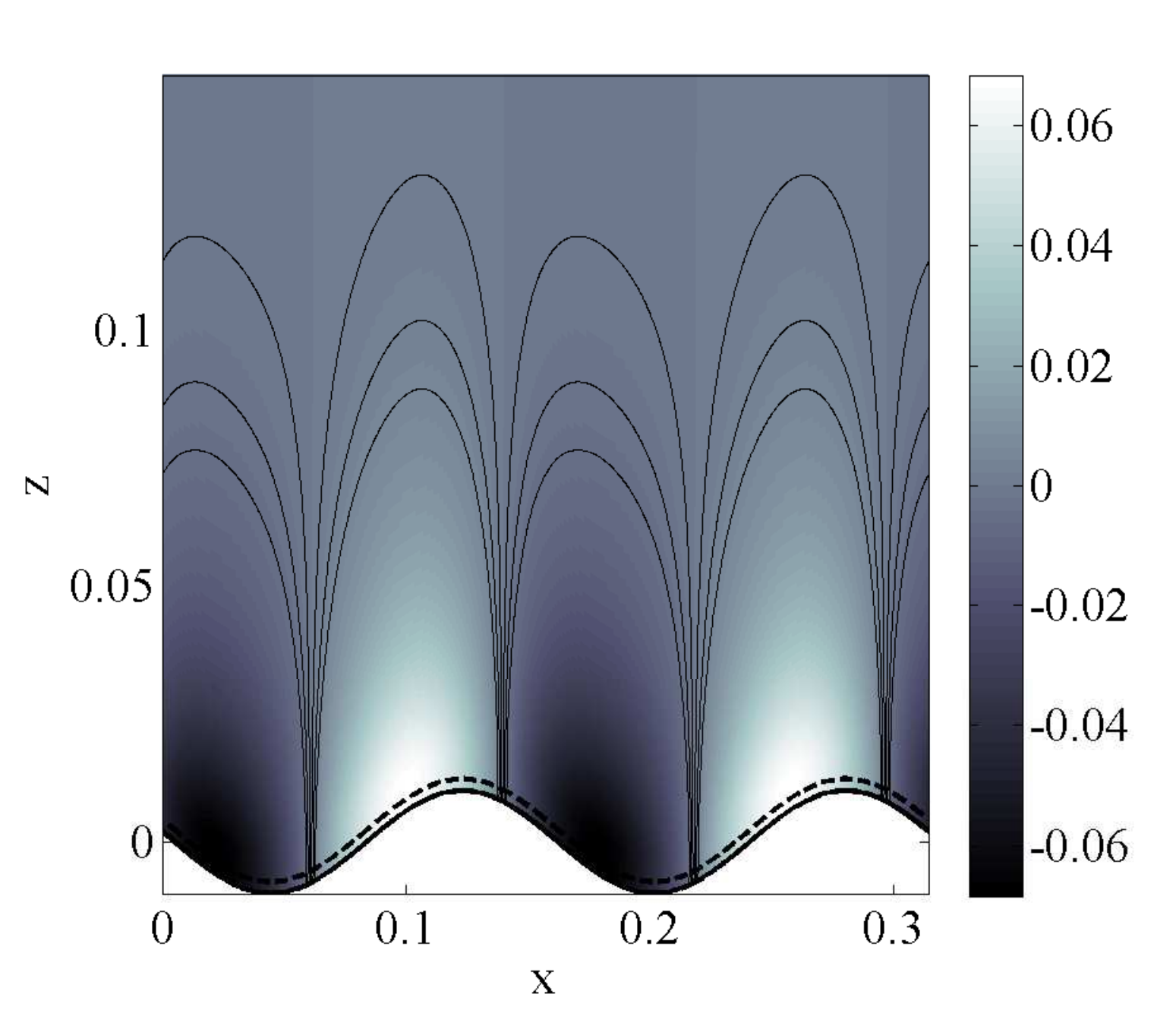}}
\caption{The pressure distribution for $\alpha=15$, (a), and $\alpha=40$, (b), normalized by $\max|u|$.  The critical layer is shown as a dotted line above the interface.}
\label{fig:pressure_deep}
\end{figure}
%
%
%
%
% to the mean flow; for the thin-film flow at lower Reynolds number, this
% angle of inclination was approximately $45^{\mathrm{o}}$.

\begin{figure}[htb]
\centering\noindent\includegraphics[width=0.45\textwidth]{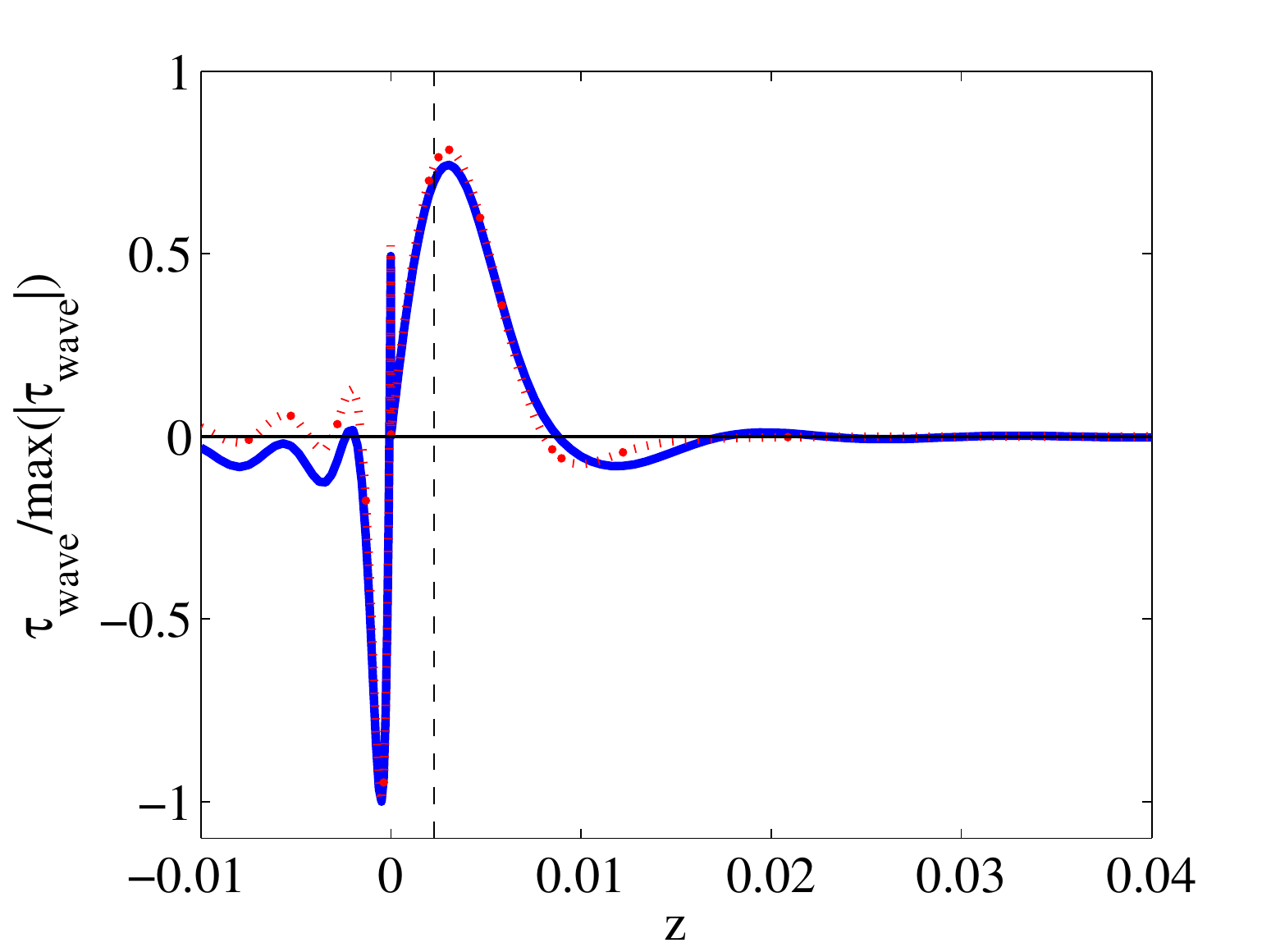}
\caption{The unit-normalized wave Reynolds stress for $\alpha=40$.  The critical layer is
at $z_{\mathrm{c}}=0.0022$, shown in the figure.  The stress function for
the non-PTS wave is given by the dotted line, while the continuous curve
indicates the PTS wave.  The PTS
stress function possesses an oscillatory structure in the gas layer.}
\label{fig:wrs20}
\end{figure}

\paragraph*{$\alpha=40$:}  At this wavenumber, the wave parameters are
\begin{eqnarray}
\lambda_{\mathrm{r}}&=&0.0043,\qquad c_{\mathrm{r}}=0.1033,\qquad\text{with
the PTS};\nonumber\\
\lambda_{\mathrm{r}}&=&0.0047,\qquad c_{\mathrm{r}}=0.1033,\qquad\text{with
no PTS};\nonumber\\
%z_{\mathrm{t}}&=&0.016;\qquad z_{\mathrm{c}}=0.0023;
z_{\mathrm{c}}&=&0.0023;
\label{eq:growth_rate20}
\end{eqnarray}
We present an energy budget for the case $\alpha=40$ in Tab.~\ref{tab:eb20}, which is calculated from
the streamfunction using the balance law formulated in Eq.~\eqref{eq:eb}.
 The energy terms are further normalized such that the total energy of
the PTS wave sums to unity.
\begin{table}[h!b!p!]
\centering
\begin{tabular}{|c||c|c|c|c|c|c|c|c|c|}
\hline
$\alpha=40$&$KIN_L$&$KIN_G$&$REY_L$&$REY_G$&$DISS_L$&$DISS_G$&$TURB$&$NOR$&$TAN$\\
\hline
\hline
   PTS&0.99&0.01&-0.04&4.15&-0.88&-2.53&0.04&-2.06&2.32\\
\hline
No PTS&0.99&0.01&-0.01&2.81&-0.62&-1.76&0&-1.01&1.61\\
\hline
\end{tabular}
\caption{The energy budget for the deep-water waves at $\alpha=40$, near
maximal growth.}
\label{tab:eb20}
\end{table}
The two energy budgets possess only small relative differences, in spite
of the large difference between the associated growth rates.  This is because
the kinetic energy possesses contributions both from $c_{\mathrm{r}}$ and
$c_{\mathrm{i}}$, and while the difference in $c_{\mathrm{i}}$ is large,
the difference in $c_{\mathrm{r}}$ is small, and $c_{\mathrm{r}}\gg c_{\mathrm{i}}$.
Thus, the precise mechanism of excess instability is not visible from the
energy budget.  Nevertheless, the energy budget does indicate that the maximum
destabilizing term is $REY_G$, and we therefore study the wave Reynolds stress:
\[
\tau_{\mathrm{wave}}^{(j)}\left(z\right)=-r_{j}\int_0^{2\pi/\alpha} \delta
u_j\left(x,z\right)\delta w_j\left(x,z\right)dx,
%REY_L+REY_G=\sum_{j=L,G}\int_{-\infty}^1 dz\frac{\mathd U_j}{\mathd z}\tau_{\mathrm{wave}}^{(j)}\left(z\right).
\]
The sum $REY_G$ is thus
\[
REY_G=\int_{0}^{1}\tau_{wave}^{G}\left(z\right)\frac{\mathd U}{dz}dz
\]
and thus the wave Reynolds stress represents the energy associated with the
power $REY_G$.  We plot the wave Reynolds stress in Fig.~\ref{fig:wrs20}.
The peak in $\tau_{\mathrm{wave}}$ next to the critical height $z_{\mathrm{c}}$
represents a net transfer of energy from the mean flow $U\left(z\right)$
into the perturbation flow.  This is present in both the PTS wave and the
non-PTS wave and indicates that the instability is a \textit{critical-layer}
or \textit{Miles} instability.  Note, however that the PTS wave possesses
an oscillation that propagates into the bulk gas flow, due to the interaction
between the turbulence and the interface.
%
%
%
% Since the term $REY_G$ is the
% largest contributor to the energy of instability (Tab.~\ref{tab:eb20}), % this extra feature in the
% energy transfer function is the source of the extra instability in the % PTS wave.
%
%
%
\begin{figure}[htb]
\centering\noindent
\subfigure[]{
\includegraphics[width=0.3\textwidth]{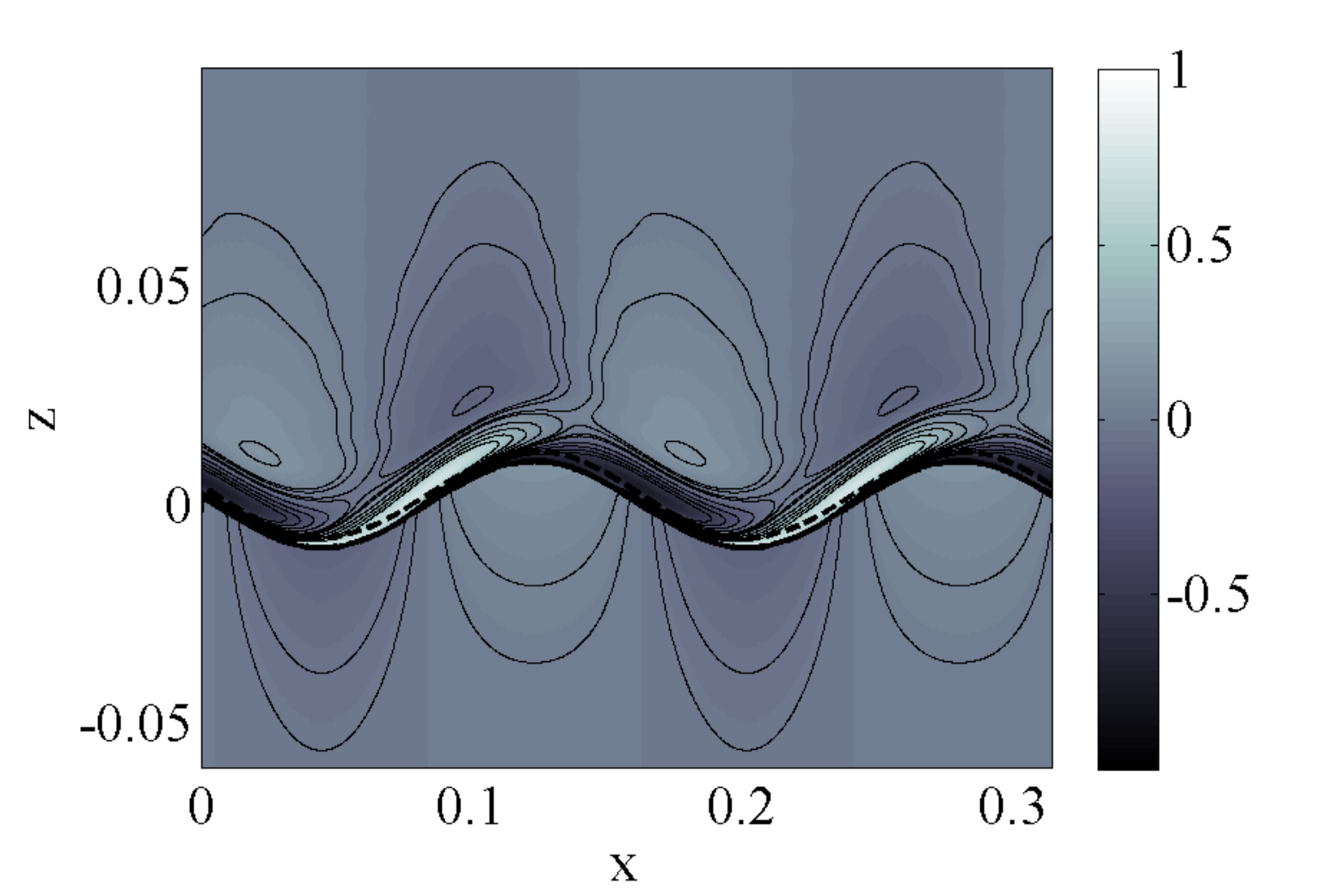}}
\subfigure[]{
\includegraphics[width=0.29\textwidth]{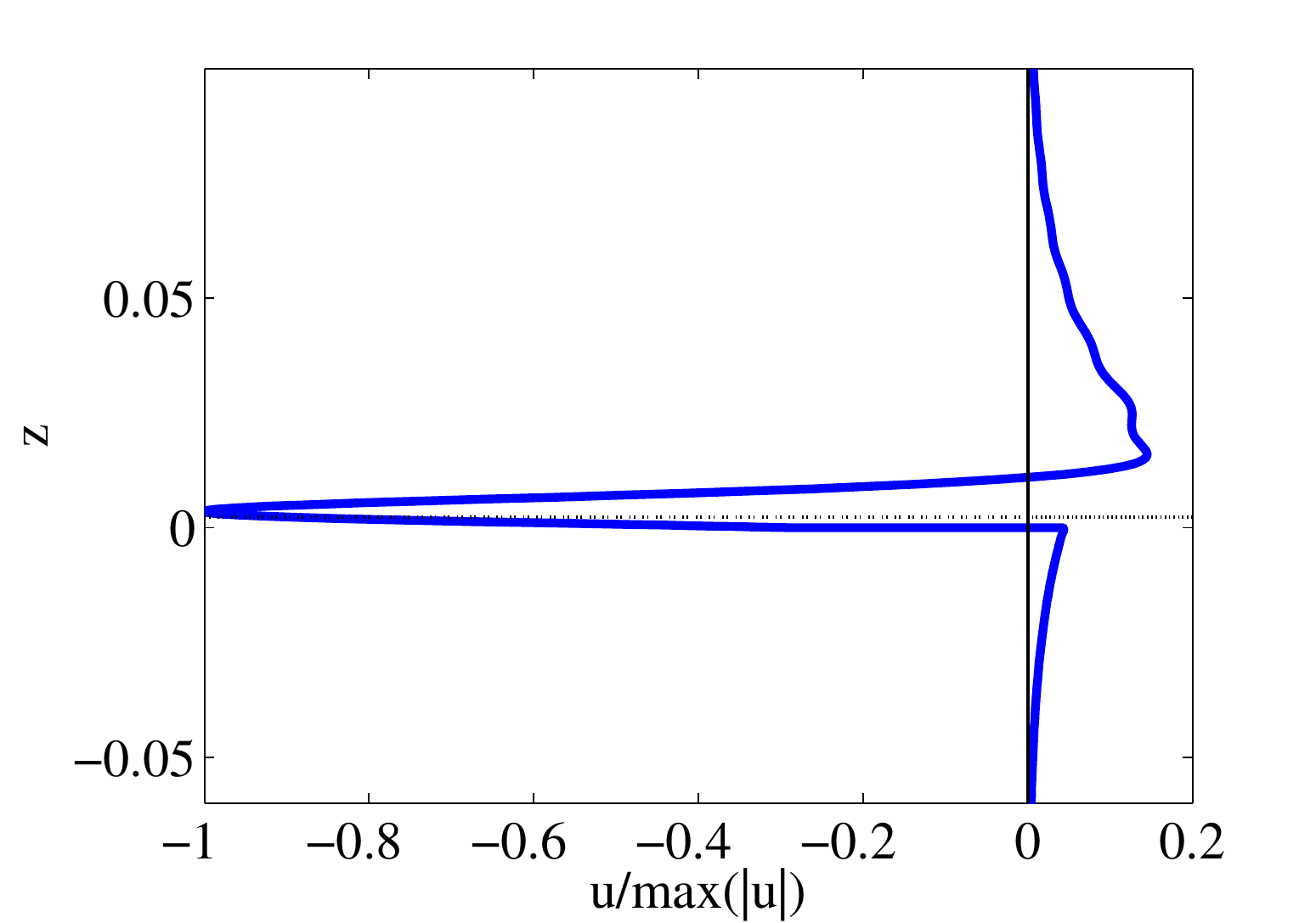}}\\
\subfigure[]{
\includegraphics[width=0.3\textwidth]{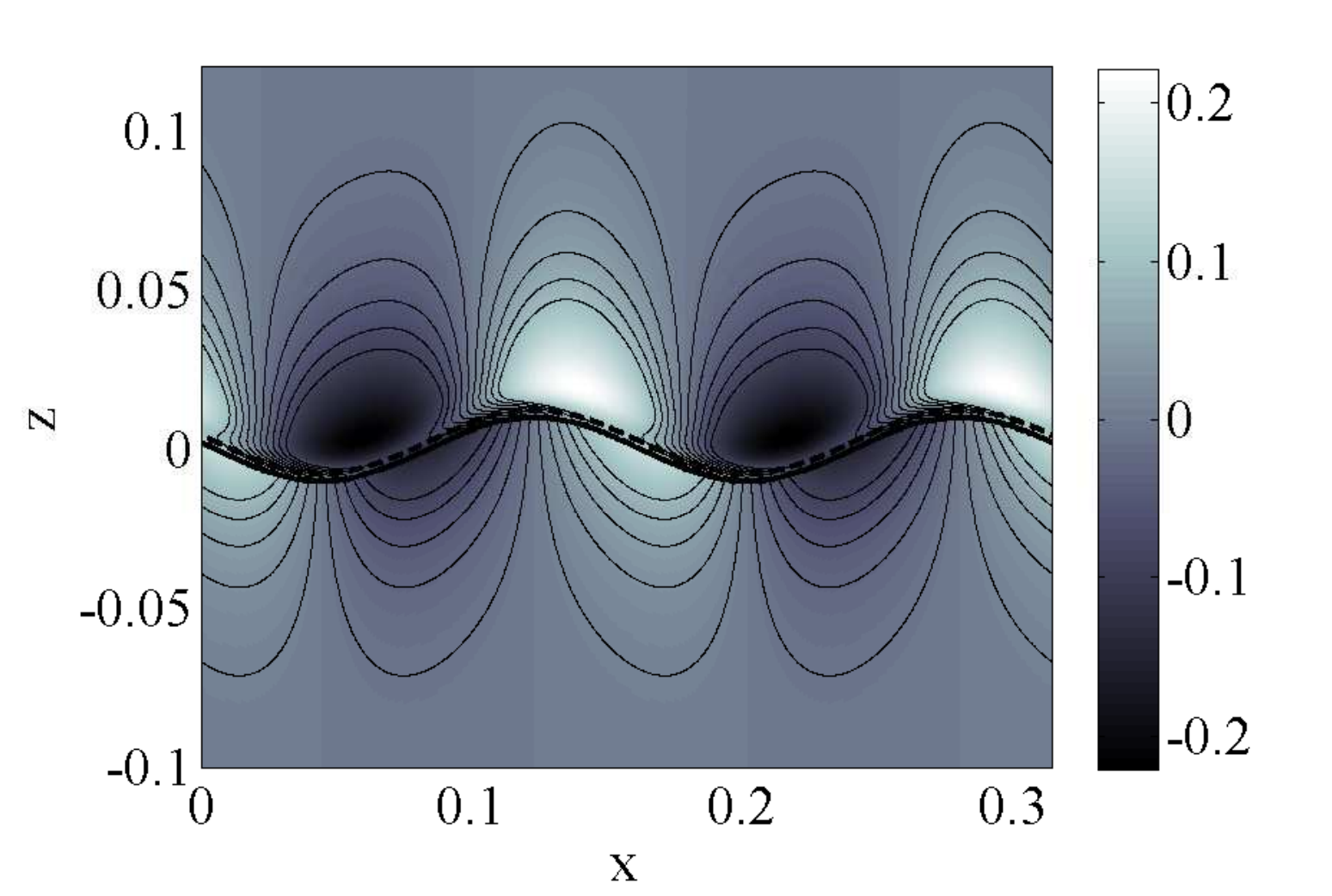}}
\subfigure[]{
\includegraphics[width=0.29\textwidth]{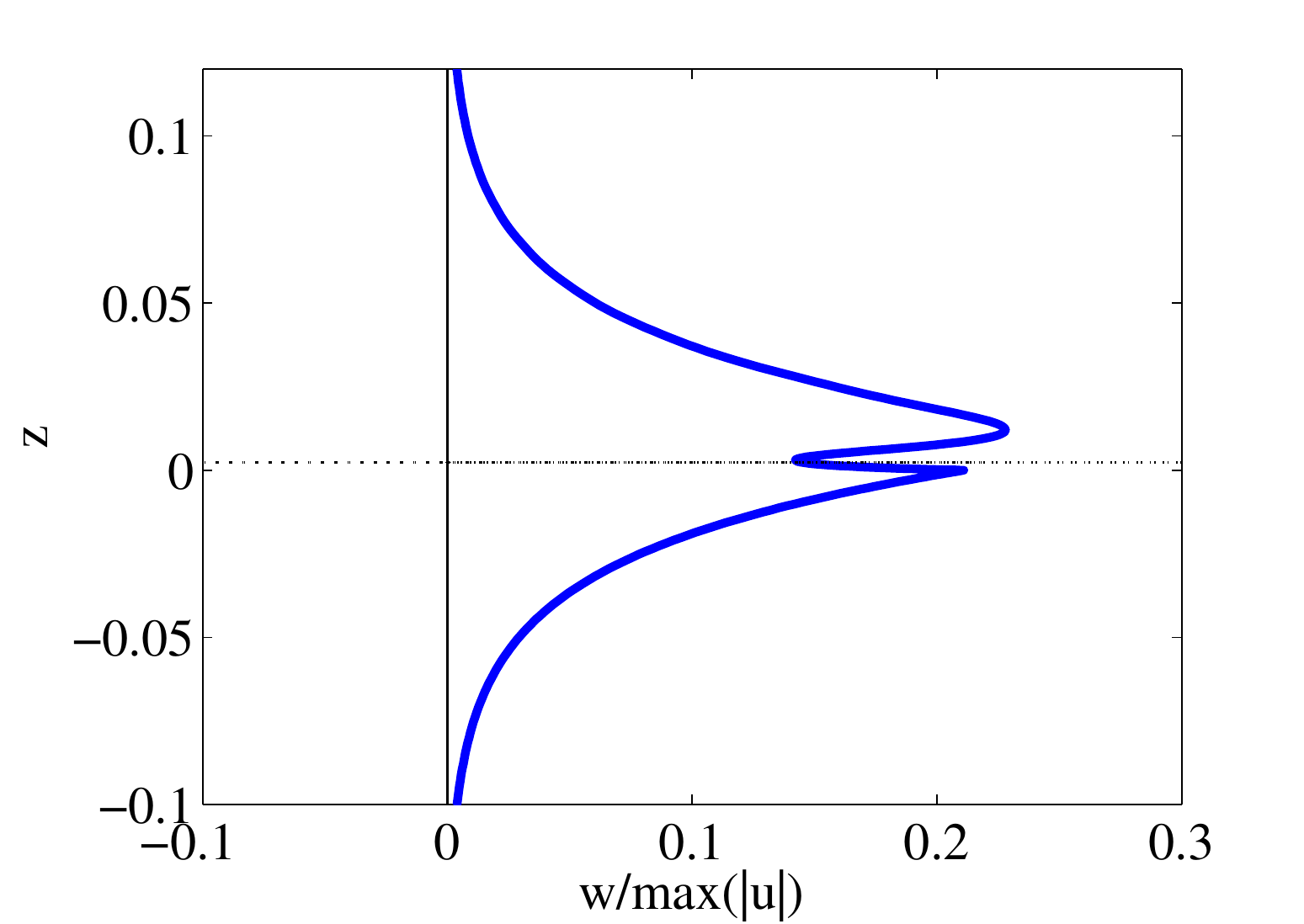}}\\
\subfigure[]{
\includegraphics[width=0.3\textwidth]{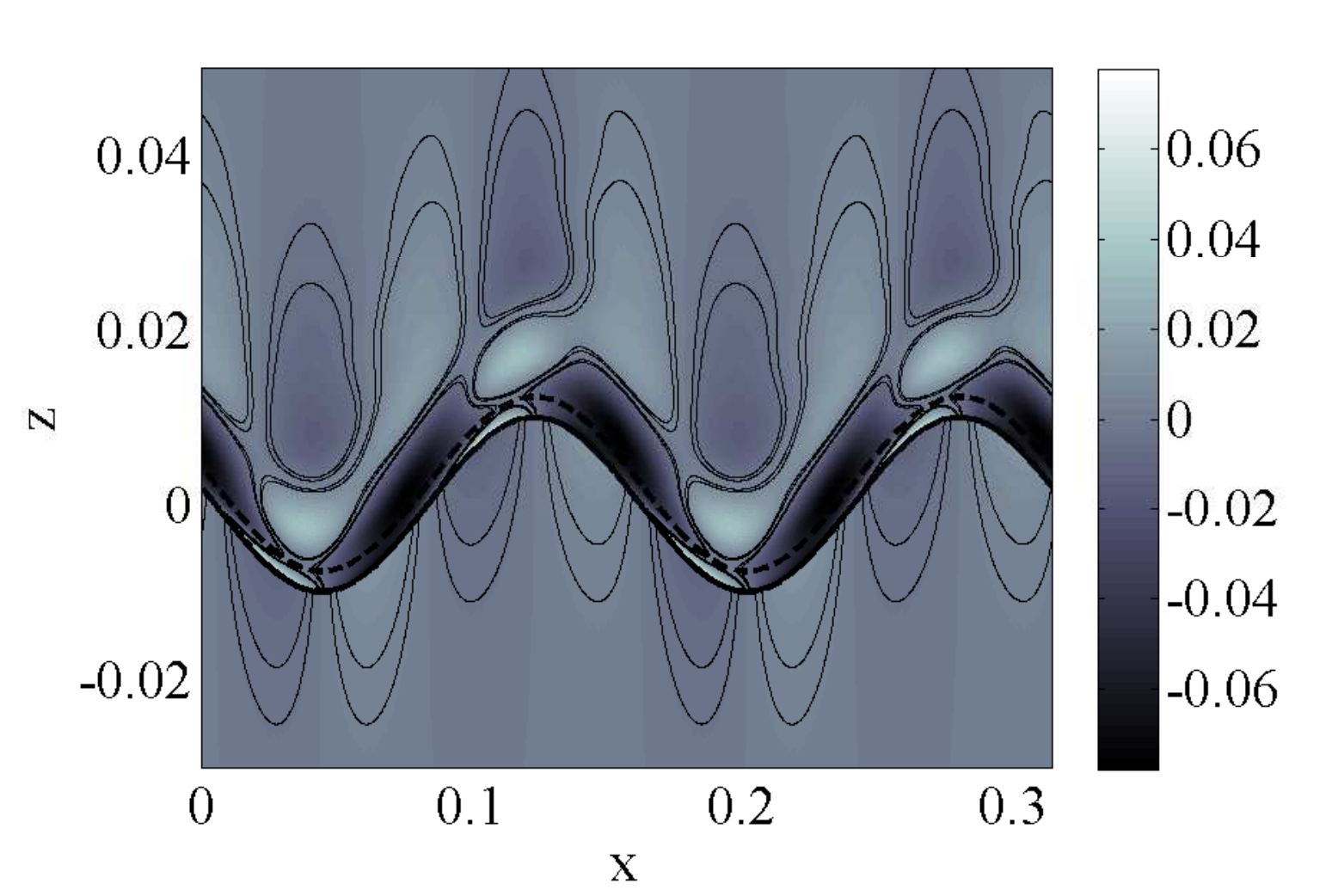}}
\subfigure[]{
\includegraphics[width=0.29\textwidth]{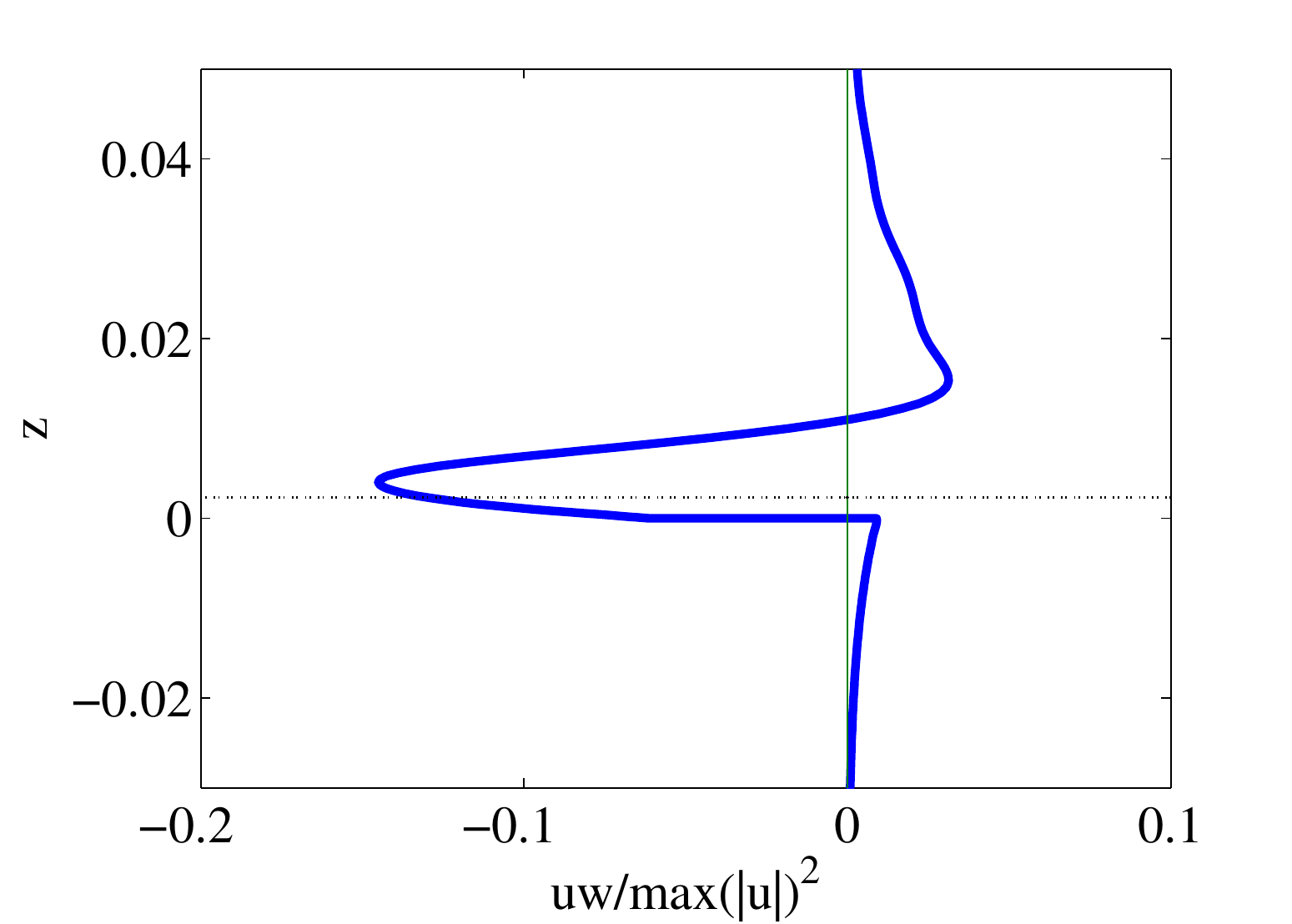}}
\caption{Flow field structure for $\alpha=40$.  The critical layer is shown as a dotted line above the interface.  The effects of the rapid
distortion on the structure of the flow field no longer visible.  The streamfunction
decays in the bulk of the gas more rapidly than in the $\alpha=10$ case (as
evidenced by a comparison between Figs.~\ref{fig:velocity20}~(d) and~\ref{fig:velocity10}~(d)).
 Subfigures~(a) and~(b) show the streamwise velocity and a profile at $x=0$;
 subfigures~(c) and~(d) show the normal velocity and a profile at $x=0$;
 subfigures~(e) and~(f) show the pre-averaged version of the wave Reynolds
 stress, namely the product $uw$.  In each case, we have normalized the velocities
 by $\max|u|$.}
\label{fig:velocity20}
\end{figure}
In contrast to the $\alpha=10$ case, the velocity fields in Fig.~\ref{fig:velocity20}
possess the same structure, regardless of the PTS, and neither the oscillation
nor the streamwise streaks are visible.  This is consistent with the predictions of the piecewise-constant model in Sec.~\ref{subsec:simple_model}.

\begin{figure}[htb]
\centering\noindent
\subfigure[]{\includegraphics[width=0.3\textwidth]{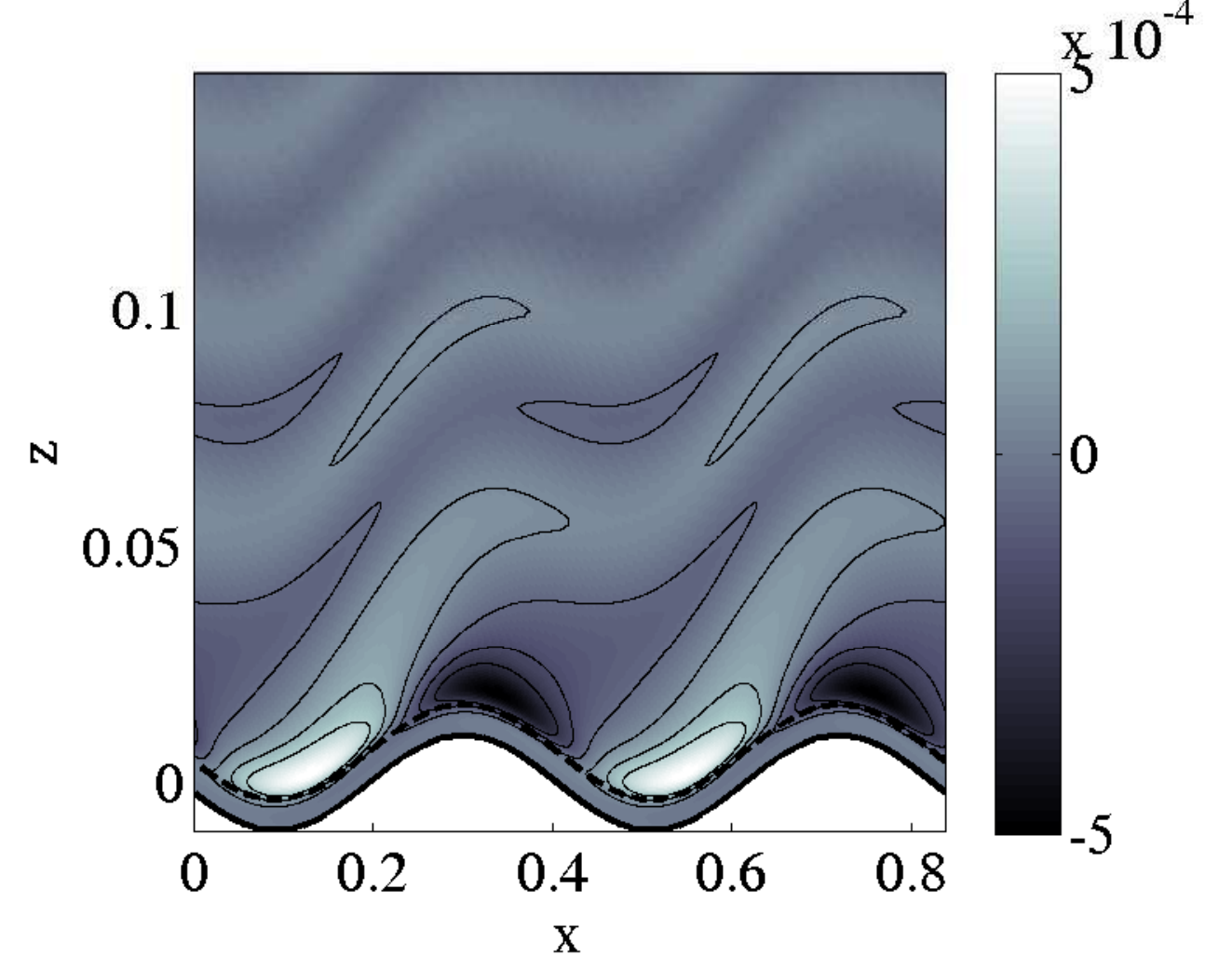}}
\subfigure[]{\includegraphics[width=0.26\textwidth]{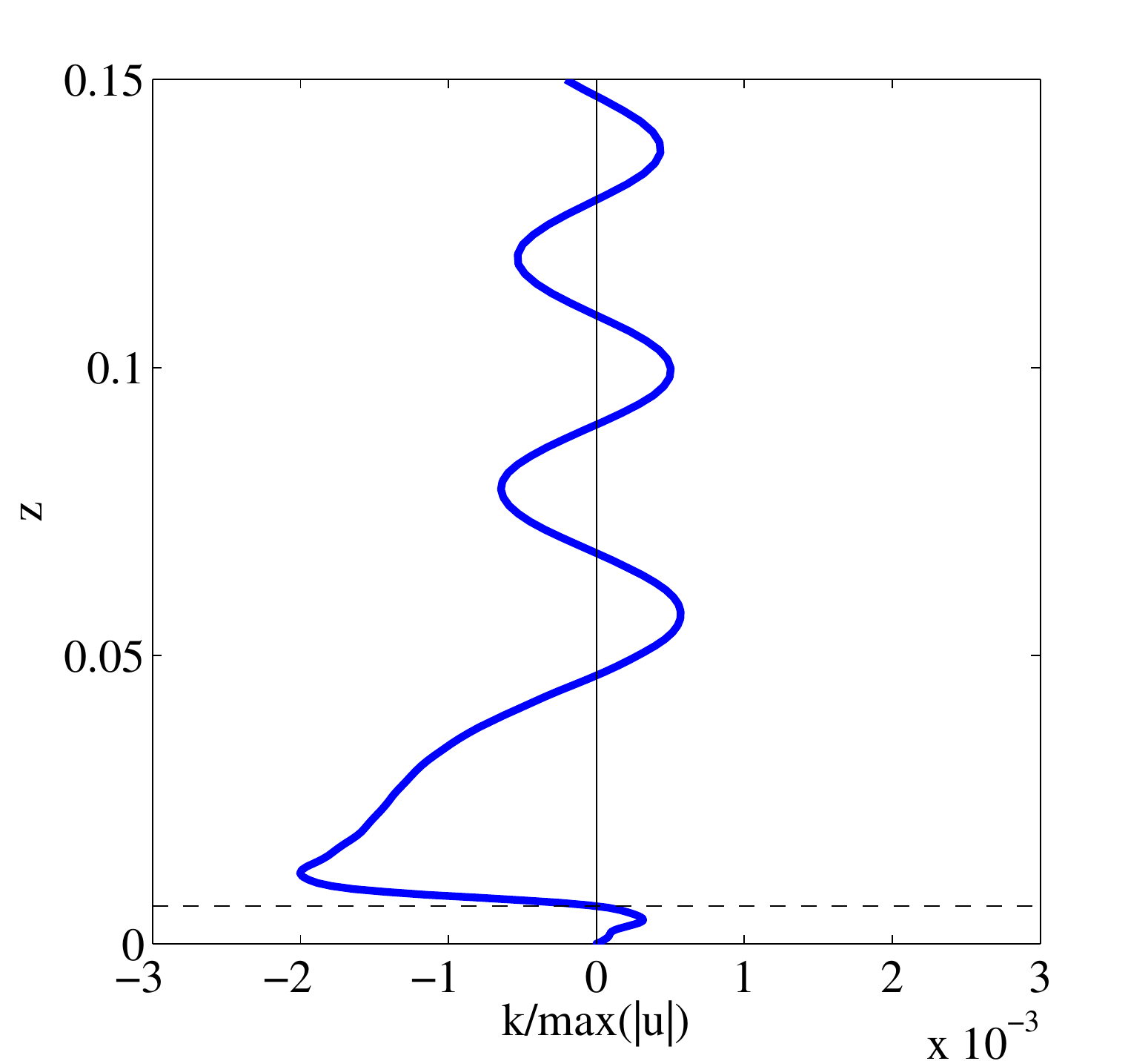}}\\
\subfigure[]{\includegraphics[width=0.3\textwidth]{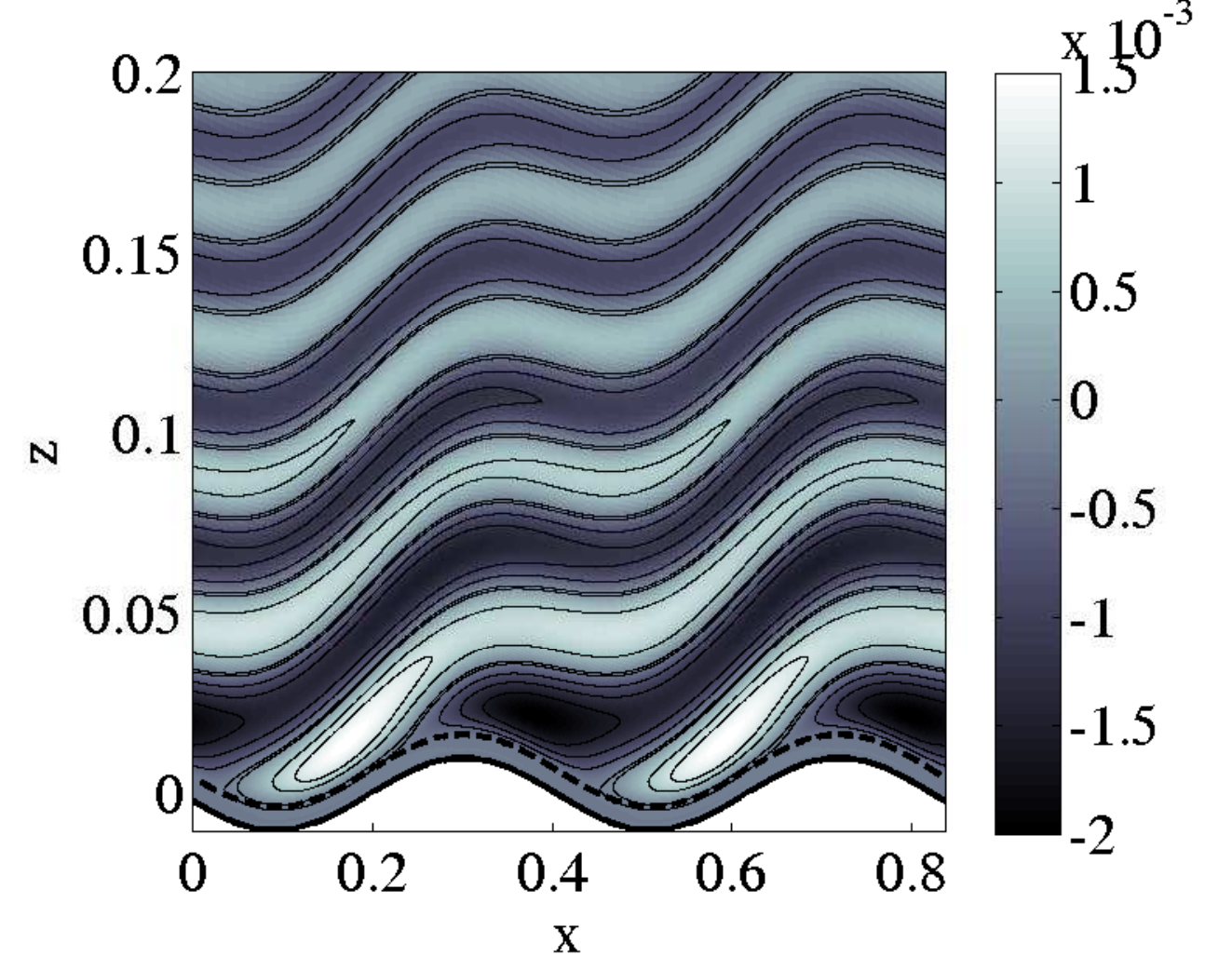}}
\subfigure[]{\includegraphics[width=0.26\textwidth]{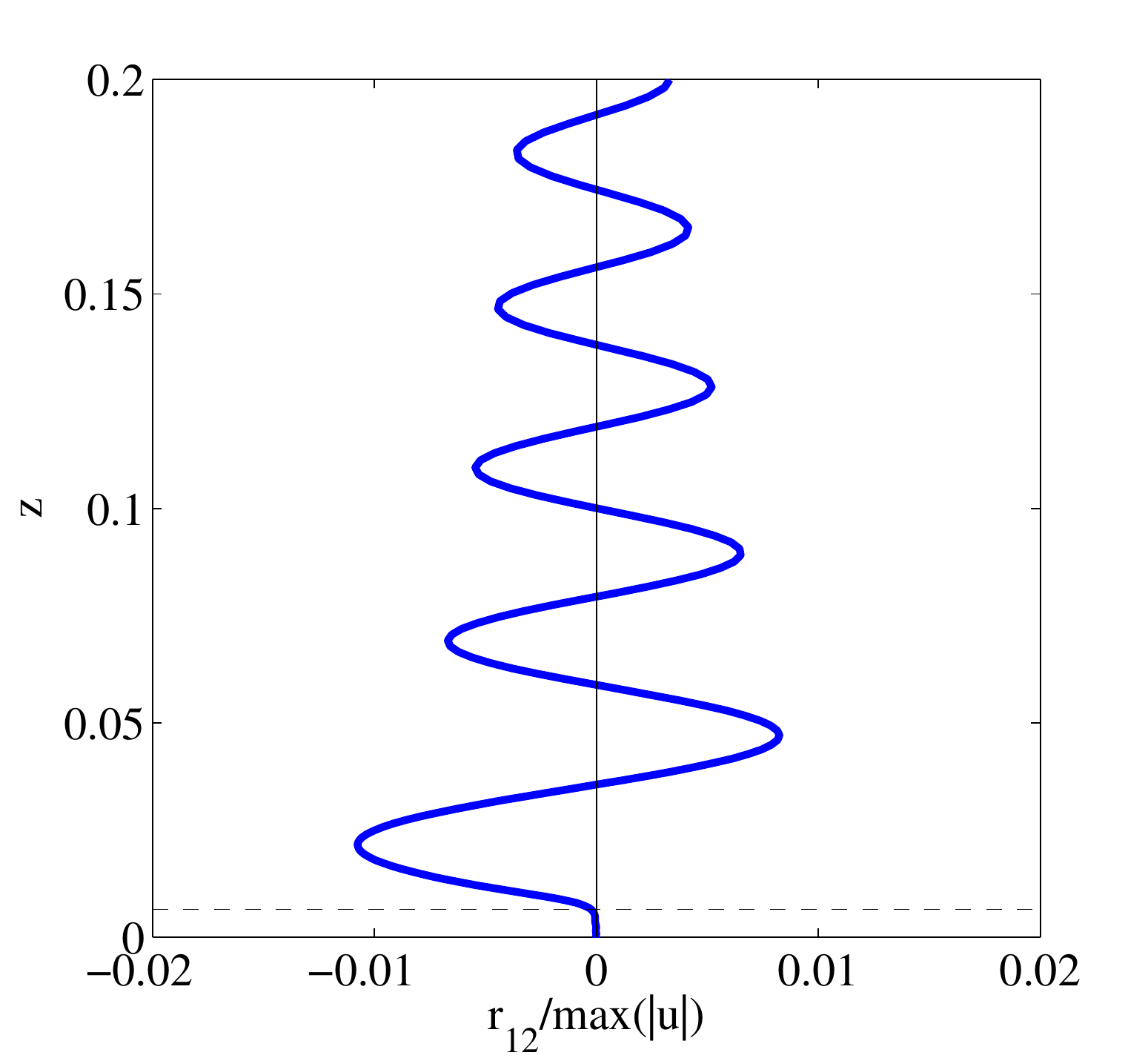}}\\
\subfigure[]{\includegraphics[width=0.3\textwidth]{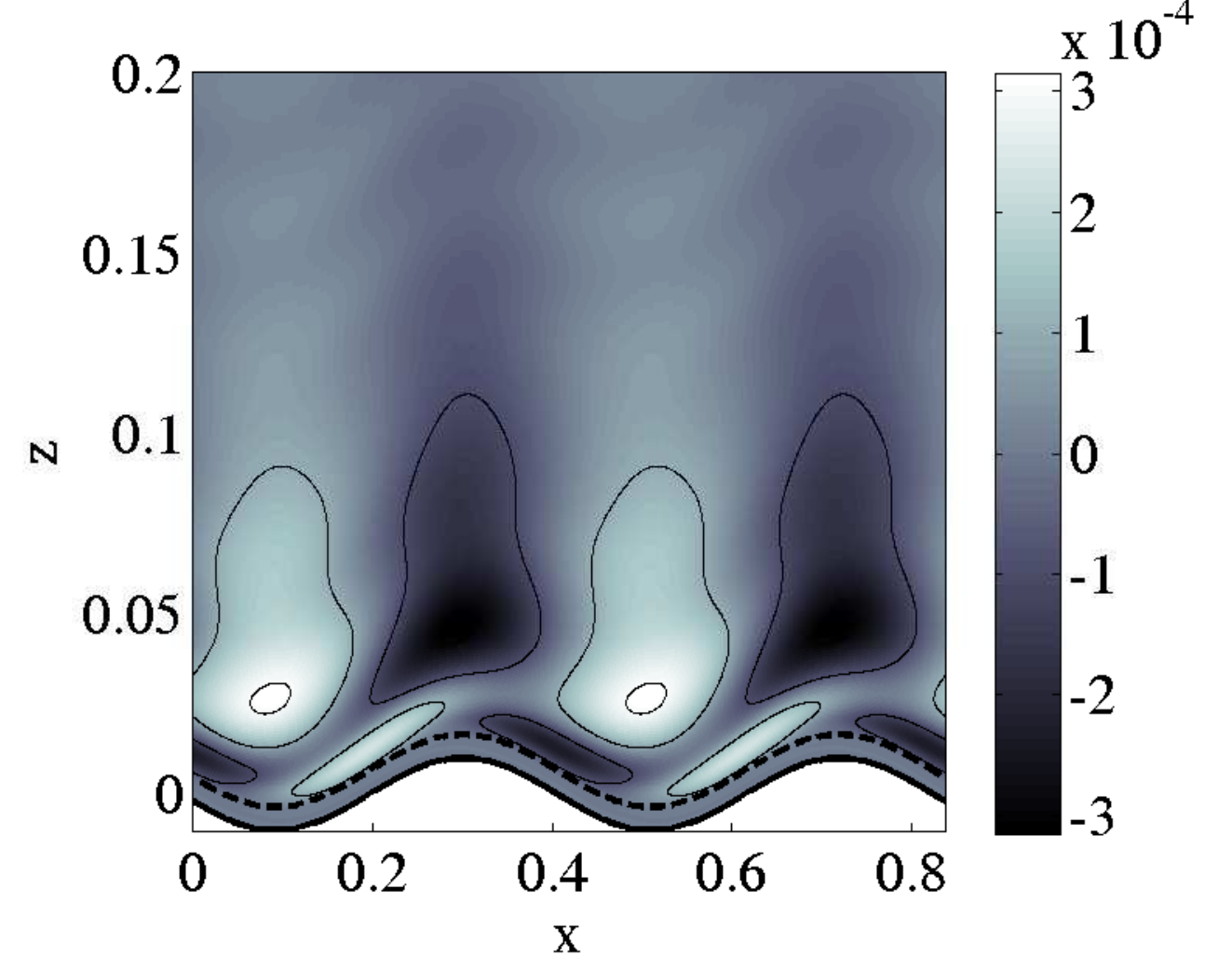}}
\subfigure[]{\includegraphics[width=0.26\textwidth]{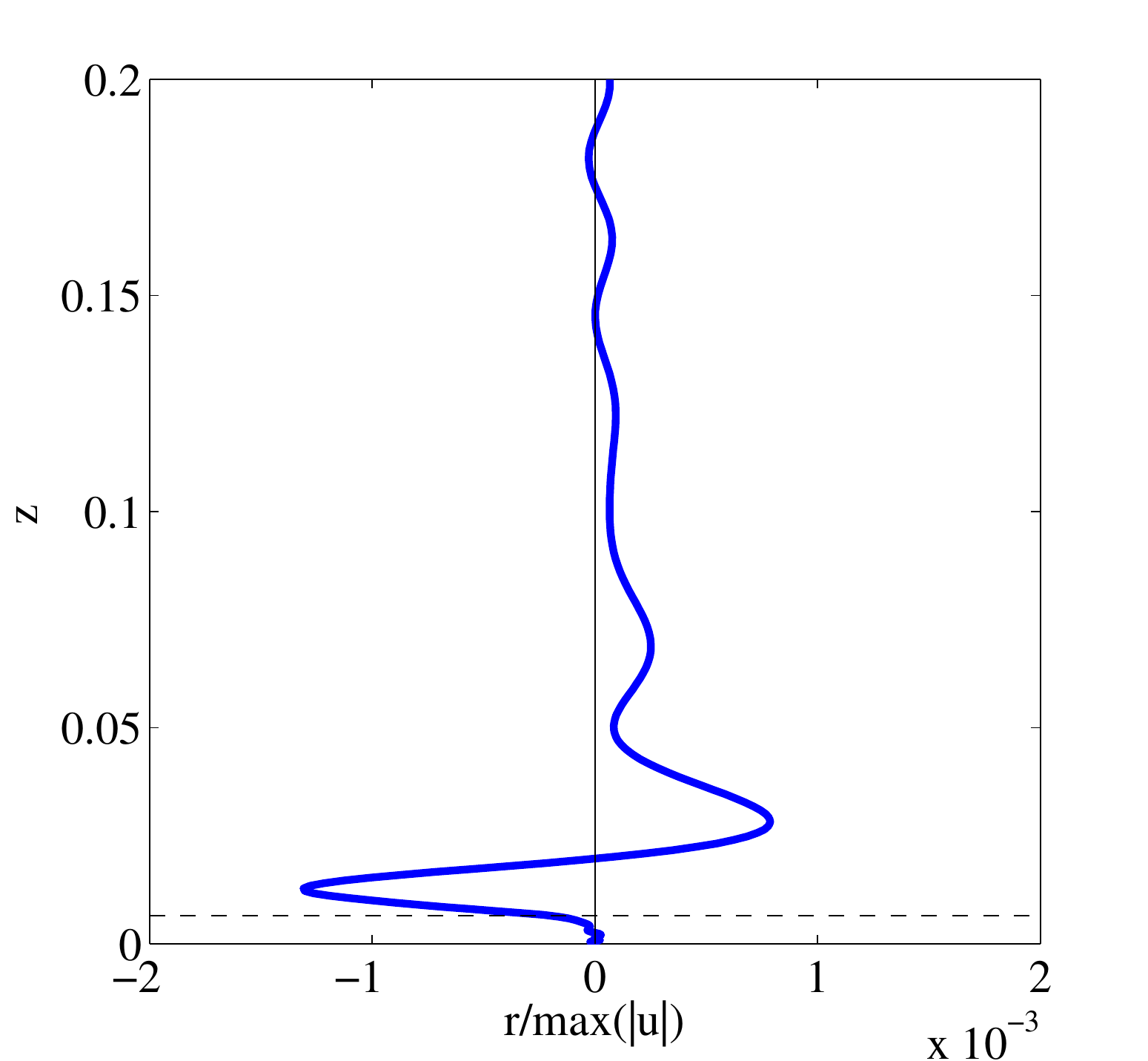}}\\
\caption{Turbulence variables at $\alpha=15$.  The critical layer is shown as a dotted line above the interface.  (a) and (b) show the turbulent kinetic energy; (c) and (d) show tangential Reynolds stress, while (e) and (f) show the normal Reynolds stress.  We have normalized the perturbed turbulent quantities by $\max|u|$, the maximum of the perturbed velocity.}
\label{fig:stress15}
\end{figure}

\begin{figure}[htb]
\centering\noindent
\subfigure[]{\includegraphics[width=0.3\textwidth]{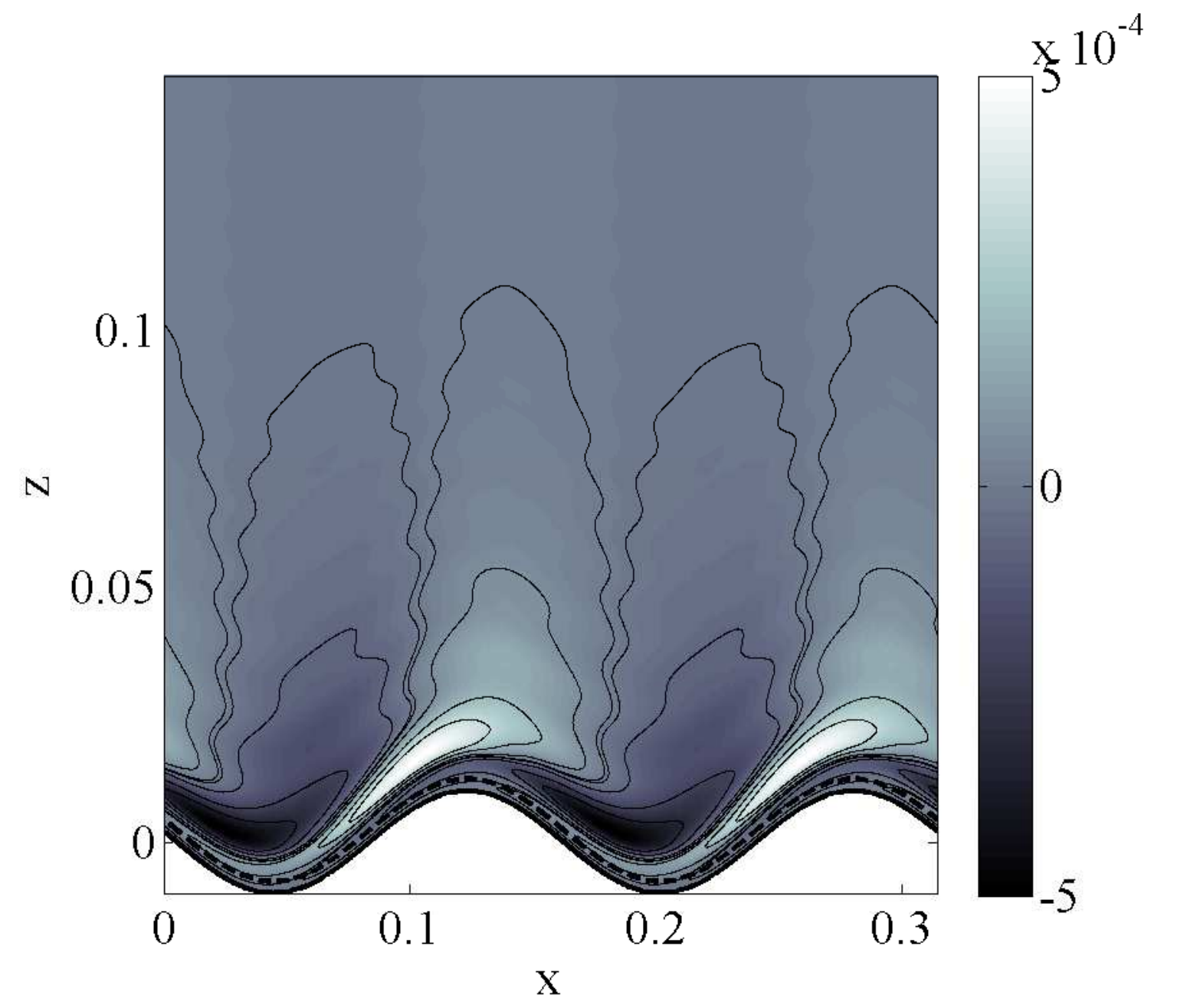}}
\subfigure[]{\includegraphics[width=0.27\textwidth]{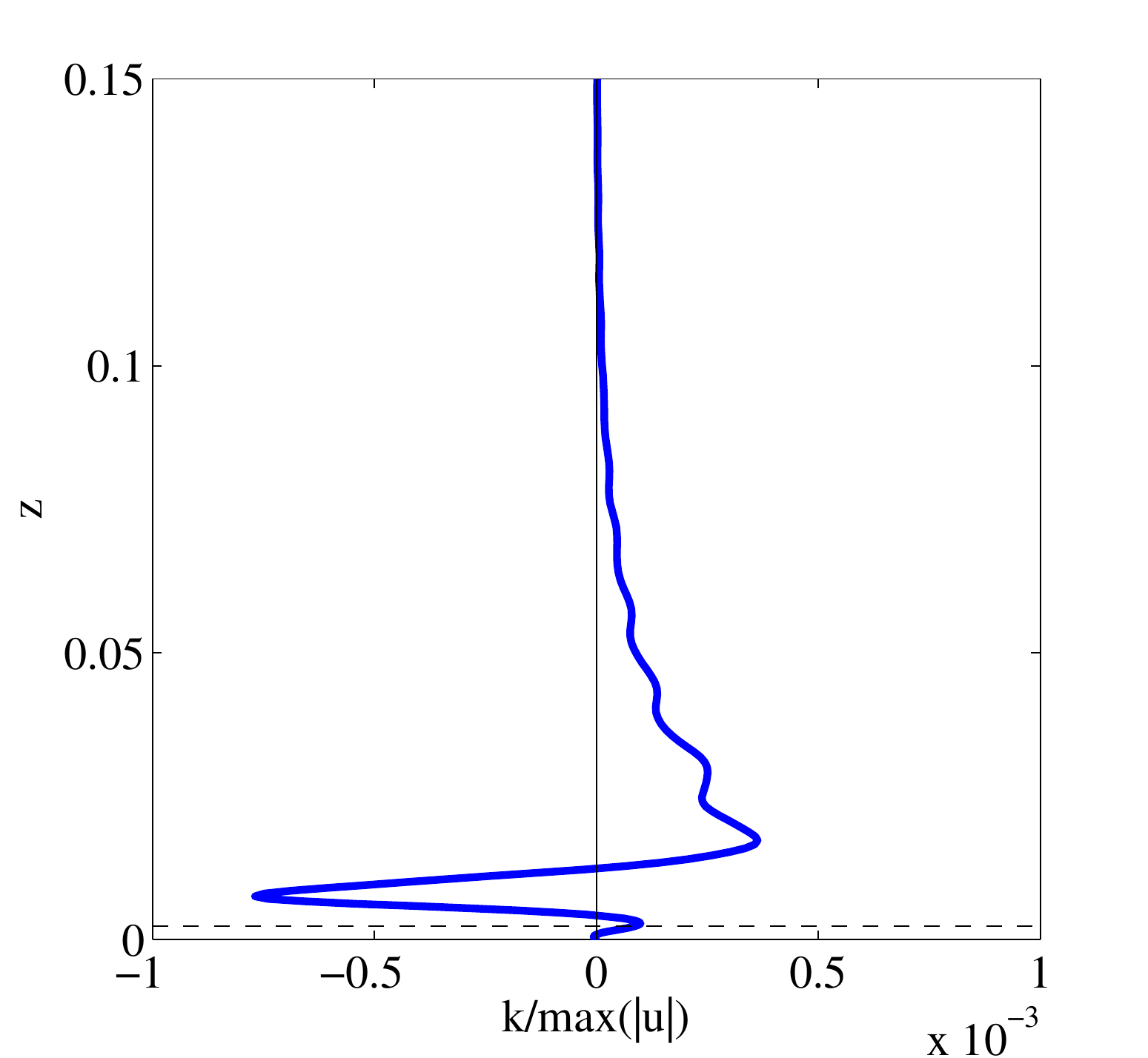}}\\
\subfigure[]{\includegraphics[width=0.3\textwidth]{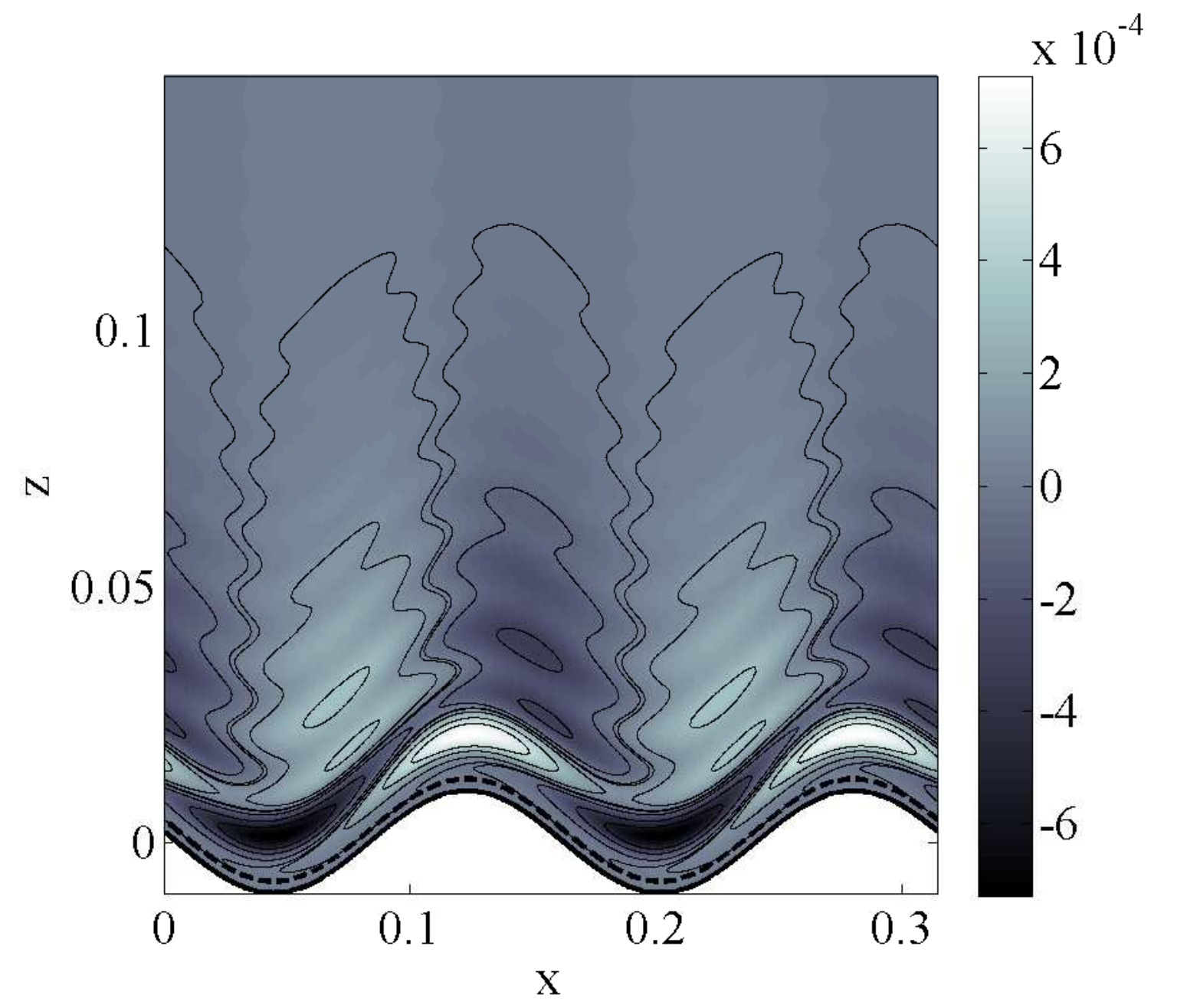}}
\subfigure[]{\includegraphics[width=0.27\textwidth]{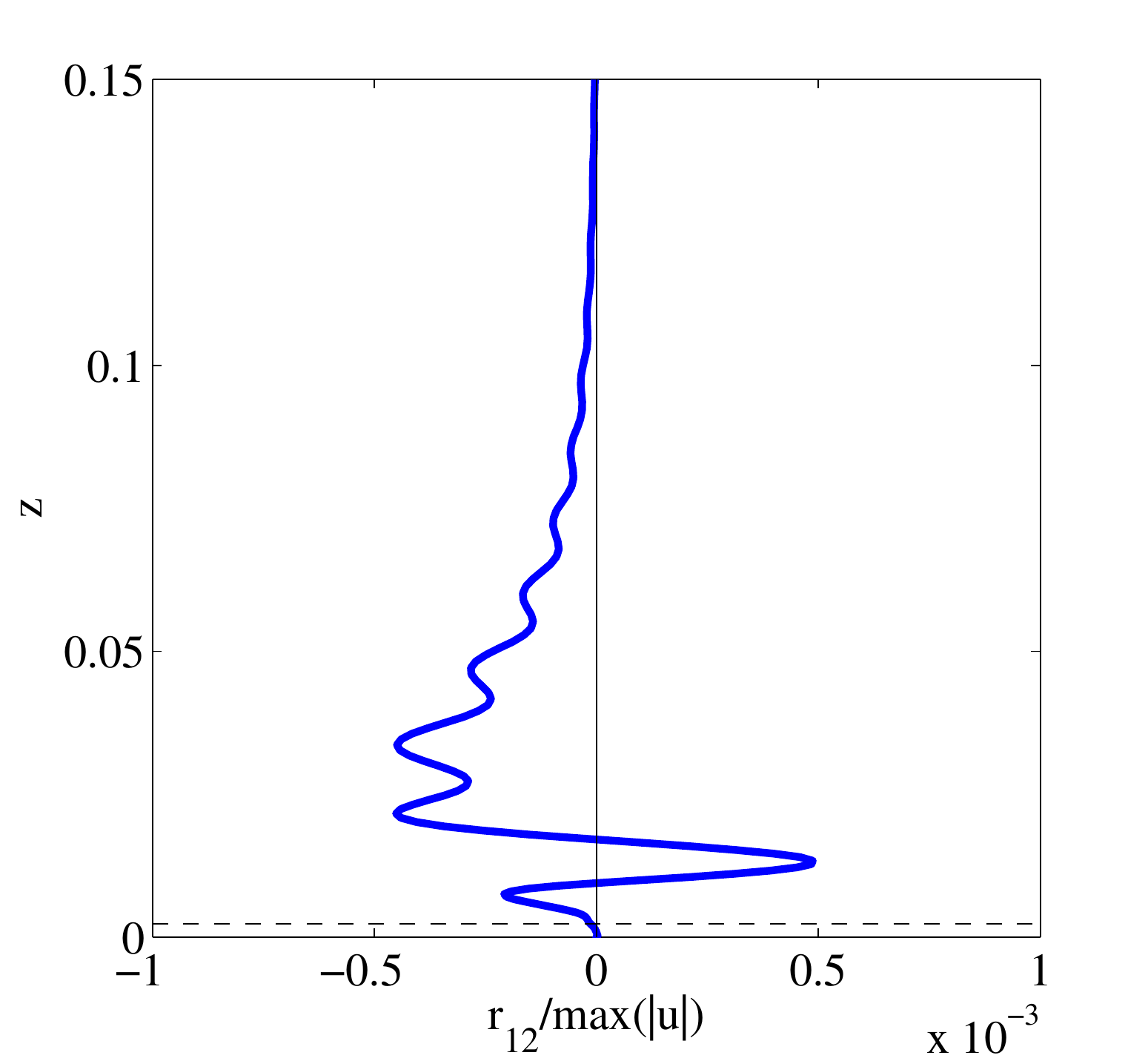}}\\
\subfigure[]{\includegraphics[width=0.3\textwidth]{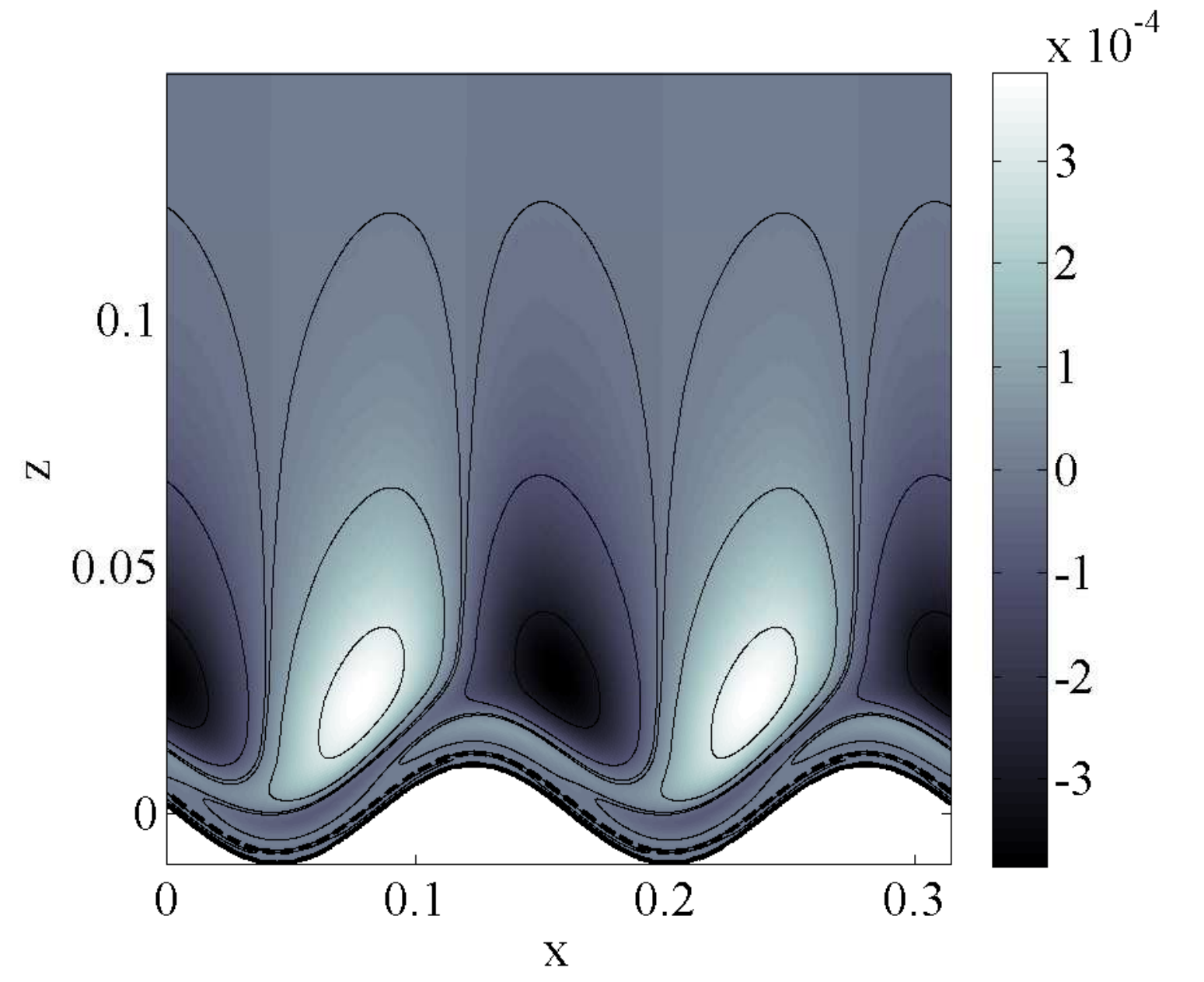}}
\subfigure[]{\includegraphics[width=0.27\textwidth]{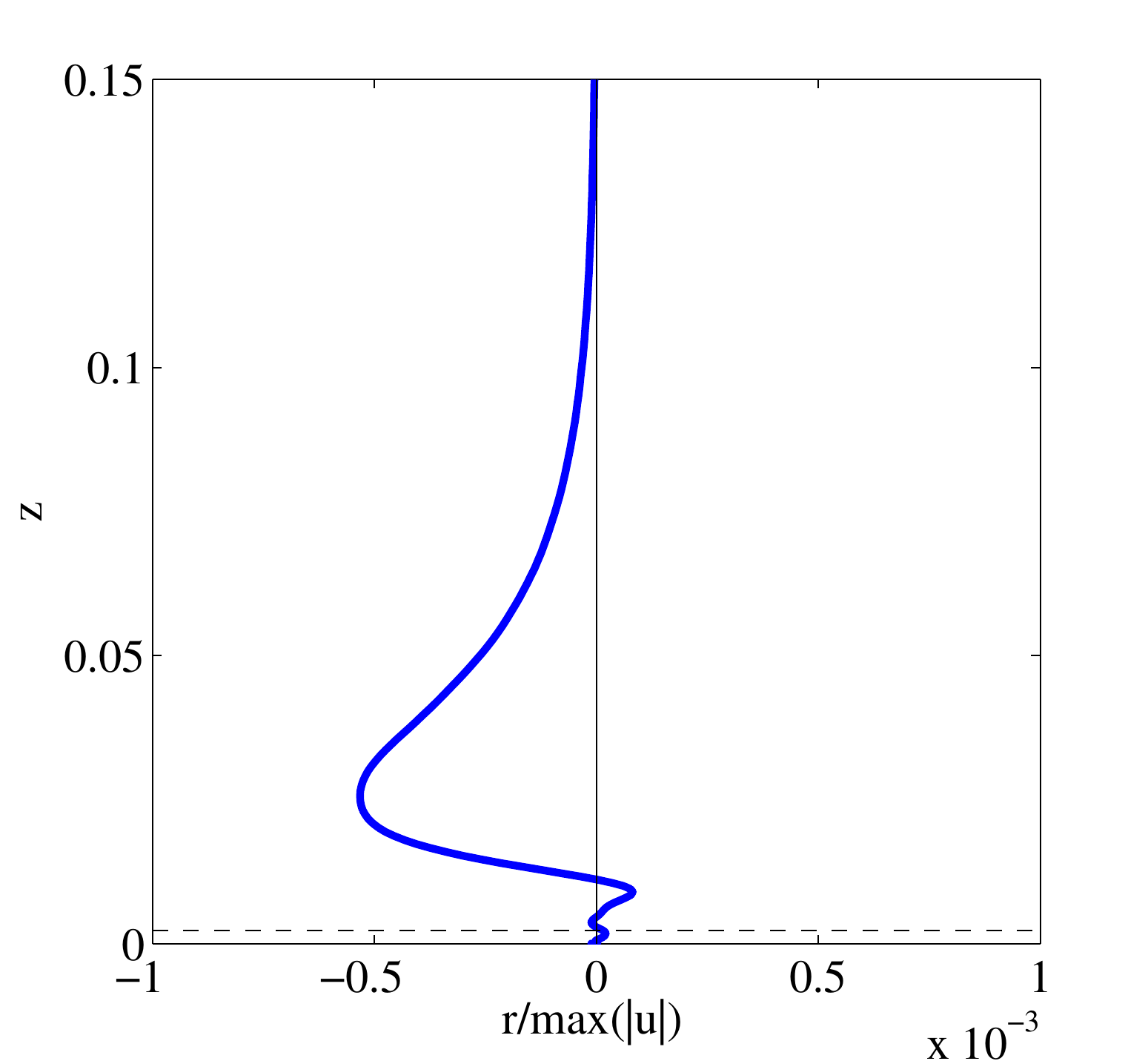}}\\
\caption{Turbulence variables at $\alpha=40$.  The critical layer is shown as a dotted line above the interface.  (a) and (b) show the turbulent kinetic energy; (c) and (d) show tangential Reynolds stress, while (e) and (f) show the normal Reynolds stress.  We have normalized the perturbed turbulent quantities by $\max|u|$, the maximum of the perturbed velocity.}
\label{fig:stress40}
\end{figure}

Finally, in Figs.~\ref{fig:stress15} and~\ref{fig:stress40} we plot the turbulent kinetic energy and Reynolds stresses at $\alpha=15$ and $40$ respectively.  These are an order of magnitude smaller than the wave Reynolds stress $uw$, confirming the importance of the latter term, relative to the turbulent variables.  Note that for the smaller of the two $\alpha$-values, the maximum values of the turbulent variables lie close to the critical layer (Fig.~\ref{fig:stress15}), while for the larger $\alpha$-value, the maximum lies far above the critical height (Fig.~\ref{fig:stress40}).  This is consistent with the growth-rate curve in Fig.~\ref{fig:growth_rate}, where the relative change in the growth rate (based on a comparison between the PTS no-PTS curves)
was larger at smaller $\alpha$-values, suggesting an interaction between the critical-layer mechanism and the wave turbulence.  For larger $\alpha$-values (in particular, close to the maximum growth rate), the relative change in the growth rate is smaller.

In summary, the behaviour of the PTS wave differs from that of the non-PTS
wave.
 The difference is both quantitative (the growth rate shifts), and qualitative
 (the flow field changes, especially at longer wavelengths).  At longer wavelengths,
 the streamwise velocity exhibits streaks
 due to the interaction of the wave and the turbulence.  At shorter wavelengths,
 this effect is reduced, although the growth
 rate is enhanced relative to the case without the PTS.

\subsection{Comparison with other work}

Since we have identified the instability as a critical-layer instability,
it is appropriate to compare our results with the theory of Miles.  Morland and Saffman~\cite{Morland1993} have developed explicit formulas that enable such a comparison.  We also develop a comparison with deep-water waves based on the work of Boomkamp and Miesen in~\cite{Boomkamp1996}.

\begin{figure}[htb]
\centering\noindent
\includegraphics[width=0.45\textwidth]{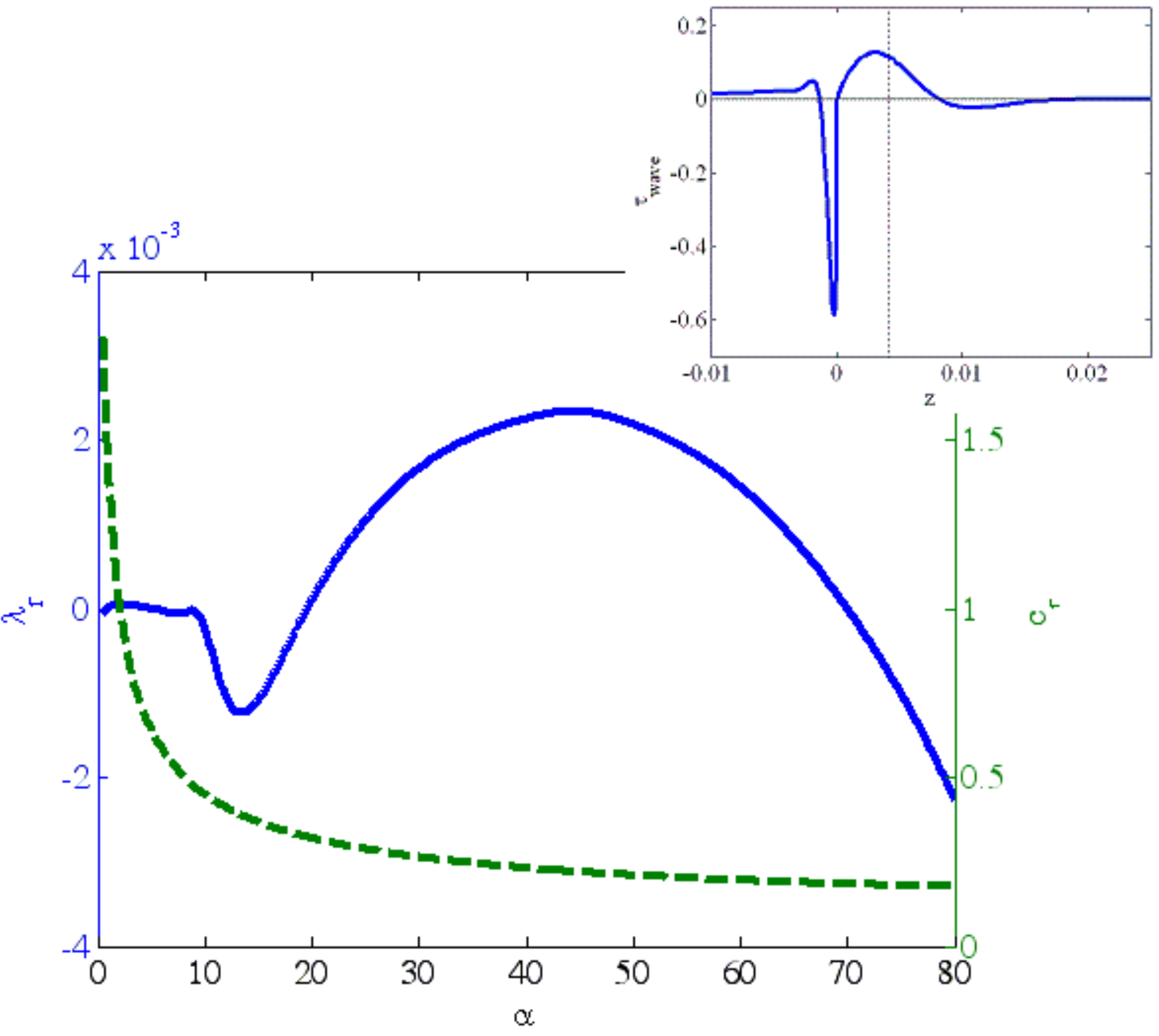}
\caption{Comparison with the work of Boomkamp and Miesen~\cite{Boomkamp1996}.  The wave Reynolds stress at $\alpha$=40 is shown as in inset, with the critical layer at $z_{\mathrm{c}}=0.0044$.  The growth-rate is designated by the solid curve, with a scale on the left-hand side, while the wave-speed is designated by the dashed-line curve, whose scale is on the right-hand side.}
\label{fig:inset_mb}
\end{figure}

\begin{figure}[htb]
\centering\noindent
\includegraphics[width=0.5\textwidth]{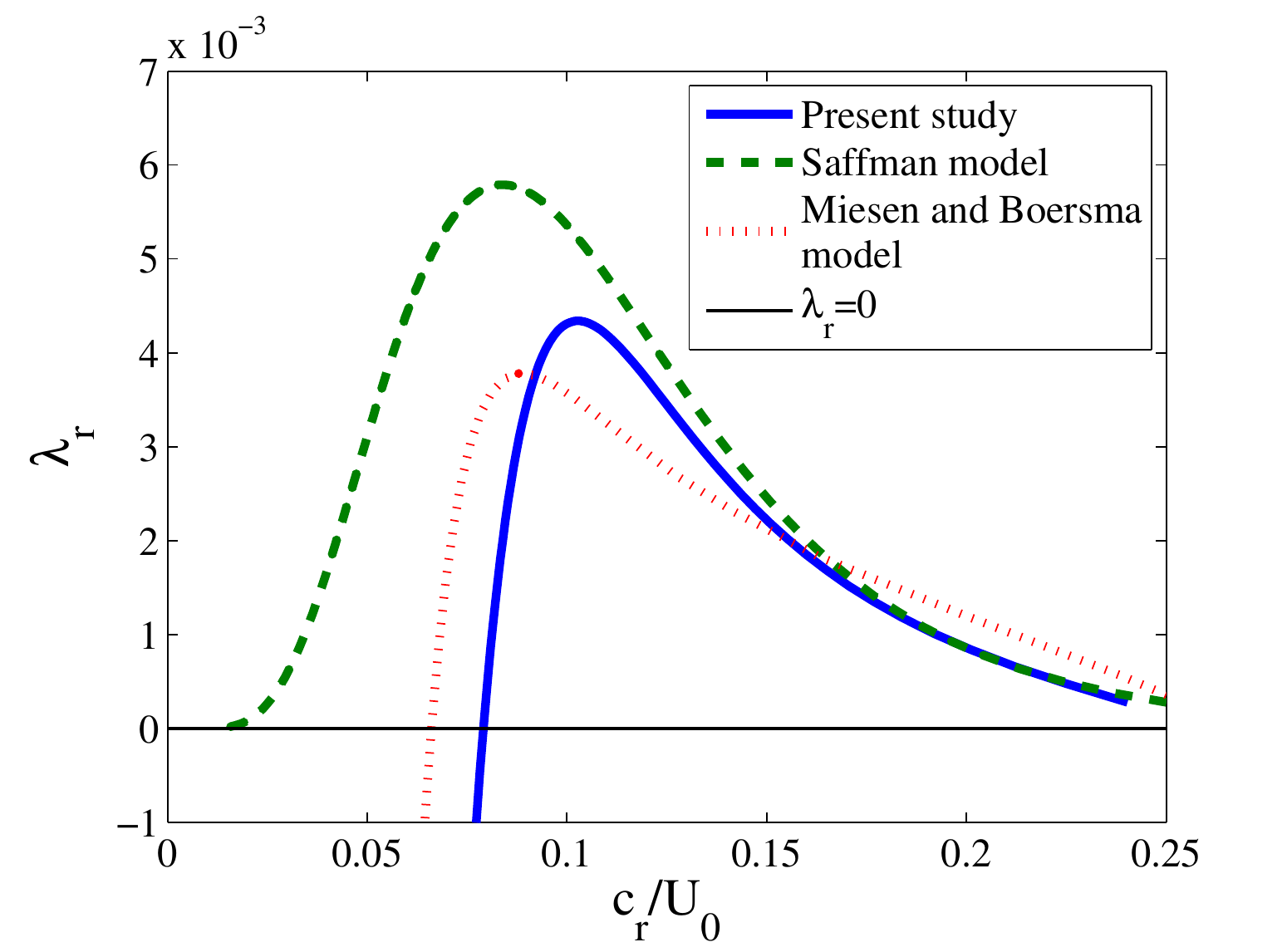}
\caption{Comparison between our work, the analytical approximation of Morland and Saffman~\cite{Morland1993}, and a reconstruction of the work of Boomkamp and Miesen~\cite{Boomkamp1996}.  There is good agreement between all three models, although the analytical formula breaks down at small wave speeds, where surface tension exerts a stabilizing influence on the growth rate.}
\label{fig:compare_miles}
\end{figure}

In the paper of Boomkamp and Miesen~\cite{Boomkamp1996}, the authors present one calculation of the wave Reynolds stress based on a boundary-layer gas profile and an exponential profile in the liquid.  We perform this calculation again and construct a dispersion curve over a range of $\alpha$-values.  This result is shown in Fig.~\ref{fig:inset_mb},
and the wave Reynolds stress calculation is given as an inset.  This calculation agrees with that given by Boomkamp and Miesen.  More interesting is the dispersion curve.  This is qualitatively similar to those given in Figs.~\ref{fig:growth_rate} and~\ref{fig:growth_rate_zoom}, although the gravity, surface-tension and Reynolds numbers are different.  Note in particular that the growth rate switches rapidly from positive to negative values at small $\alpha$-values.  This is not a finite-size effect, since it persists upon increasing the size of the computational domain and upon grid refinement.  Having confirmed a qualitative similarity between the present work and a reconstruction of that of Boomkamp and Miesen, we endeavour to produce a more exact comparison.  This we do by changing the parameter-values in the reconstructed work: specifically, we take $Fr=500$, $S=0$, $Re=10^5$, and normalize the base-state velocity such that $U_G\left(\tfrac{1}{2}\right)=\tfrac{1}{2}$.  This facilitates an accurate comparison between our work and that of Boomkamp and Miesen.  We present the comparison in Fig.~\ref{fig:compare_miles}, together with a comparison with the Miles formula.

Now the critical-layer theory of Miles~\cite{Miles1957} involves the solution of the Rayleigh equation with boundary
conditions at infinity, and at a wavy, impermeable wall, which is supposed
to represent the interface of two fluids with a large density contrast.
For this model problem, Morland and Saffman~\cite{Morland1993} have derived
an explicit but approximate formula for the growth rate in the case of
the exponential profile
\[
U=U_\infty\left(1-e^{-2z/\Delta}\right),
\]
where for comparison we take $U_\infty=U_0/2$, and $\Delta=Re/Re_*^2$.
Their approximate formula for the growth rate is
\begin{equation}
\lambda_{\mathrm{r}}=\frac{16\pi\alpha^2\Delta U_\infty^2}{rc_0\left(2+\alpha\Delta\right)^2\left(4+\alpha\Delta\right)^2}\left(1-\frac{c_0}{U_\infty}\right)^{2+\alpha\Delta},\qquad
c_0=\sqrt{\frac{g}{\alpha}}.
\label{eq:gr_rate_exp}
\end{equation}
We plot $\lambda_\mathrm{r}$ against $c_0$ for this simplified system, and
compare the result with the relationship we have obtained between $\lambda_\mathrm{r}$
and $c_{\mathrm{r}}$.  The results are shown in Fig.~\ref{fig:compare_miles}.
There is excellent agreement among the three curves shown, although
the analytical curve corresponding to the formula~\eqref{eq:gr_rate_exp} breaks down at small wave speeds, where surface tension stabilizes the system.
Nevertheless,
the form of curves in Fig.~\ref{fig:compare_miles} is the same in each case,
which, in addition to the energy budgets and Fig.~\ref{fig:wrs20}, provides
confirmation of the critical-layer nature of the instability.

There are also some DNS results available in this field for comparison.
In particular, we focus on the work of Sullivan \textit{et al.}~\cite{Sullivan1999} and compare our results
with DNS results found therein.  While a direct comparison is not possible,
since the work of Sullivan \textit{et al.} is for flow over a wavy wall,
there our some qualitative similarities between the two two studies.

Sullivan \textit{et al.} identify slow-wave and fast-wave cases.  Thus, we make a comparison between our Fig.~\ref{fig:velocity_thif60} and Fig.~17 in the work of Sullivan \textit{et al.}, which is for slow waves, and for a critical layer that that plays no role in the dynamics.  There is good qualitative agreement between these to sets of figures, although evidence of rapid distortion is absent in both cases.  Next, we compare a fast-wave scenario, namely our Figs.~\ref{fig:velocity10} and~\ref{fig:velocity20} with Fig. 19 in Sullivan \textit{et al.}.  Again, there is good qualitative agreement between these sets of figures, in particular for the phase relationships at the interface (the corresponding results for the phase relationships of the pressure field also show good agreement).  There is some evidence of rapid distortion in the secondary extrema in the streamwise velocity in Fig.~19~(a) of Sullivan \textit{et al.}, although this is not conclusive: this comparison implies that our model over-estimates this effect.  That said, these secondary oscillations cannot be reproduced by our linear theory if we neglect the PTS.

\subsection{Transition to the viscosity-contrast instability}
\label{pageref:duality}

\begin{figure}[htb]
\centering\noindent
\subfigure[]{
\includegraphics[width=0.45\textwidth]{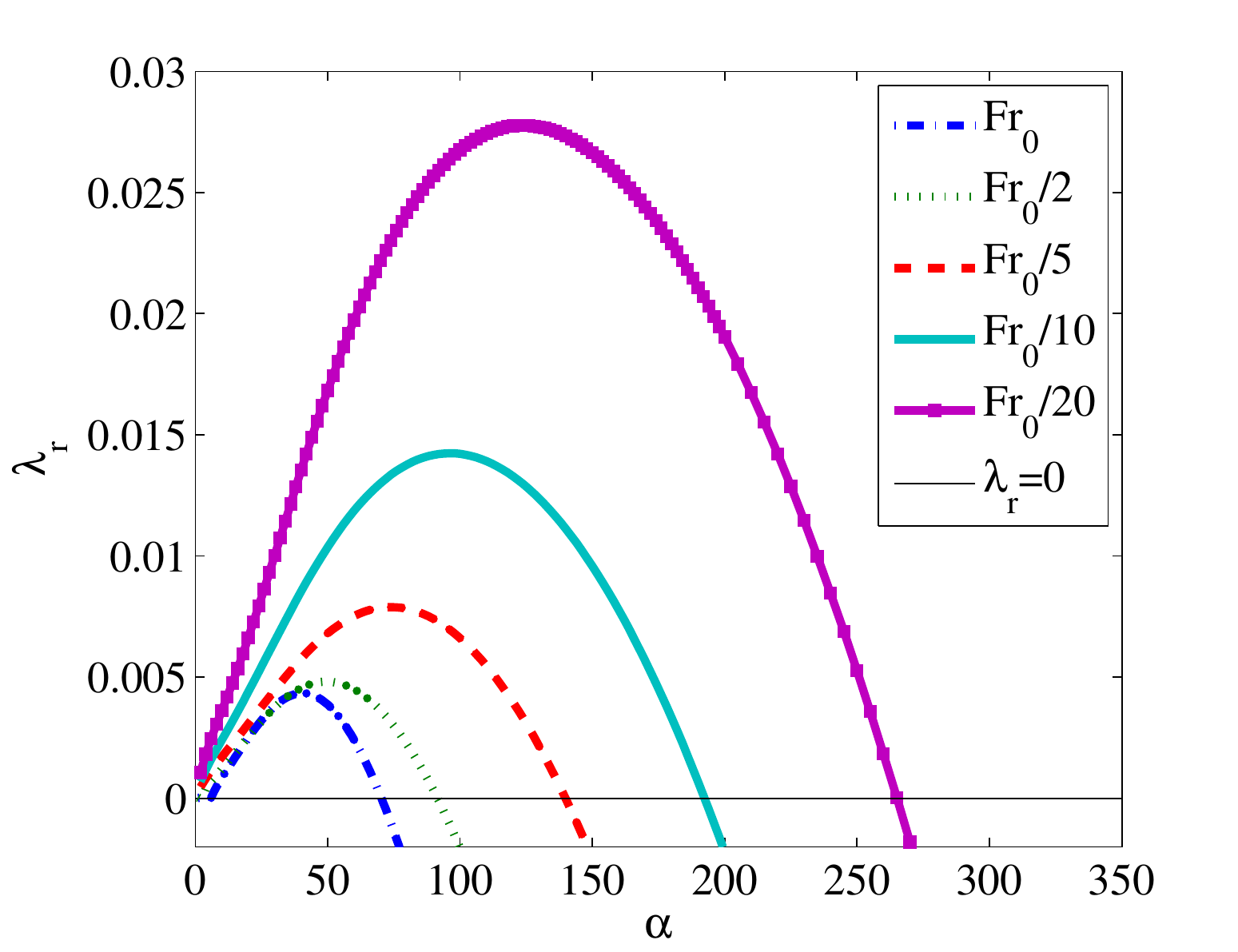}}
\subfigure[]{
\includegraphics[width=0.45\textwidth]{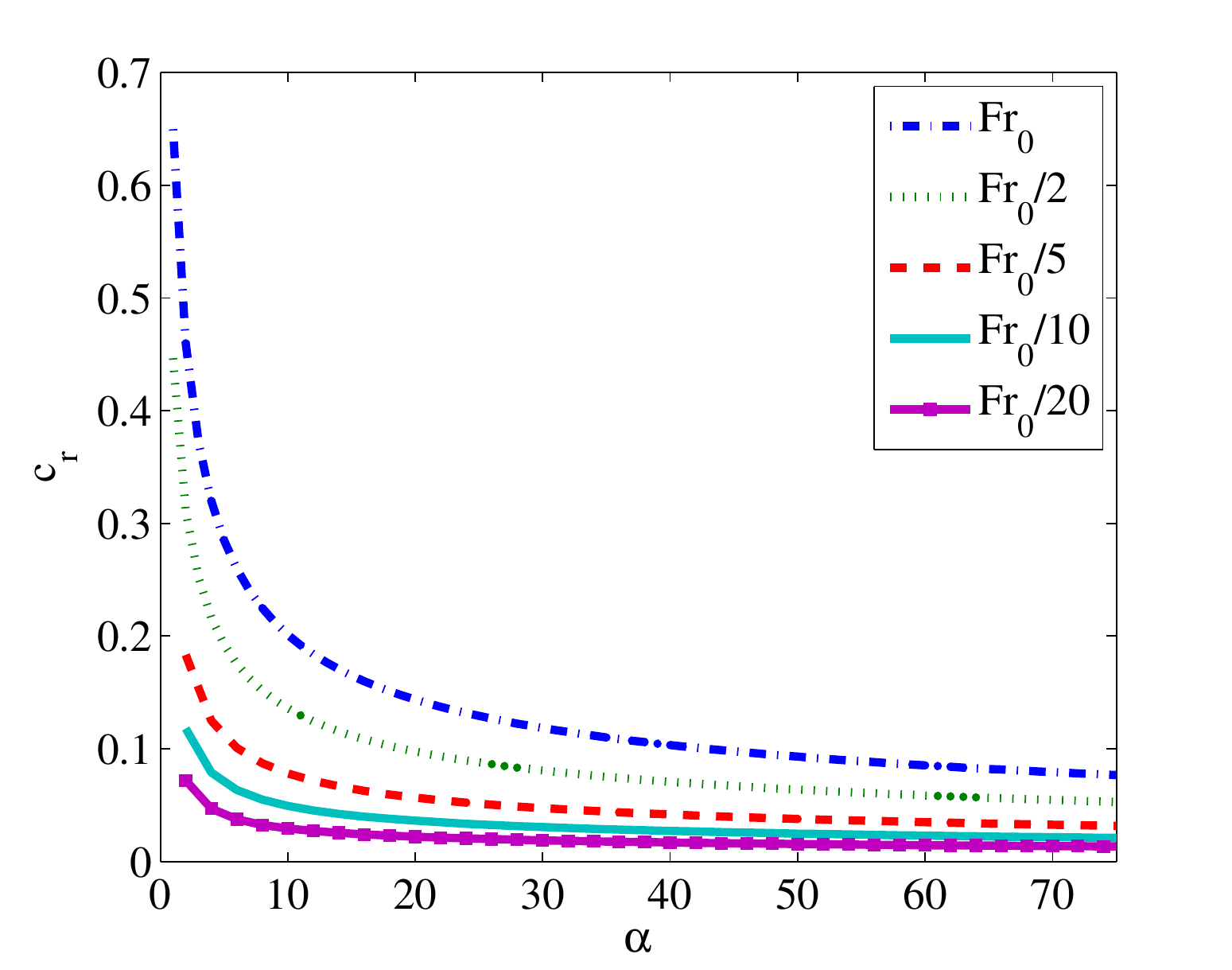}}
\caption{The growth rate and wave speed in the absence of the PTS, as a function
of gravity number, for $Fr_0=500$ and $Re=10^5$.  The instability changes character as the gravity number
is decreased.}
\label{fig:froude}
\end{figure}

We have investigated two types of interfacial instability, and the effects of the PTS thereon.  On the energy-budget
side, these waves are distinguished by the energy term that produces the
instability: either the viscosity-jump mechanism, or the critical-layer
mechanism.  On the turbulence side, they are distinguished by the wave speed:
the viscosity-jump waves are slow, while the critical-layer waves are fast.
 This separation of speeds leads to distinct properties with respect to the
 PTS.  We want to find a transition from
 one regime to another: this can be achieved by changing
 the gravity number $Fr$ (inverse Froude number).
We study the deep-water waves again.  We obtain the dispersion curve for
a range of gravity numbers, and at fixed Reynolds number, and reduce the gravity
number.  This can be achieved either by changing the degree of density stratification,
the gas-layer thickness, or by changing the shear velocity $U_0$:
\[
Fr=\frac{g\left(\rho_L-\rho_G\right)d_G}{\rho_GU_0^2}.
\]
The dispersion curves are shown as a function of gravity number, expressed
as a fraction of $Fr_0=500$ in Fig.~\ref{fig:froude}.
 The growth rate changes shape as the gravity number is decreased: the high-gravity
 number shape is similar to that observed in this section for critical-layer
 waves, while the low-gravity number shape is similar to that observed in
 Sec.~\ref{sec:thif} for viscosity-stratified waves.  The plot of wave speed
 in Fig.~\ref{fig:froude}~(b)
 indicates that the high-gravity number waves are fast while the low-gravity
 number waves are slow.  Going to larger $Fr$-values stabilizes the interface completely, although we have not shown this effect in Fig.~\ref{fig:froude}.
For each dispersion curve, we obtain the energy budget associated with maximum
growth, given in Tab.~\ref{tab:eb_transition}.
We observe a transition from critical-layer to viscosity-driven
instability, with decreasing gravity number.  The term $TAN$ represents
only a small positive contribution at high $Fr$-values, while at low $Fr$-values,
it is the dominant positive contribution.  The presence of the PTS does little to modify this partition of energy between the tangential and $REY_G$ terms.
%
%
% At high $Fr$-values, the ratio $TAN/REY_G$ shifts slightly (up to $6\%$ % at $Fr=Fr_0/2$) due to the PTS, since these stresses re-distribute the % energy of instability.  Thus,
% the transition is modified by the presence of the PTS, albeit marginally.
% At lower $Fr$-values, however, this shift is insignificant ($<1\%$).

This mechanism for modifying
the character of the instability could be applied to the thin-film case,
with one reservation.  To move the critical layer sufficiently far into the
bulk gas domain such that $U''\left(z_{\mathrm{c}}\right)$ is significant,
it is necessary to increase the Reynolds number.  This in turn will destabilize
the liquid through an internal mode, which necessitates the turbulent modelling
both of the base liquid flow, and of the liquid PTS, which is beyond the scope of the present work.  However, by naively retaining the linear profile in the liquid, we have observed a switch between critical-layer and viscosity-stratified waves in the thin-film case, through a suitable modification of the gravity and Reynolds numbers ($Re=10^6$, $Fr=0.1$, maximum growth rate at $\alpha=25$).
%
%
% The upshot of this is a sigmoidal shape for the
% liquid flow~\cite{Biberg2007} (rather than linear).  Once this rather trivial % modification
% has been effected, it is possible self-consistently to increase the Reynolds
% number, and to switch between critical-layer and viscosity-stratified waves
% in the thin-film case, through a suitable modification of the gravity and % Reynolds numbers ($Re=10^6$,
% $Fr=0.1$, maximum growth rate at $\alpha=25$).
%
%
%
%
%
%
%
%
%
%
\begin{table}[h!b!p!]
\centering
\begin{tabular}{|c||c|c|c|c|c|c|c|c|c|}
\hline
$Fr$&$REY_L$&$REY_G$&$DISS_L$&$DISS_G$&$TURB$&$NOR$&$INT$\\
\hline
\hline
$Fr_0$&0.00&1.75&-0.39&-1.10&0.00&-0.58&1.00\\
\hline
$\tfrac{1}{2}Fr_0$&-0.01&0.42&-0.32&-0.74&0.00&-0.17&1.00\\
\hline
$\tfrac{1}{5}Fr_0$&0.00&0.07&-0.19&-0.72&0.00&-0.04&1.00\\
\hline
$\tfrac{1}{10}Fr_0$&0.00&-0.02&-0.14&-0.74&0.00&-0.04&1.00\\
\hline
$\tfrac{1}{20}Fr_0$&0.00&-0.07&-0.09&-0.77&0.00&-0.05&1.00\\
\hline
\end{tabular}
\vskip 0.2in
\begin{tabular}{|c||c|c|c|c|c|c|c|c|c|}
\hline
$Fr$&$REY_L$&$REY_G$&$DISS_L$&$DISS_G$&$TURB$&$NOR$&$INT$\\
\hline
\hline
$Fr_0$&-0.02&1.79&-0.38&-1.09&0.02&-0.89&1.00\\
\hline
$\tfrac{1}{2}Fr_0$&-0.01&0.42&-0.32&-0.74&0.00&-0.23&1.00\\
\hline
$\tfrac{1}{5}Fr_0$&-0.01&0.07&-0.19&-0.72&0.00&-0.12&1.00\\
\hline
$\tfrac{1}{10}Fr_0$&0.00&-0.02&-0.14&-0.74&0.00&-0.07&1.00\\
\hline
$\tfrac{1}{20}Fr_0$&0.00&-0.07&-0.09&-0.77&0.00&-0.05&1.00\\
\hline
\end{tabular}
\caption{Energy budget detailing the transition from critical-layer to viscosity-stratified
waves, as a function of gravity number, where $Fr_0=500$ and $Re=10^5$.  The budgets have been normalized such that $TAN$=1 in each case.  In the first table, we have included the PTS; in the second table, they are set to zero.}
\label{tab:eb_transition}
\end{table}
%
%
% Finally, we note the effects of increasing the gravity number beyond $Fr_0=500$.  Fig.~\ref{fig:froude} shows that the % maximum growth rate decreases monotonically with increasing Froude number.  This is confirmed
% in Fig.~\ref{fig:froude1}, where we show the growth rate and wave speed at $Fr=1000$.  Increasing $Fr$ beyond
% this value % stabilizes the interface altogether.

\section{Conclusions}
\label{sec:conc}

We have investigated the stability of an interface separating a liquid layer
from a fully-developed turbulent gas flow.  The linear-stability analysis involves the study of the dynamics of a wave on the interface, and this
wave interacts with the turbulence and induces perturbation turbulent stresses
(PTS), which modify the stability properties of the system.   Using a separation-of-domains
technique, we derived a model for the PTS, based on the Orr--Sommerfeld equation
for the streamfunction.  We also developed a model of the base flow that
takes near-wall (interface) regions into account, and provides the friction
velocity $U_{*\mathrm{i}}$ as  a function of Reynolds number.

We have applied our model in two distinct cases.  For flow over a thin viscous
film, and at moderate values of the Reynolds number, the turbulence modelling does not materially affect the growth rate,
although the structure of the velocity field is modified: the streamwise
velocity develops streaks that extend into the bulk gas layer, as observed
in DNS~\cite{Sullivan1999}.  On the other hand, for deep-water waves and at high Reynolds numbers, the
maximum growth rate is shifted upwards by the PTS, and the flow structure
is again modified, especially at longer wavelengths, when the spatial extent
 of the streamfunction extends into the rapid-distortion domain.  The waves
 in the thin film and the deep channel differ are slow and fast respectively,
 compared with the shear velocity at the upper plate.  They can also be classified
 respectively as viscosity-stratified, or critical-layer waves.  The waves observed
 in both cases can, however, be brought into co-incidence by a modification
 of the gravity number.  By decreasing the gravity number in the deep-water
 waves, we have observed a transition from the critical-layer to the viscosity-stratified
 waves.  The transition has, effectively, been effected by modifying the
 shear velocity, the gas-layer depth, or the degree of stratification.  This
 suggests that a detailed parameter study will be useful in understanding the different
 mechanisms that generate instability, an approach we develop in \'O N\'araigh \textit{et al.}~\cite{Onaraigh2_2009}

\subsection*{Acknowledgements}

%This work has been undertaken within the Joint Project on Transient Multiphase
%Flows and Flow Assurance.  The authors wish to acknowledge the contributions
%made to this project by the UK Engineering and Physical Sciences Research
%Council (EPSRC), grant number EP/E021468/1, and the following: - Advantica;
%BP Exploration; CD-adapco; Chevron; ConocoPhillips; ENI; ExxonMobil; FEESA;
%IFP; Institutt for Energiteknikk; Norsk Hydro; PDVSA (INTERVEP); Petrobras;
%Scandpower PT; Shell; SINTEF; Statoil and TOTAL. The authors wish to express
%their sincere gratitude for this support.

This work has been undertaken within the Joint Project on Transient Multiphase Flows and Flow Assurance. The Authors wish to acknowledge the contributions made to this project by the UK Engineering and Physical Sciences Research Council (EPSRC) and the following: - Advantica; BP Exploration; CD-adapco; Chevron; ConocoPhillips; ENI; ExxonMobil; FEESA; IFP; Institutt for Energiteknikk; PDVSA (INTEVEP); Petrobras; PETRONAS; Scandpower PT; Shell; SINTEF; StatoilHydro and TOTAL. The Authors wish to express their sincere gratitude for this support.

L.\'O.N. would also like to thank K. Tong and M. Wong for their
assistance in carrying out the numerical studies.

% \bibliographystyle{unsrt}
% \bibliography{turbulence_bibliography}
% 

\end{document}